# Study and Development of a Data Acquisition & Control (DAQ) System Using TCP/Modbus Protocol

***Sourangsu Banerji***
*Department of Electronics & Communication Engineering,*
*RCC-Institute of Information Technology,*
*Under West Bengal University of Technology,*
***February, 2013.***

# PROJECT REPORT:

# STUDY AND DEVELOPMENT OF A DATA ACQUISITION & CONTROL (DAQ) SYSTEM USING TCP/MODBUS PROTOCOL

**SOURANGSU BANDYOPADHYAY,**
**DEPARTMENT OF ELECTRONICS & COMMUNICATION ENGINEERING**
**RCC-INSTITUTE OF INFORMATION TECHNOLOGY.**

**VARIABLE ENERGY CYCLOTRON CENTRE,**
**A UNIT OF BHABHA ATOMIC RESEARCH CENTRE,**
**DEPARTMENT OF ATOMIC ENERGY,**
**1/AF Bidhan Nagar, Kolkata-700064.**

# DECLARATION

I hereby declare that the project work entitled" **STUDY AND DEVELOPMENT OF A DATA ACQUISITION & CONTROL (DAQ) SYSTEM USING TCP/MODBUS PROTOCOL**" submitted to the Variable Energy Cyclotron Centre, Kolkata is a record of an original work done by me under the guidance of Mr. Tamal Bhattacharya, Head, Cryogenic Instrumentation Centre, VECC and this project work is submitted in the fulfillment of a winter project carried out from December 26th 2012 to January 25, 2013. The results embodied in this project report have not been submitted to any other University or Institute for the award of any degree or diploma.

Sourangsu Bandyopadhyay

# CERTIFICATE

This is to certify that Mr. Sourangsu Bandyopadhyay a student of Department of Electronics & Communication Engineering, RCC-Institute of Information Technology, has undergone a Project work titled " **STUDY AND DEVELOPMENT OF A DATA ACQUISITION & CONTROL (DAQ) SYSTEM USING TCP/MODBUS PROTOCOL**" from December 26th 2012 to January 25, 2013 under my supervision.

Signature of the Project Guide

# ACKNOWLEDGEMENT

Engineering is not only a theoretical study but it is a implementation of all we study for creating something new and making things more easy and useful through practical study. It is an art which can be gained with systematic study, observation and practice. In the college curriculum we usually get the theoretical knowledge. Along with a B.Tech. degree, it is necessary to construct a bridge between the educational life and industry. I consider myself extremely fortunate to obtain the opportunity to do a winter project at one of the prestigious institutions in India, Variable Energy Cyclotron Centre (VECC). No project, big or small can be accomplished without proper guidance and support of able mentors. So, I would like to take this opportunity to convey my sincere gratitude to all those who helped me to materialize this project.

Firstly, I would like to thank Mr. Tamal Bhattacharya for his invaluable suggestions and encouragement. Secondly, I would like to thank Mr. Tanmoy Das for his able guidance and help. This enabled me to bridge the gap between theoretical and practical knowledge. His suggestions at every step of my project will remain an asset for me. I would also like to thank Dr. P.Y. Nabhiraj for giving me the chance to do a winter project at this premier institution.

Lastly, I would also like to thank the entire staff of VECC for cooperating with us during this tenure.

# ABSTRACT

The aim of the project was to develop a HMI (Human-Machine Interface) with the help of which a person could remotely control and monitor the Vacuum measurement system. The Vacuum measurement system was constructed using a DAQ (Data Acquisition & Control) implementation instead of a PLC based implementation because of the cost involvement and complexity involved in deployment when only one basic parameter i.e. vacuum is required to be measured. The system is to be installed in the Superconducting Cyclotron section of VECC. The need for remote monitoring arises as during the operation of the K500 Superconducting Cyclotron, people are not allowed to enter within a certain specified range due to effective ion radiation. Using the designed software i.e. HMI the following objective of remote monitoring could be achieved effortlessly from any area which is in the safe zone. Moreover the software was designed in a way that data could be recorded real time and in an unmanned way. The hardware is also easy to setup and overcomes the complexity involved in interfacing a PLC with other hardware. The deployment time is also quite fast. Lastly, the practical results obtained showed an appreciable degree of accuracy of the system and friendliness with the user.

# Contents

# 1

# Vacuum Basics

## 1.1 Pressure

Pressure is defined in DIN Standard 1314 as the quotient of standardized force applied to a surface and the extent of this surface (force referenced to the surface area). Even though the Torr is no longer used as a unit for measuring pressure, it is nonetheless useful in the interest of "transparency" to mention this pressure unit: 1 Torr is that gas pressure which is able to raise a column of mercury by 1 mm at 0 °C. (Standard atmospheric pressure is 760 Torr or 760 mm Hg.) Pressure p can be more closely defined by way of subscripts:

### 1.1.1 Absolute pressure $P_{abs}$

Absolute pressure is always specified in vacuum technology so that the "abs" index can normally be omitted.

### 1.1.2 Total pressure $P_t$

The total pressure in a vessel is the sum of the partial pressures for all the gases and vapors within the vessel.

### 1.1.3 Partial pressure $P_i$

The partial pressure of a certain gas or vapor is the pressure which that gas or vapor would exert if it alone were present in the vessel.
*Important note:* Particularly in rough vacuum technology, partial pressure in a mix of gas and vapor is often understood to be the sum of the partial pressures for all the non-condensable components present in the mix – in case of the "**partial ultimate pressure**" at a rotary vane pump, for example.

### 1.1.4 Saturation vapor pressure $P_s$

The pressure of the saturated vapor is referred to as saturation vapor pressure ps. ps will be a function of temperature for any given substance.

### 1.1.5 Vapor pressure $P_d$

Partial pressure of those vapors which can be liquefied at the temperature of liquid nitrogen (LN2).

### 1.1.6 Standard pressure $P_n$

Standard pressure Pn is defined in DIN 1343 as a pressure of Pn = 1013.25 mbar.

**1.1.7 Ultimate pressure $P_{end}$**
The lowest pressure which can be achieved in a vacuum vessel. The so called ultimate pressure pend depends not only on the pump's suction speed but also upon the vapor pressure pd for the lubricants, sealants and propellants used in the pump. If a container is evacuated simply with an oil-sealed rotary (positive displacement) vacuum pump, then the ultimate pressure which can be attained will be determined primarily by the vapor pressure of the pump oil being used and, depending on the cleanliness of the vessel, also on the vapors released from the vessel walls and, of course, on the leak tightness of the vacuum vessel itself.

**1.1.8 Ambient pressure $P_{amb}$** or **(absolute) atmospheric pressure**
**Overpressure $P_e$ or gauge pressure**
(Index symbol from "excess")

$$Pe = Pabs – Pamb$$

Here positive values for Pe will indicate overpressure or gauge pressure; negative values will characterize a vacuum.

**1.1.9 Working pressure Pw**
During evacuation the gases and/or vapors are removed from a vessel.
Gases are understood to be matter in a gaseous state which will not, however, condense at working or operating temperature. Vapor is also matter in a gaseous state but it may be liquefied at prevailing temperatures by increasing pressure. Finally, saturated vapor is matter which at the prevailing temperature is gas in equilibrium with the liquid phase of the same substance. A strict differentiation between gases and vapors will be made in the comments which follow only where necessary for complete understanding.

**1.1.10 Particle number density n** (cm-3)
According to the kinetic gas theory the number n of the gas molecules, referenced to the volume, is dependent on pressure p and thermodynamic temperature T as expressed in the following:

$$p = n \cdot k \cdot T \quad (1.1)$$

n = particle number density
k = Boltzmann's constant

At a certain temperature, therefore, the pressure exerted by a gas depends only on the particle number density and not on the nature of the gas. The nature of a gaseous particle is characterized, among other factors, by its mass mT.

**1.1.11 Gas density** ρ (kg · m-3, g · cm-3)
The product of the particle number density n and the particle mass mT is the gas density ρ:

$$\rho = n \cdot mT \quad (1.2)$$

**1.1.12 The ideal gas law**
The relationship between the mass mT of a gas molecule and the molar mass M of this gas is as follows:

$$M = NA \cdot mT$$

# 1.2 Vacuum
Vacuum, simply defined, is a volume devoid of matter. The vacuums achieved in "vacuum systems" used in physics and in the electronics industry are far from being absolutely empty inside, however. Even at the limits of pumping technology, there are hundreds of molecules in each cubic centimeter of volume. Still, compared to the atmospheric density of 2.5 × 10^19 molecules/cm3, the relative crowding is much less in the vacuum system! Vacuum is officially defined by the American Vacuum Society as a volume filled with gas at any pressure under atmospheric. For purposes of interesting physics, the "real" vacuum range does not begin until about 1/1000 of an atmosphere.

Vacuum technology is an extremely important tool for many areas of physics. In condensed matter physics and materials science, vacuum systems are used in many surface processing steps. Without the vacuum, such processes as sputtering, evaporative metal deposition, ion beam implantation, and electron beam lithography would be impossible. A high vacuum is required in particle accelerators, from the cyclotrons used to create radio nuclides in hospitals up to the gigantic high-energy physics colliders such as the LHC. Vacuum systems are also used in many precision physics applications where the scattering and forces induced by the atmosphere would introduce a major background to precise measurements.

### 1.2.1 Vacuum Units

The vacuum field is plagued by a super-sufficiency of units for measuring pressure, and it is worth a moment's study to familiarize oneself with the various units used in different texts and manuals. For the purposes of the class, we will use primarily the SI unit of pressure, the Pascal. Another widely used unit is the Torr or millimeter of mercury, based on the traditional mercury manometer technique for measuring pressure. One atmosphere is 1.01 × 10^5 Pa or 760 Torr. Other units and their inter-conversions are given in Table 1.

| $X$ \ $Y$ | Pascal | Torr | atm | bar | psi |
|---|---|---|---|---|---|
| Pascal | 1 | 0.0075 | $9.87\times10^{-6}$ | $1\times10^{-5}$ | $1.45\times10^{-4}$ |
| Torr | 133.3 | 1 | 0.00132 | 0.00133 | 0.0193 |
| atm | $1.01\times10^{5}$ | 760 | 1 | 1.01 | 14.7 |
| bar | $1\times10^{5}$ | 750 | 0.987 | 1 | 14.5 |
| psi | 6895 | 51.7 | 0.068 | 0.069 | 1 |

Table 1: Conversion factors between various system of pressure units. (1X = nY )

The terms "high vacuum", "ultrahigh vacuum", "low vacuum" and the like are often used in vacuum textbooks and laboratory discussions. The borders between the regions are somewhat arbitrary, but a general guideline is given in Table 2.

| Vacuum range | Pressure Range (Pa) | | |
|---|---|---|---|
| Low | $10^{5}$ | – | $3.3\times10^{3}$ |
| Medium | $3.3\times10^{3}$ | – | $10^{-1}$ |
| High | $10^{-1}$ | – | $10^{-4}$ |
| Very high | $10^{-4}$ | – | $10^{-7}$ |
| Ultra high | $10^{-7}$ | – | |

Table 2: Vacuum ranges, as given in Dictionary for Vacuum Science and Technology.

## 1.3 Theory of Gas at Low Pressure

In beginning physics and chemistry classes, ideal gases are introduced with the ideal gas equation

$$pV = NRT$$

This equation explains much of the behavior of gas at low pressure. For example, the basis of the most simple pressure gauge depends on gases at equal temperature so that the relation p1V1 = p2V2 holds for two volumes sharing a movable but impermeable wall (such as a column of mercury). We can derive a number of additional useful results by considering a kinetic theory of the gas, where the molecules are taken independently.

## 1.4 Kinetic Theory of Gases

The underlying assumption of the kinetic theory is that gases are composed of molecules which are in constant motion. This motion is directly related to the temperature of the gas. In their motion, the molecules may collide with the walls of the container and thereby transfer momentum to the walls. This momentum transfer can be averaged over many collisions and observed as the pressure of the gas.
Consider just one molecule bouncing between the walls of a container. The molecule has a mass m and is traveling with a velocity v along the container. All collisions are considered to be elastic, so the transfer of momentum during each collision is

$$(mv) = mv - m(-v) = 2mv$$

Since the container is of length L, the particle will travel from one end to the other in a time L/v at which point it will transfer 2mv of momentum to the far wall. If we assume the walls to have an area A, then the force F on the two end walls of the box is

$$\begin{aligned} F &= \frac{\Delta p}{\Delta t} = \frac{2mv^2}{L} \\ p &= \frac{F}{2A} = \frac{mv^2}{LA} \\ &= \frac{mv^2}{V} \end{aligned}$$

If we now consider N particles in the box and allow them to hit any side, the pressure on a face should be

$$p = N\frac{m\overline{v_\perp^2}}{V}$$

the average v^2 of a particle perpendicular to the surface. Since the motion is random, there is no distinguishing between the directions, and by the Pythagorean Theorem,

$$\overline{v^2} = \overline{v_x^2} + \overline{v_y^2} + \overline{v_z^2} = 3\overline{v_\perp^2}$$

A second way of writing the ideal gas equation is p = nkBT, where kB is the Boltzmann constant and the density n = N/V. Thus

$$\begin{aligned} p &= \frac{n}{3} m\overline{v^2} \\ nk_B T &= \frac{n}{3} m\overline{v^2} \\ 3k_B T &= m\overline{v^2} \end{aligned}$$

Now, if we can express the average kinetic energy of a gas molecule E¯ as ½*(mv^¯2), just as for any kinetic energy. If we combine this expression for the average kinetic energy with equation 1, we obtain

$$\bar{E} = \frac{3}{2} k_B T$$

This shows that the average kinetic energy for all molecules in a gas is the same and it is directly related to the temperature.

### 1.4.1 Velocity Distribution

Despite the fact that the average kinetic energy is the same for all molecules, these molecules do collide and exchange energy and achieve a distribution of velocities away from the average. The derivation of the velocity distribution is complicated, but the result of Maxwell and Boltzmann (the Maxwell-Boltzmann velocity distribution) is

$$\frac{1}{n}\frac{\mathrm{d}n}{\mathrm{d}v} = f_v = \frac{4}{\sqrt{\pi}}\left(\frac{m}{2k_B T}\right)^{\frac{3}{2}} v^2 \exp\left(\frac{-mv^2}{2k_B T}\right) \quad (2)$$

The maximum of this distribution (shown in Figure 1) is at $v_p = \sqrt{\frac{2k_B T}{m}}$, while the average velocity[1] is obtained by integrating

$$\bar{v} = \frac{\int_0^\infty v f_v \,\mathrm{d}v}{\int_0^\infty f_v \,\mathrm{d}v} = \frac{2}{\sqrt{\pi}} v_p \approx 1.128 v_p \quad (3)$$

---

[1] $v_p$ is not the same as the square root of the average of $v^2$! That value is the RMS average ($\frac{1}{N}\sum v^2$ instead of $\left(\frac{1}{N}\sum v\right)^2$) and can be obtained directly from kinetic theory as $\sqrt{\frac{3k_B T}{m}}$.

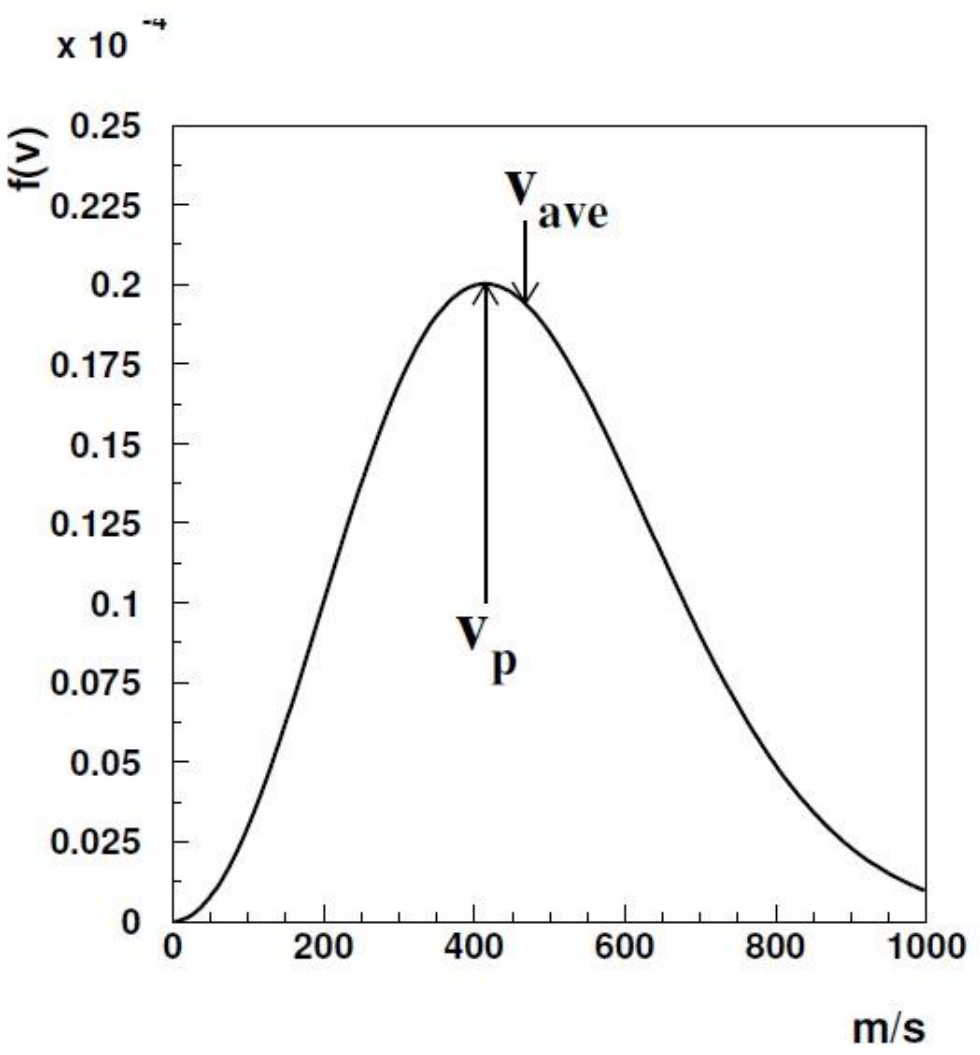

Figure 1 Maxwell Boltzmann distribution. The most probable velocity is Vp, while Vave is the average velocity.

This is the value which should be used for calculating transport properties, though not for kinetic energy purposes.

If one is interested in the incidence rate on any given surface, one should consider the one-dimensional velocity distribution, rather than the full three dimensional version. Carrying this calculation through, we obtain an average surface flux of

$$\phi = \frac{n}{2\sqrt{\pi}}\sqrt{\frac{2k_BT}{m}} = \frac{n\bar{v}}{4}$$

## 1.4.2 Mean Free Path

Since the molecules suffer collisions with each other, there is a distance which each molecule travels between collisions. We would expect this mean free path to be a function of the density, the mean velocity, and the size of the molecules themselves. We model the molecules as spheres with diameter d. If we consider all molecules except one to be fixed, we see that the molecule travels a distance v t in a time t. If its center moves within d of any other molecule, the two will collide, so the molecule sweeps out a free volume of d^2v t. If we have a molecular density n, the space for any given molecule is 1/v so this is the free volume a molecule should on average sweep out before colliding. This yields the result

$$\frac{1}{n} = \pi d^2 v\tau$$

where $\tau$ is the mean time between collisions and the mean free path $\lambda$ is

$$\lambda = v\tau = \frac{1}{n\pi d^2} = \frac{k_B T}{\pi d^2 p}$$

We have ignored the motion of the other molecules in this process. Their motion complicates the calculation, but if we apply the Maxwell-Boltzmann velocity distribution(in a non-trivial calculation), we obtain just a factor of $\frac{1}{\sqrt{2}}$, so

$$\lambda = \frac{k_B T}{\sqrt{2}\pi d^2 p}$$

For quick reference, we can apply the values for nitrogen (air) at room temperature and we find that

$$\lambda_{\text{air,300K}} = \frac{6.7}{P}\ \text{mm} \cdot \text{Pa}$$

So we see that at atmospheric pressure, the mean free path is roughly 66 nm, while at 10−3 Pa, the mean free path is 6.7 meters!

Fundamentally, many interesting applications of vacuum technology depend on the mean free path being much larger than other dimensions in the system. For example, evaporative deposition of metal films onto substrates is efficient only when the mean free path is significantly larger than the distance between the evaporator element and the substrate – often a distance on the order of 10 centimeters. The Knudsen number is a dimensionless number which captures this concept:

$$K_n = \frac{\lambda}{d}.$$

Therefore, in a situation with $K_n \ll 1$, a gas molecule will suffer many collisions in transport between two walls of the container, while when $K_n \gg 1$, the collisions between gas molecules can be neglected. The regime where $K_n \gg 1$ is sometimes called the *molecular flow* regime, while $K_n \ll 1$ is sometimes called the *viscous flow* regime.

### 1.4.3 Surface Interactions

An infinite volume filled with gas is not very interesting, and in any case it would take very long time to pump down! In a real vacuum system, there are many solid surfaces: chamber walls, vacuum pump parts, and research samples for example. These surfaces can interact with the volume of the system in several ways which include adsorption, absorption, and evaporation.

The gas molecules within a vacuum chamber will frequently impact on the surfaces. When a gas molecule impacts a surface it may “reflect” off the surface elastically, but often it will briefly adhere to the surface because of van der Waals forces and then escape again. In these cases, the escaping gas molecule may go in any direction and will have a thermal energy characteristic of the temperature of the surface. Gas molecules may remain on the surface for a long period of time, in which case they are considered to be adsorbed on the surface. The gradual release of adsorbed gas (“outgassing”) is a major issue in high vacuum systems, but the application of high temperatures (> 400◦ C) to the chamber will greatly increase the desorption rate. Therefore, high vacuum systems are generally “baked-out” at high temperature and moderate vacuum for several days before first use.

Absorption is the diffusion of gas molecules into a solid material. Once a surface is clean from adsorbed gas, the diffusion of the absorbed gas back to the surface becomes the major inflow. The rate of diffusion can also be increased by high temperatures and the baking-out process aids in the removal of much absorbed gas as well. Fundamentally, a perfectly sealed vacuum system is limited in pressure by the diffusion of gas from the out- side surface to the inside surface – a process called permeation. This process is extremely slow for most gasses except hydrogen, but even hydrogen is rarely a limiting factor until pressures of 10^−9 Pa are reached

A more important source of gas in the system can be the surface materials themselves if they have vapor pressures which are comparable to the pressures in the vacuum system. For this reason, only certain types of materials should be used for vacuum system elements. The most commonly-used materials are stainless steel, aluminum, and glass.

## 1.4.4 Viscosity

If two surfaces are moving parallel to each other in a fluid, the fluid can transfer force between the two surfaces. If we consider two planes separated by a distance y, one fixed and the other moving with a speed Ux, then the viscosity ( ) is defined as

$$\frac{F}{A} = \eta \frac{u}{y}$$

for the fluid forces.

In the regime where $K_n \ll 1$, the momentum transfer takes place from layer to layer of the gas molecules before transferring to the stationary plate. An analysis of this using kinetic theory yields a value

$$\eta = 0.4999 nm\bar{v}\lambda.$$

If we substitute the expressions for $\lambda$ and $v$, we obtain

$$\eta = \frac{0.4999\sqrt{4mkT}}{\pi^{3/2}d^2}.$$

We notice that this expression is independent of pressure or density.

In the regime where Kn >> 1, the viscous force does not depend on the spacing between the plates, but rather the number of molecules available to carry momentum from one plate to the other. Remembering the microscopic view of gas-surface interactions, the gas molecules will briefly adhere to the surface of each plate, coming to rest, and then escape again with a random velocity vector relative to the plate. Each pair of collisions will transfer a small amount of momentum from one plate to the other, so the viscous force thus depends on the flux of gas molecules landing on the surfaces and their mass. The viscous force is thus changed to

$$\begin{aligned}\frac{F}{A} &= m\phi\frac{u_x}{\beta}\\ &= \frac{nm\bar{v}}{4}\frac{u_x}{\beta}\\ &= p\sqrt{\frac{m}{2\pi kT}}\frac{u_x}{\beta}\end{aligned}$$

Where B (read beta) is related to the slipping of gas molecules on the surface and is approximately unity for most gas-material pairs. We notice that, in the molecular flow regime, the viscous force is proportional to pressure and this principle has been used to construct pressure gauges.

## 1.4.5 Thermal Conductivity

The conduction of heat (thermal energy) between two parallel plates with a temperature difference T requires the same analysis as the transfer of momentum which gave the viscous force. For high pressures, the result is

$$\frac{H}{A} = \frac{9\gamma-5}{4}\eta c_v\frac{\Delta T}{y}$$

where $c_v$ is the gas specific heat at constant volume and $\gamma = \frac{c_p}{c_v}$. As in the viscous force case, this is independent of pressure.

At low pressures, the thermal conductivity of the gas begins to depend on the number of molecules available to carry the heat, so

$$\frac{H}{A} = \alpha K p \Delta T$$

where $\alpha$ measures how efficiently the gas molecules are thermalized on each surface and $K$ depends on the gas molecule velocity and heat capacity. As we will see later, these are often bundled into a single proportionality constant.

## 1.4.6 Flow

A simple vacuum system is shown in Figure 2. A working chamber V is connected by way of a pipe with diameter d to a pump. In designing a system, it is important to know how the various pipe and pump parameters will affect the behavior of the system. In particular, it is useful to understand the flow of gas through the system and how quickly the system should reach a specified pressure.

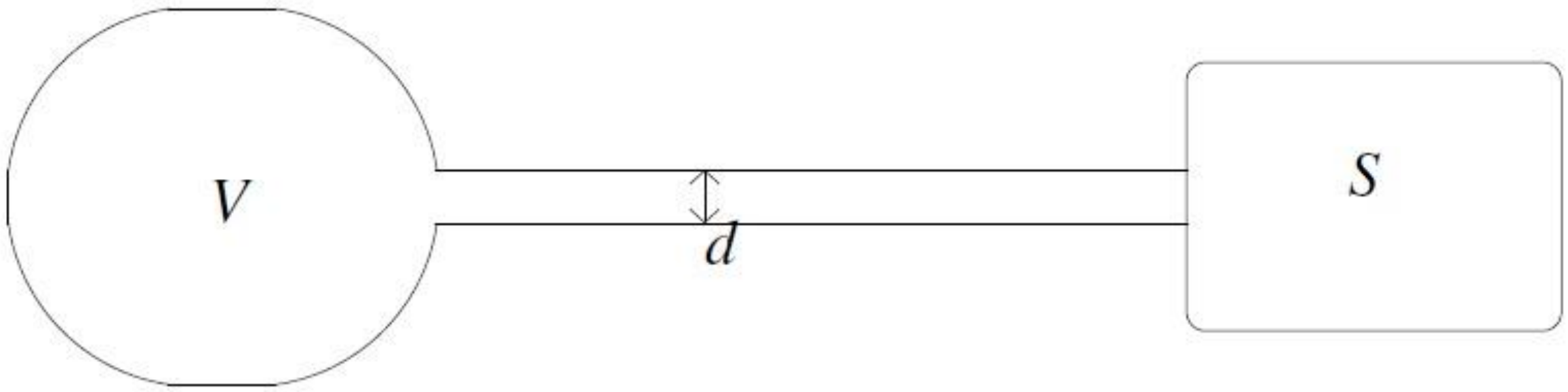

Figure 2 Simple vacuum system with a working chamber connected by a long pipe to a single pump.

Flow in a vacuum system can be modeled in a very similar way to electrical circuits, subject to some care. Instead of resistance, the reciprocal concept (conductance) is used and the voltage difference is replaced by the pressure difference. Ohm's law thus becomes

$$C = \frac{Q}{p_2 - p_1}.$$

The units of conductance are generally L/s or m^3/s.

We can combine conductance in many cases to create a global conductance, although care must be taken that the elements being connected do not have internal structure which, taken together, would limit the flow more than when taken separately. Assuming we are considering simple pipes, a series of pipes will have a conductance

$$\frac{1}{C_T} = \frac{1}{C_1} + \frac{1}{C_2} + \frac{1}{C_3} + \dots$$

while parallel pipes connecting the same volumes will have a net conductance

$$C_T = C_1 + C_2 + C_3 + \dots$$

General conductance can be quite difficult to calculate, and often Monte Carlo methods must be used for complex geometries. However, for long straight pipes the conductance can be determined in a straightforward way. In the viscous regime, a long pipe (` > 20d) will have a conductance

$$C(\mathrm{L/s}) = \frac{\pi}{128\eta} \frac{d^4}{\ell} \frac{p_1 + p_2}{2}$$

where the constant of proportionality is ≈ 1.38 × 10^6 for air at 0◦ C. In the molecular flow regime, a long tube will have a conductance of

$$C = \frac{\pi}{12}\bar{v}\frac{d^3}{\ell}$$

while a thin aperture's conductance is simply the flux divided by the density times the area of the aperture

$$C = \frac{\bar{v}}{4}A.$$

The conductance of short pipes will fall between these two limits.
A pump also has an effective conductance, which is generally called its pumping speed S:

$$S = \frac{Q}{p}$$

where the pressure is measured at the pump's inlet.
The conductance of the piping and the pumping speed can be combined to determine the rate at which the system can be pumped down. If we assume the system to be at constant temperature

$$\begin{aligned} Q &= c_v T \frac{dN}{dt} \\ &= c_v V \frac{dp_1}{dt} \\ Q &= C(p_2 - p_1) \\ \frac{dp_1}{dt} &= \frac{C}{c_v V}(p_2 - p_1) \end{aligned}$$

If the conductance of the pipe is less than the pumping speed at the, then pressure at the pump inlet p2 to be constant, then

$$p_1 = p_2 + (p_1 - p_2)e^{-\frac{C}{c_v V}t}$$

so the pressure in the chamber decreases exponentially towards the pump's inlet pressure. If the pumping speed is less than the conductance of the pipe, p2 will also be a function of time and a system of coupled differential equations is the result.

# 2

# Vacuum Measurement System

A vacuum system typically consists of one or more pumps which are connected to a chamber. The former produces the vacuum; the latter contains whatever apparatus requires the use of the vacuum. In between the two may be various combinations of tubing, fittings and valves. These are required for the system to operate but each introduces other complications such as leaks, additional surface area for outgassing and added resistance to the flow of gas from the chamber to the pumps. Additionally, one or more vacuum gauges are usually connected to the system to monitor pressure.

## 2.1 Vacuum Terminology

The language of vacuum is extensive and what follows only covers the bare minimum. However, these are the terms and concepts that will be found to be the most valuable to the beginning vacuum experimenter. Understand these and you will be off to a good start.

### 2.1.1 Mean Free Path

Reduction in pressure results in a lower density of gas molecules. Given a certain average velocity for each constituent molecule of air at a given temperature (at room temperature this is about 1673 km/hr) an average molecule will travel a certain distance before it interacts (collides) with another at any given pressure. This average distance between collisions is the mean free path. At 1 Torr in air this distance is about 0.005 cm, a value that scales directly with pressure. Thus the mean free path would be 5 cm at 1 mTorr and 50 meters at 0.001 mTorr. The lengthening of mean free path at low pressures is a key enabler for devices such as vacuum tubes and particle accelerators as well as for processes such as vacuum coating where microscopic particles such as electrons, ions or molecules must traverse considerable distances with minimal interference.

### 2.1.2 Flow

Gases at very low pressures behave very differently from gases at normal pressures. As a reduction in pressure occurs in a vacuum system, the gas in the system will pass through several flow regimes. At higher pressures the gas is in viscous flow where the gas behaves much like a liquid. Viscous flow includes turbulent flow, where the flow is irregular, and laminar, where the flow is regular with no eddies. Moving deeper into the vacuum environment, Knudsen or transition flow occurs when the mean free path is greater than about one-hundredth of the diameter of the tubing. Full molecular flow, where molecules behave independently, begins when the mean free path exceeds the tubing diameter. Which flow regime the gas is in is dependent upon several factors including tube diameter and pumping speed.

To summarize, when the ratio of the average mean free path in a tube to the radius of the tube is less than 0.01, the flow is viscous. When the ratio is greater than 1.00 the flow is molecular. Transition (or Knudsen) flow exists between the viscous and molecular flow regimes and we have a behavior that is bit of both. One of the factors which determine pump applicability is the flow regime it needs to operate in. Mechanical pumps are not effective in the molecular region whereas diffusion pumps are.

## 2.1.3 Backstreaming

It is always hoped that the flow of gas and vapor in a vacuum system is away from the chamber, through the pump, and out to the atmosphere. However, this is not the case in molecular flow where molecules behave as individuals with some of them going against the main flow direction. This is not a good situation to have when there are undesirable things downstream of the chamber (like pump oil) that we would prefer not to have get into the experimental area. This is one reason why diffusion pumps always have some sort of baffle or trap - otherwise fairly large quantities of oil vapor will migrate out of the pump and into the chamber.

## 2.1.4 Pumping Speed and Throughput

The speed of a pump is the volume of gas flow across the cross section of the tubing per unit time. The standard units are liters/second. Since the density of a gas changes with pressure (i.e. the mass or number of molecules of gas in a given volume) an important measure is mass flow or throughput which is the product of pressure and speed with the units of Torr-liters/second. If you think of the vacuum system as an electrical circuit, throughput is like current flow and it is constant everywhere in the circuit. The various elements of the system (lines and pumps) are analogous to resistances except instead of voltage drops there are pressure differentials. In putting together a vacuum system you want minimal pressure differentials in the connecting lines and maximum throughput everywhere.

A simple example will pull this together. Consider a small diffusion pump that has a rated inlet speed of 100 liters/second at 0.0001 Torr (0.1 mTorr). The throughput would be 100 x 0.0001 or 0.01 Torr-liters/sec. Now, connected to the outlet of the diffusion pump we have a mechanical forepump which is capable of maintaining a pressure of 0.1 Torr. Given the fact that throughput at the diffusion pump inlet must equal throughput at the outlet and that there is a pressure of 0.1 Torr at that outlet, the minimum speed of the forepump must be 0.1 liters/sec, a speed easily met by even very small mechanical pumps. On the other hand, if the diffusion pump inlet pressure is 0.01 Torr (10 mTorr) - say just after the pump is started or if it is working against a very gassy load - the forepump would have to have a speed of 10 liters/sec to allow the diffusion pump to work at full speed. This would be a large pump.

To summarize all of this, at high diffusion pump inlet pressures, the speed most likely will be constrained by the speed of the forepump. At low inlet pressures there is so little mass flow that a very small forepump can keep pace with even a large high vacuum pump. In fact, in a tight system you can shut off the forepump once a low enough pressure has been reached simply because so little mass remains in the system.

## 2.1.5 Conductance of Tubing

As mentioned above, the tubing in a vacuum system can represent a significant resistance. When one end of a tube is connected to a pump, that end of the tube will have a higher pumping speed than will the other end. For viscous flow, as would be the nominal case for roughing lines (i.e. mechanically pumped), the conductance, C, is dependent upon gas pressure and viscosity and, at room temperature and air, is (for a tube diameter of D cm, length of L cm and at an average pressure of P Torr):

$$C = 180 \times D^4/L \times P \text{ (liters/sec)}$$

An example would be a foreline of 2 cm diameter and 60 cm long. At one end is a venerable CencoMegavac pump; the other end is connected to the outlet of a diffusion pump. Referring to the manufacturer's literature for the pump we find that the pumping speed of the roughing pump is 0.5 liter/sec at 100 mTorr, the maximum recommended foreline pressure of the diffusion pump. Plugging in the numbers, we find that the line conductance is 4.8 liters/sec. Thus, the line is not limiting the capabilities of the forepump.

Interestingly, pressure is not a factor in the molecular flow regime where, for example, a diffusion pump would operate. Here we have:

$$C = 12 \times D^3/L \text{ (liters/sec)}$$

An example here would be a 2 inch (5 cm) diffusion pump which has a specified inlet pumping speed of 100 liters/sec. The pump is connected to a small experiment chamber through 60 cm of 2.5 cm diameter tubing. Inserting the numbers, we find a line conductance of only 3.1 liters/sec. This may be adequate for the small chamber but it certainly throttles the pump significantly. If a 5 cm line were substituted (same length) the conductance would rise to 25 liters/sec. In either case, the most important thing to bear in mind is that conductance is strongly influenced by the tube diameter. 1 cm to the third or fourth power is a whole lot less than 3 cm to the same powers. The bottom line is: go for fat tubes, and keep them short, particularly in high vacuum lines.

### 2.1.6 Outgassing and Vapor Pressure.

Assuming that a system is tight, as the pressure gets lower most of the load is from gases evolving from the surfaces of the materials in the system. This becomes significant below pressures of around 100 mTorr. Outgassing will be the main limiting factor with regard to the ultimate pressure which any particular system may reach, assuming that leaks are absent. Leaks may be either real leaks, like holes in the chamber, or virtual leaks that are caused by gas escaping from, for example, screw threads within the system or porous surfaces that contain volatile materials. The level of outgassing is reduced by keeping the system clean and dry and with a proper selection of materials. If the construction of a system is appropriate to the practice, adsorbed layers of water vapor and other gases may be evolved by heating the system in an oven or with a hot air gun to a temperature of at least 150 °C and usually more. For most of the applications that we will be discussing this level of cleaning is not required. However, the system components should be kept clean (no fingerprints or other grime), dry and, as much as possible, sealed off from room air (a major source of moisture).

Related to outgassing are the vapor pressures of the materials used in the system. All materials evolve vapors of their constituent parts and these vapors will add to the gas load in a system. Water is the worst commonly encountered material and is a good example of what vapor pressure means. At 100 °C, the vapor pressure of water is 1 atmosphere (760 Torr). Under those circumstances, when the vapor pressure is equal to the surrounding pressure, we know what happens - the water boils. At room temperature, the vapor pressure of water drops to 17.5 Torr and it will boil at that pressure. Water is not a good material to have in high vacuum systems. Other materials having high vapor pressures include some plastics, particularly those with volatile plasticizers, and metals such as mercury, lead, zinc and cadmium. Low vapor pressure materials include glass, copper, aluminum, stainless steel, silver, some other plastics and some synthetic rubbers. As vapor pressure is a function of temperature, some higher vapor pressure materials, e.g. zinc bearing brass, are quite acceptable in many applications as long as excessive temperatures are not encountered.

## 2.2 Production of Vacuum

The most common appearance of the word "vacuum" in normal usage refers to vacuum cleaners, which function by pulling a jet of air through a cleaning head and the carpet. The viscous flow of the air carries along particles which are filtered out into a bag. The vacuum in vacuum cleaners is generally produced by a rapidly spinning fan. This type of pump has a very high pumping speed S, but the ultimate low pressure is strongly limited by the large interblade spacing of the fan. These parameters, pumping speed and lowest pressure are two important features in any vacuum pump. An additional parameter for high vacuum pumps is the exhaust pressure. Often a high vacuum pressure pump cannot pump against atmospheric pressure but can pump against a lower pressure, so a backing pump is required to maintain a low pressure at the exhaust of the main pump after a roughing pump has reduced the pressure in the vacuum chamber to a "rough vacuum".

### 2.2.1 Mechanical Pumps (Roughing and Backing)

For low to medium vacuum applications, a mechanical pump may be sufficient and mechanical pumps in various forms are used as roughing and backing pumps in many systems.
Piston pumps are the easiest type of roughing pump to diagram (Figure 3).

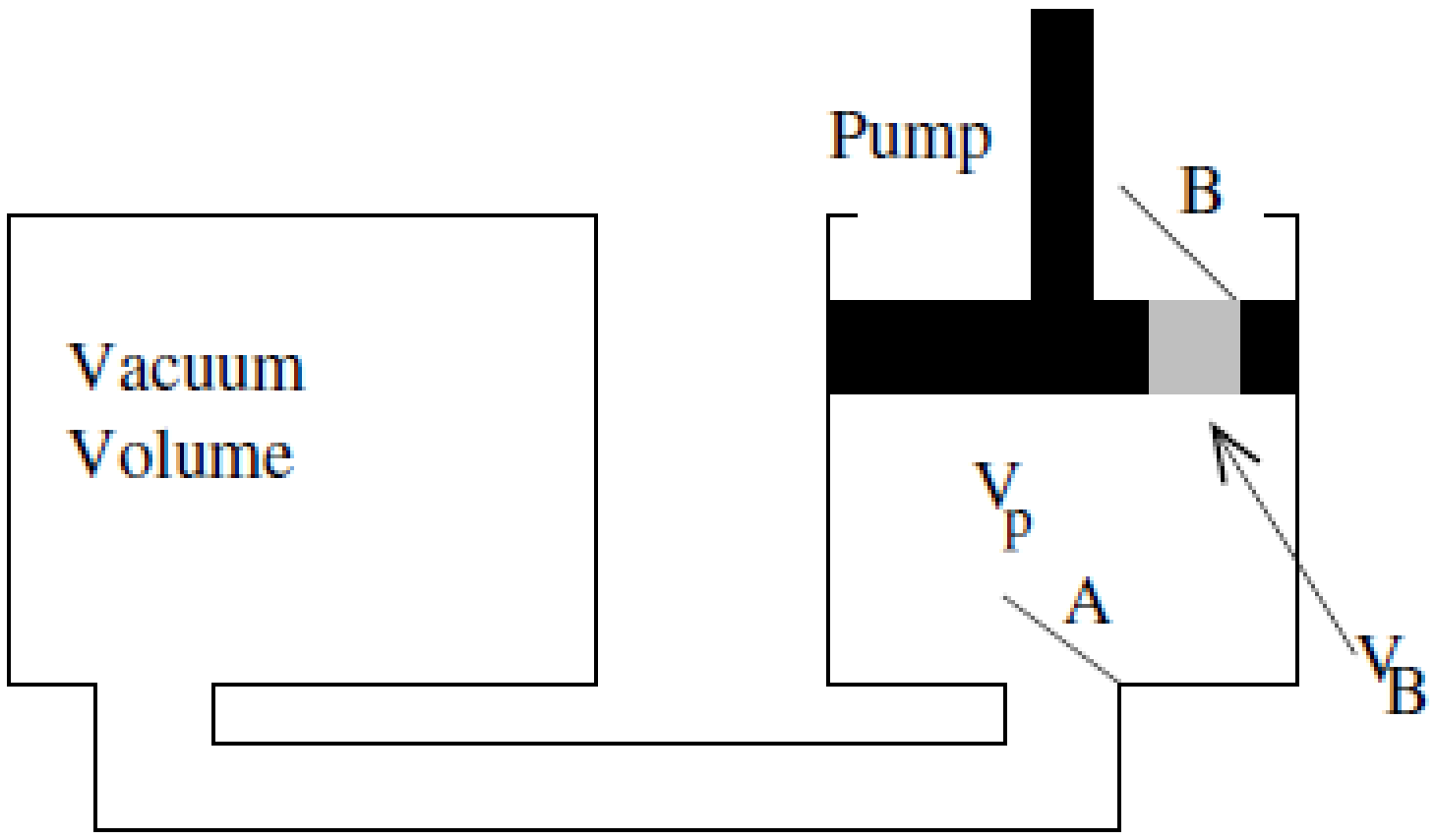

Figure 3 Piston Pump. The shaded region VB is the pump's dead volume.

The pump is formed from a piston sealed into a cylinder with two valves, one on the entrance to the cylinder (A) and one on the piston (B). When the piston moves up, the external pressure difference forces valve B closed. When the pressure in the cylinder is less than the pressure in the vacuum volume, valve A opens and the piston fills with gas. Then, when the piston compresses, valve A closes and then once the pressure in the cylinder rises above atmospheric, valve B opens and the piston is exhausted to atmosphere. The pump is limited by the dead volume which the piston cannot close. In the figure, the dead volume is labeled VB. The ultimate low pressure is Patm VB/Vp, where Vp is the volume of the pump cylinder at maximum expansion. Typical pumps in practice may have VB/Vp = 0.08. Beyond the dead volume limit, the sealing of the cylinder sets an additional low pressure limit even for low dead-volume pumps. As the pressure falls, the speed of the pump will also decrease and the pressure will exponentially approach the limiting pressure, but not reach it.

Both piston and rotary-action pumps (described below) have sealing oil, which limits the minimum pressure to the vapor pressure of the oil. The diaphragm pump is designed to eliminate the need for sealing oils, which makes for a cleaner vacuum system. The diaphragm pump is very similar to a piston pump, except the piston is replaced by a flexible diaphragm. Diaphragm pumps are quite robust and are widely used as backing and roughing pumps. They are able to achieve pressures of ~ 500 Pa.

There are several types of rotary action pumps, of which the rotating-vane pump is the most common. The rotating-vane pump consists of a cylindrical rotor set off center in a cylindrical volume (stator). The rotor has two spring-loaded vanes mounted opposite each other which separate the stator into two moving sections. The action of the pump is shown in Figure 5. The advantage of the rotating-vane pump over piston-style pumps is partly in speed, since the rotating-vane carries out two pumping cycles overlapped, while the piston pump has separate filling and exhaust phases. Rotating-vane pumps generally operate at 350-700 RPM.

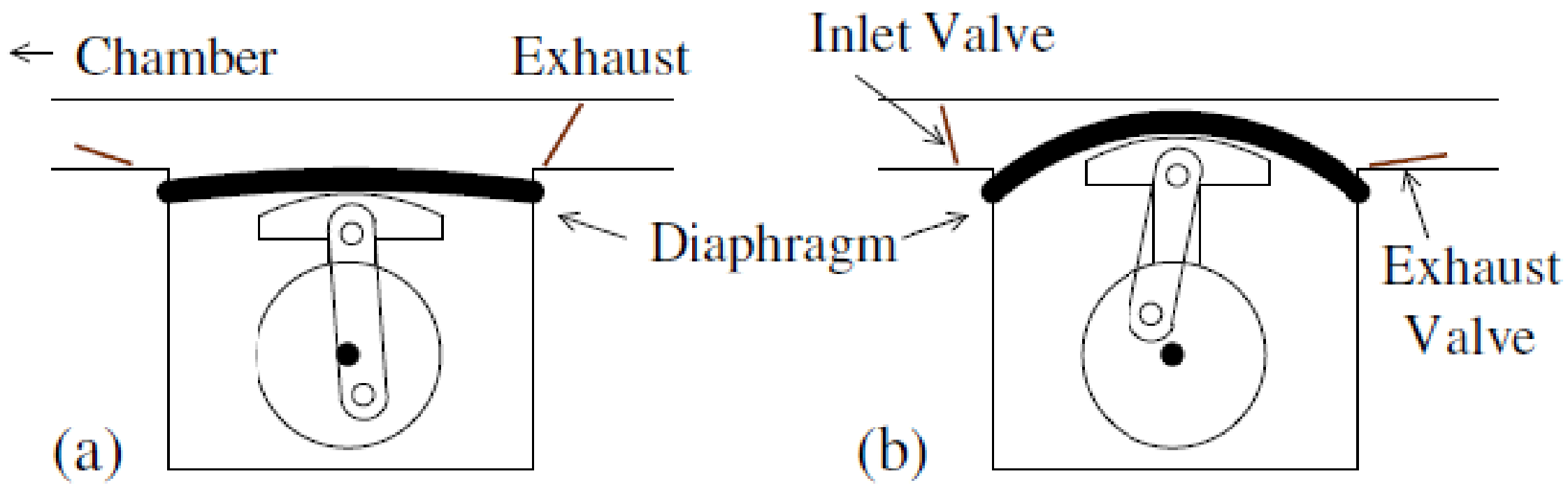

Figure 4 Diaphragm pump. (a) shows the arrangement of the valves while the pumping volume is filled and (b) as the pumping volume is exhausted. Usually the valves are controlled by mechanical linkages to the main motor driving the diaphragm.

The minimum pressure is limited by the dead volume ratio, as for the piston pump, but pressures as low as 10 Pa can be achieved with rotating vane pumps. Other variations on the rotating vane pump include a pump with a single sliding-vane attached to the stator and pressing against the rotor and the rotating plunger pump.

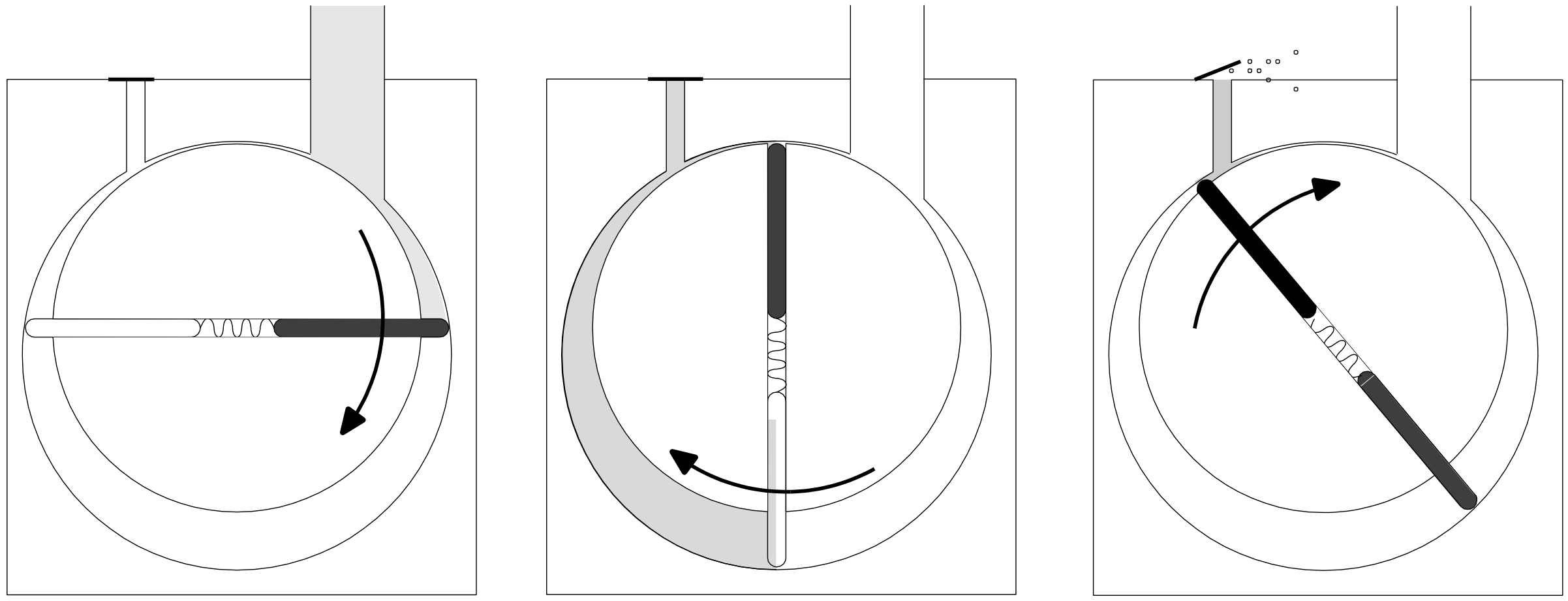

Figure 5: Rotating vane pump. The gas is shown for only one cycle. **(a)** Gas from the vacuum volume enters the cylinder. **(b)** The rotor has turned $270^{o}$ from (a) and the gas from the vacuum volume is now on the exhaust side of the cylinder. **(c)** The rotor has turned an additional $125^{o}$ and the gas has been forced out the exhaust.

### 2.2.2 Medium and High Vacuum Pumps

High vacuum pumps are used once the system has moved from the viscous regime to the molecular or kinetic regime. Many of these pumps depend on the transfer of momentum from fast moving objects to gas molecules while others depend on the capturing gas molecules in a cryogenic trap or absorbing them into a material.

### 2.2.2.1 Roots Pumps

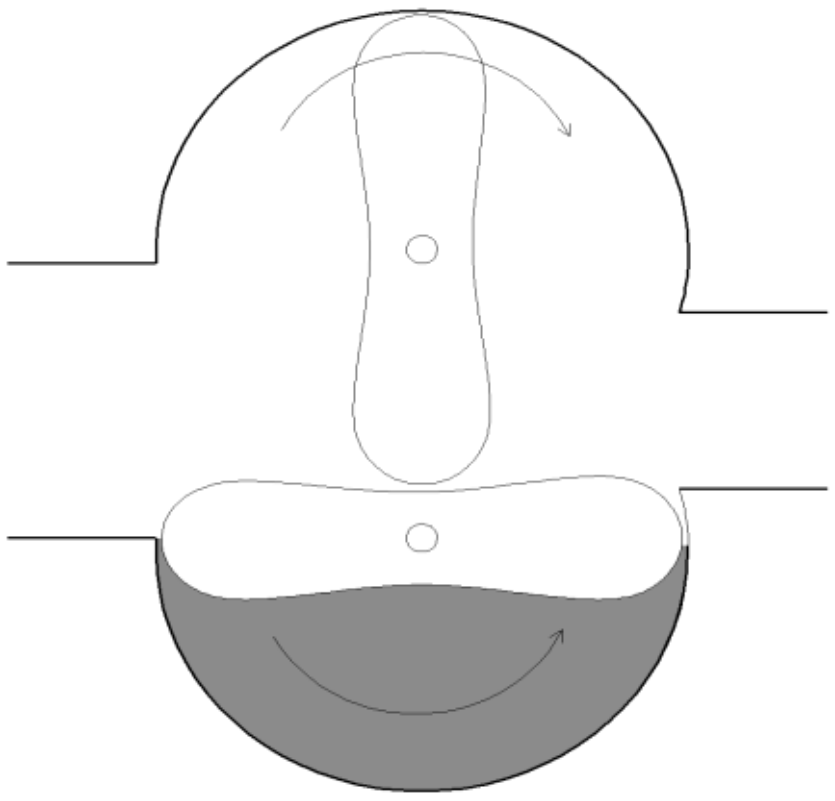

Figure 6 Roots pump. Several Roots pumps may be used in series to achieve a large total pressure drop. The pumping capacity of a Roots pump is generally quite large.

A Roots pump or Roots blower is a medium vacuum pump consisting of two lobed rotors mounted on parallel shafts, as shown in Figure 6. The rotors have clearance between themselves and the housing which is fairly substantial (0.2 mm) and which is not lubricated or sealed by oil. This allows for high rotation speeds (≈ 3500 RPM) without heating. Roots pumps are capable of high pumping speeds and low pressures, down to 0.01 Pa. A Roots pump will generally be combined with a rotary-vane backing pump capable of keeping the output pressure (backing pressure) at 100 Pa or less.

### 2.2.2.2 Molecular Drag Pumps

Molecular drag pumps are medium vacuum pumps which achieve ∼10−3 Pa level vacuums. Molecular drag pumps work by imparting momentum to the gas molecules through collision with a quickly spinning rotor (remember the discussion of viscous force). The pump is designed so the momentum transfer tends to send the molecules towards the exhaust. For example in the molecular drag pump shown in Figure 7, the sides of the pump housing are grooved in a helical pattern leading down to the exhaust. The moving surface must have a velocity close to the average gas molecule velocity for the pump to be efficient, so the rotor spins at high speeds, with 27,000 RPM being quite typical.

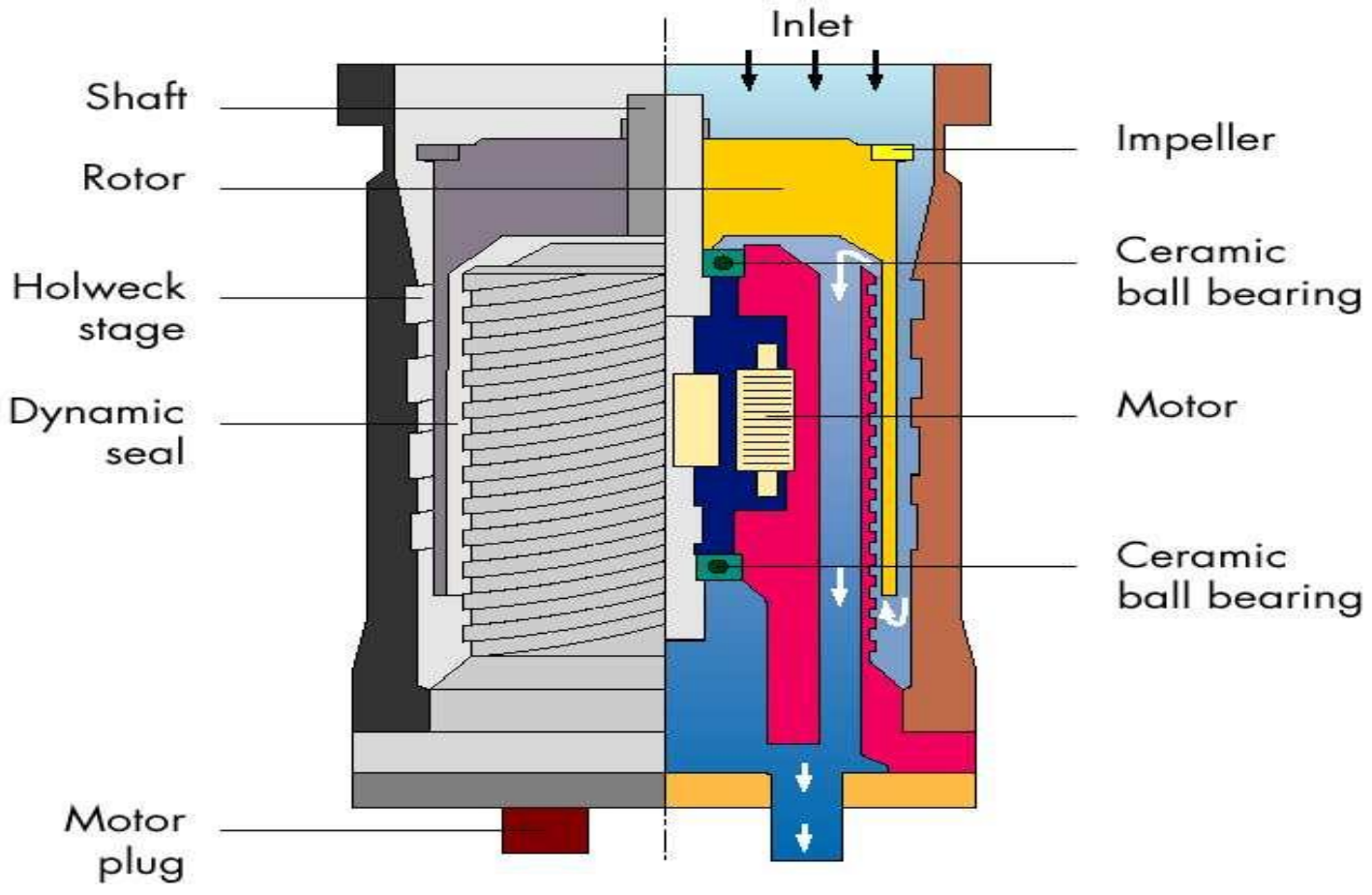

Figure 7 Complete Molecular Drag Pump

#### 2.2.2.3 Turbomolecular Pumps

Turbomolecular pumps are an extreme form of molecular drag pump capable of producing pressures of 10^−7 Pa. A turbomolecular pump is composed of a repeating series of bladed turbines angled in opposite directions. One set of blades, the rotor, rotates at very high speed (up to 60,000 RPM) to achieve near-molecular velocities at the edge and the other set is fixed (stator). A rough diagram of the operation of the turbomolecular blades is shown in Figure 8. Each stage of the pump can achieve a compression ratio of about 5, so a nine stage turbopump should reach a compression of 5^9 = 2 × 10^6. The ultimate pressure limit is set by the back-diffusion of hydrogen gas, which has a very low mass and has the highest average velocity of any gas at a given temperature (equation 3). Many turbomolecular pumps include a molecular drag stage at the end of the pump to increase the acceptable backing pressure on the pump to ∼ 10^3 Pa.

#### 2.2.2.4 Diffusion Pumps

Diffusion pumps work by spraying hot oil from nozzles close to the vacuum chamber down to a reservoir at the base of the pump. When the oil encounters an air molecule, it tends to deflect it towards the bottom of the pump.

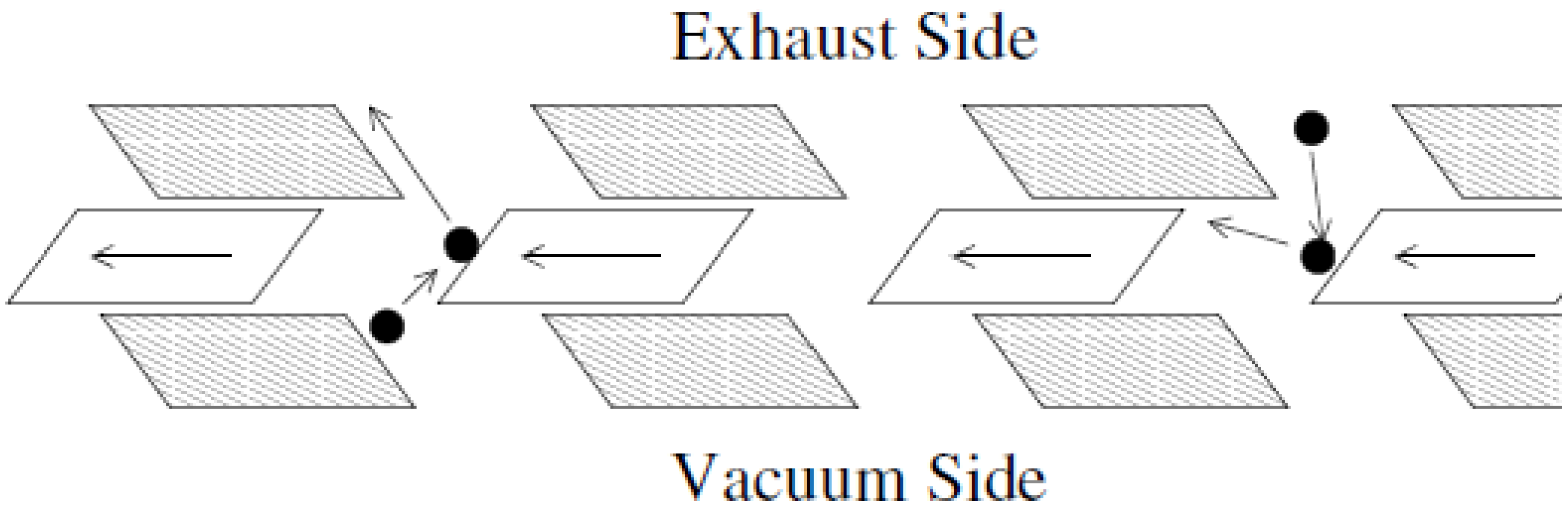

Figure 8 Operation of a turbomolecular pump. The stators are hashed and the rotor is shown in outline. A turbomolecular pump might contain nine or more stages of rotor and stator.

Near the bottom of the pump there is an exhaust line connected to a backing pump which removes the gas molecules accumulated by the oil spray from the system. To avoid oil getting into the main vacuum system, it is necessary to use a cold trap which condenses the oil vapor. The oil then drips back into the main pumping system. Diffusion pumps are by nature very messy and fairly temperamental, but they are quite simple and inexpensive compared to turbomolecular pumps. Diffusion pumps are limited in pressure by the presence of oil vapors and molecules derived from the chemical decomposition (cracking) of the pump oil, which may occur (for example) in a hot-cathode vacuum gauge.

### 2.2.3 Capture Pumps

Most pumps work by moving gas molecules from the vacuum volume into an exhaust volume, often eventually the atmosphere. Capture pumps simply trap the gas molecules and hold them, removing them from the volume but not exhausting them. Although capture pumps have limited pumping capacity, they can pump over very long time periods without power being used.

### 2.2.3.1 Cryopumps and Sorption Pumps

Cryopumps work by condensing gas on extremely cold manifolds. Using liquid nitrogen and liquid helium as refrigerants, cryopumps can reduce the partial pressure of all gasses except helium to 10^−8 Pa. Simple cryopumps use pools of coolant for pumping, but more modern models use integrated compressor loops to cool a limited volume of refrigerant repeatedly. Cryopumps can easily saturate with too much frozen gas too quickly if started at a high pressure (> 1 Pa or so), so a good roughing pump is essential. Sorption pumps capture gas molecules by using materials with large surface areas (such as zeolite) and are often combined with cryogenic techniques for best performance. Cryopumping action is often a benign byproduct in superconducting systems which require low temperatures for normal operation, and such distributed cyropumping may be explicitly considered in the design for systems such as particle accelerators.

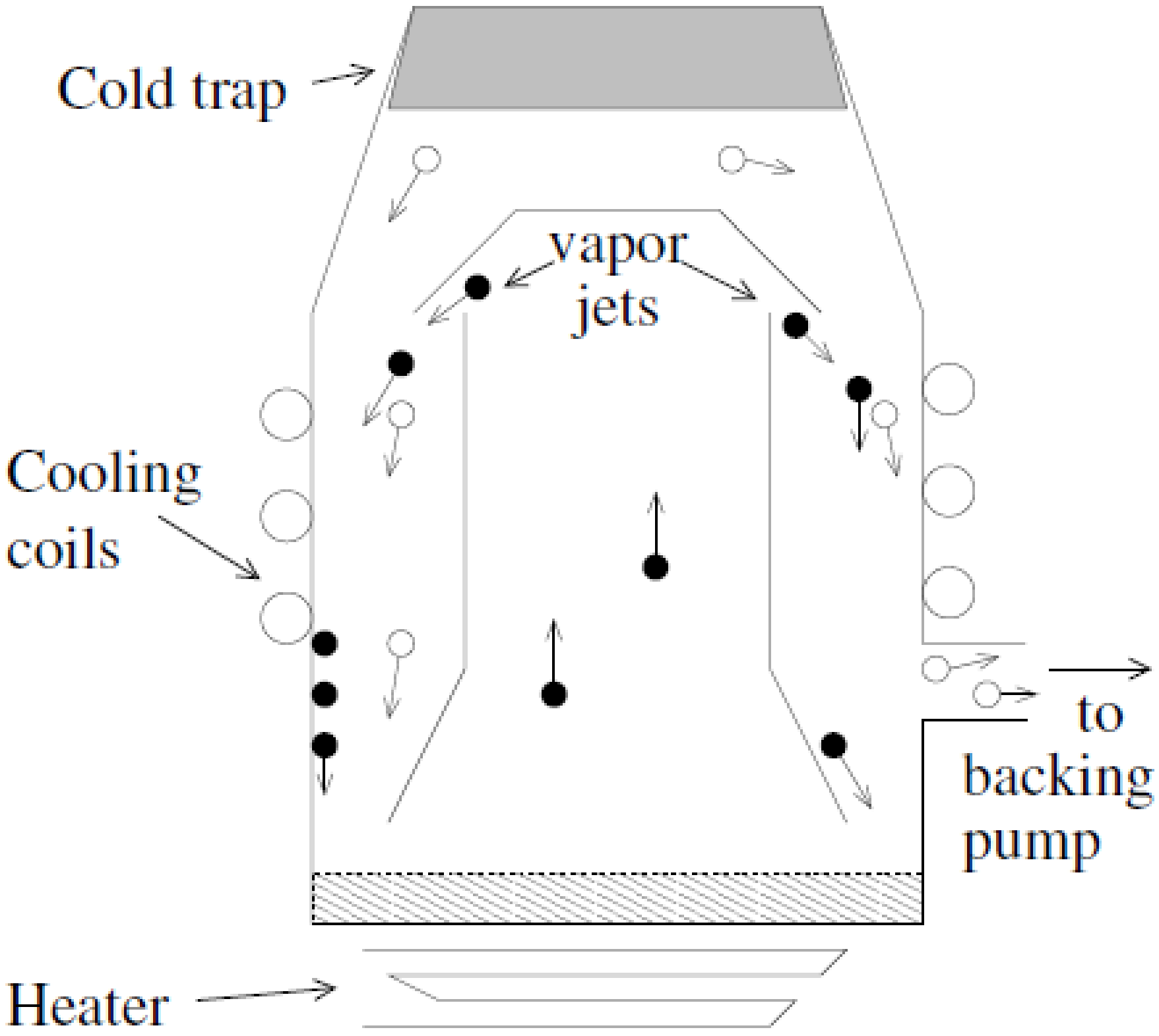

Figure 9 Diffusion Pump

### 2.2.3.2 Getter Pumps

Getter pumps are similar to cryopumps, but use chemical action to capture the gas molecules. Getters are used in photomultiplier tubes and other sealed systems to remove gases which penetrate the system or desorbs from the surface after the device is disconnected from active pumps. Usually getters are activated by heating, which pumps the gas from the surface into the bulk of the getter. This prevents the getter from being saturated while in storage or during initial pump down. In the form called the "non-evaporative getter" or NEG, getters are used in particle accelerators where very large but extremely narrow (low flow rate) volumes are evacuated. In such volumes, the pressures in the pipe are decreased by providing a larger pump (only by the ultimate pressure of the pump) and a distributed pumping system is essential.

NEGs are inexpensive, simple, and small which makes them ideal for distributed pumping over the 27 km circumference of the LHC accelerator ring at CERN.

### 2.2.3.3 Ion Pumps

The performance of getter pumps is very dependent on the chemical activity of the gases in the volume, so they tend to work extremely well for chemically active gases such as hydrogen, oxygen, and nitrogen, but rather poorly for noble gases. Ion pumps function much like getter pumps, but add an additional capture mechanism. Ion pumps produce beams of electrons between a cathode and anode, which are constrained by a magnetic field to move in helices, greatly increasing their path length and the chance of ionizing a molecule of the gas. The ionized gas molecules are accelerated to the cathode. Generally the impact of the gas molecules on the cathode will eject cathode atoms which sputter onto the surface of a secondary cathode and bury gas molecules, trapping them there. This process has three advantages over a getter pump. First, the electrons sweep out the volume actively, rather than waiting for the gas molecules to collide with the surface. Second, the electrons ionize noble gas atoms which are accelerated by the field and embed themselves deep inside the bulk of the cathodes, effectively trapping them. In the end, the major gas remaining tends to be argon, since it is heavier than helium and does not accelerate to as high a velocity in the field when ionized, so it is not buried as deeply. Finally, the ions reaching the surface of the secondary electrode constantly renew the surface, which improves the gettering action of the surface. Some ion pumps exhibit instabilities where the trapped gas atoms escape, and more advanced ion pump designs are intended to reduce these instabilities. One of these designs, the triode ion pump, is shown in Figure 10.

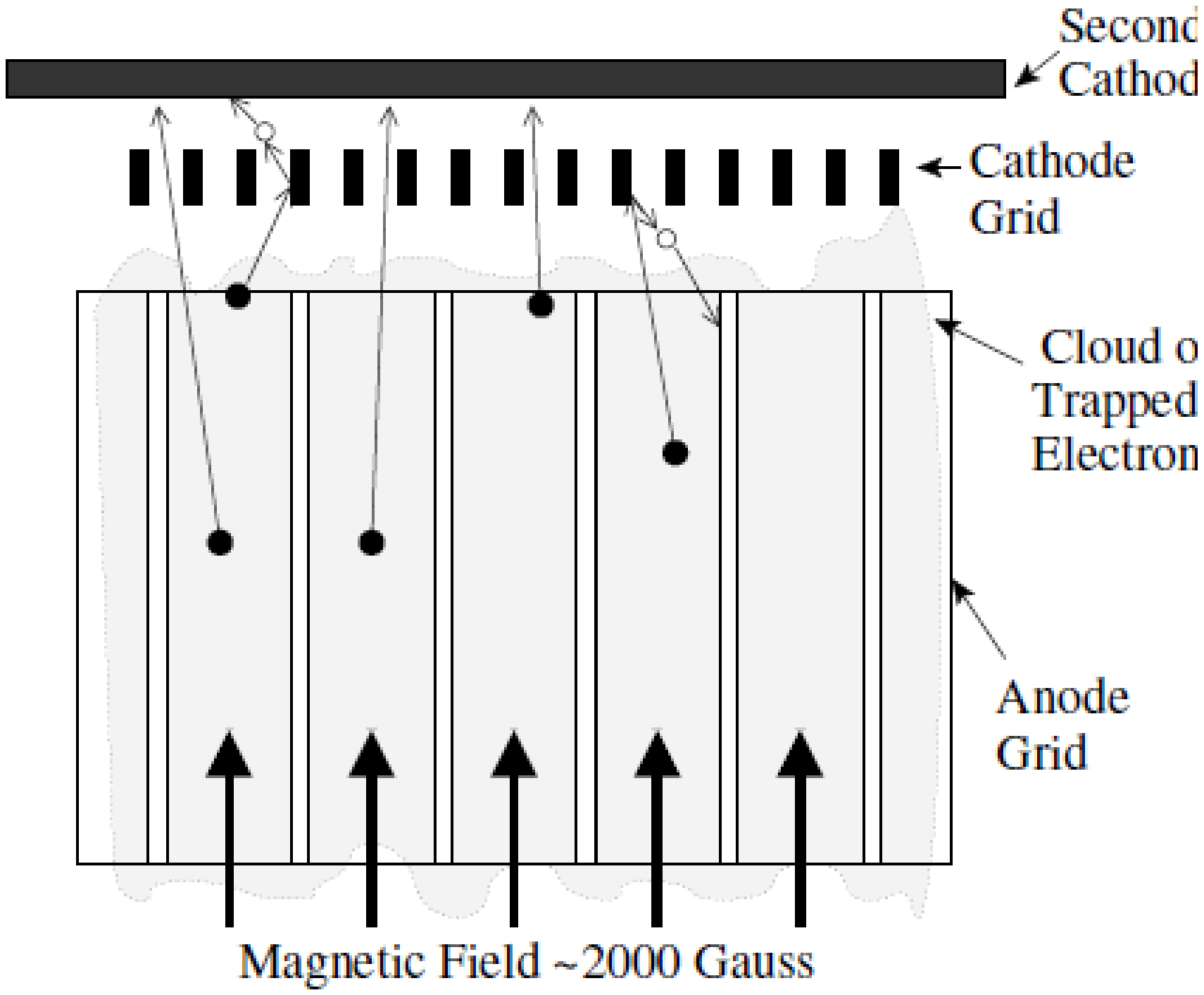

Figure 10 Ion triode pump

## 2.3 Vacuum Gauges

For many purposes, it is necessary to measure the level of vacuum achieved in a vacuum system. Given the very large range of pressures produced in vacuum systems (up to 19 orders of magnitude in some systems) there is no single gauge capable of measuring the full range of pressures. Most vacuum systems must have at least two different types of gauges, or even three.

### 2.3.1 Direct Pressure/Force Methods

The most straightforward method of measuring pressure is to measure the force exerted on the wall of the container or on the surface of a liquid. Generally, this technique can be used down to a minimum pressure of 1 to 100 Pa, depending on the specific gauge. One advantage of the direct force measurement is that is independent of the types of gases.

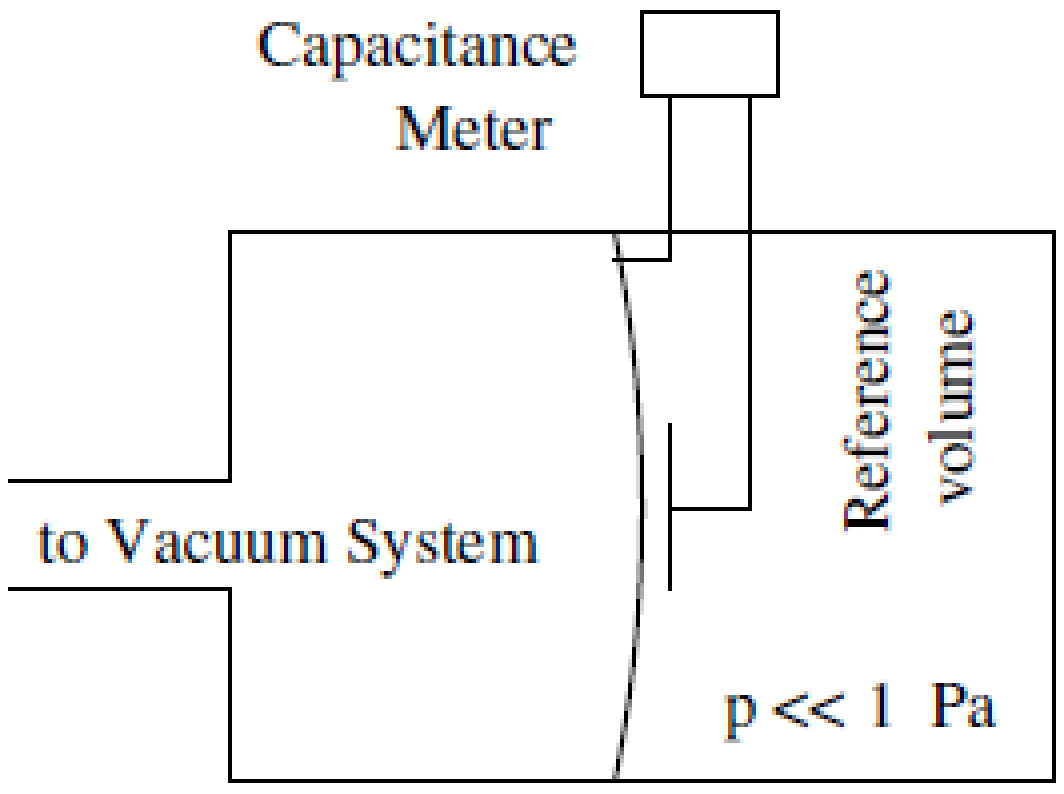

Figure 11 Capacitive Manometer Pressure Gauge

Mechanical force-measurement gauges usually involve measuring the deflection of a diaphragm supported between the vacuum system and either atmosphere (Bourdon gauge) or a chamber evacuated to a pressure significantly less than 1 Pa. This deflection can be indicated by a mechanical linkage, which limits the reading to about 100 Pa, or by electrical or optical methods. The most commonly-used readout system is based on adding a second plate close to the diaphragm and measuring the capacitance of the two plates. A diagram of such a gauge is shown in Figure 11. The capacitive technique is useful to at least 1 Pa. Other gauge developers have used deflection of a light beam or measurement of the mutual inductance of the two coils placed near to the diaphragm to measure the position of the diaphragm.

The classic pressure gauge is the column of mercury in the Earth's gravitational field. The difference in pressure between two volumes of gas can be read as $P = mgh$, where is the density of the mercury or oil used the gauge. The difference in heights between the columns on two sides can be read by eye to about 100 Pa or 1 Torr (1 mm of Hg). Using magnification, 1 Pa can be read, but variations in the glass refraction, in the capillary action of the surface, temperature, and other factors increase the error. Ultimate precisions of 10^−3 Pa have been achieved in such systems using oil instead of mercury by ultrasonic interferometry. The difficulty of such measurements relative to other measurement techniques keeps such systems as mere curiosities. In any case, the environmental dangers of mercury had recently led to its rejection for most routine measurement purposes.

## 2.3.2 Thermal Conductivity Gauges

Once a volume changes from the viscous regime to the free-molecular regime, the heat transport capacity of the gas falls in proportion to the number of gas molecules in the volume, as described above. The measurement of this effect is the basis of the thermal conductivity gauge. For a wire in a cylindrical envelope, the heat transport Qc is

$$Q_c = K_q p \left(T_s - T_e\right) 2\pi r_s \ell \tag{4}$$

where $p$ is the pressure of the gas, $T_e$ is the temperature of the envelope, and $T_s$ is the temperature of the heated wire or source. Geometrically, $r_e$ is the radius of the envelope, $r_s$ is the radius of the wire, and $\ell$ is the length of the wire. The term $K_q$ is the thermal conductivity of the gas and is gas-dependent. For nitrogen (air, essentially), the thermal conductivity

$$K_q \approx 1.2 \times 10^{-4} \frac{\mathrm{W}}{\mathrm{cm}^2 \cdot \mathrm{K} \cdot \mathrm{Pa}}$$

Using Equation 4 and keeping $T_e$ constant, there is a simple relation between $Q_c$, $T_s$, and the pressure $p$. For thermal gauges, the various parameters cannot be accurately obtained by calculation, so gauges are calibrated against a standard gauge of another type instead.

Thermal gauges become effective once $K_n > 1$, which implies that the mean free path is larger than $r_e$. For a gauge to be effective, the heat loss due to the gas must dominate any conduction down the leads or radiation. The ultimate low pressure limit for thermal gauges is set by the radiative heat loss

$$Q_r = K_r \left(T_s^4 - T_e^4\right) 2\pi r_s \ell$$

which rises very rapidly with increasing temperature.

In the Pirani gauge, one of the most widely-used gauges, the measurement is carried out by heating a filament and measuring its temperature as a function of the input power. The temperature of the wire can be determined from its resistance: for most metals, resistance increases with increasing temperature. The gauge can be operated in either constant temperature or constant current mode. In the constant current mode, the temperature of the wire provides the measurement, while in constant temperature mode, the current going to the gauge head is modulated to keep Ts constant. In a constant temperature gauge, the pressure is proportional to the power I^2*R. The constant temperature mode has the advantage of avoiding burning out the element, since the maximum temperature is limited. Additionally, the low-pressure limit for a constant temperature gauge is lower than for a constant power gauge because of the rapid growth of T 4 at high temperatures. Pirani gauges can generally measure down to about 0.1 Pa.

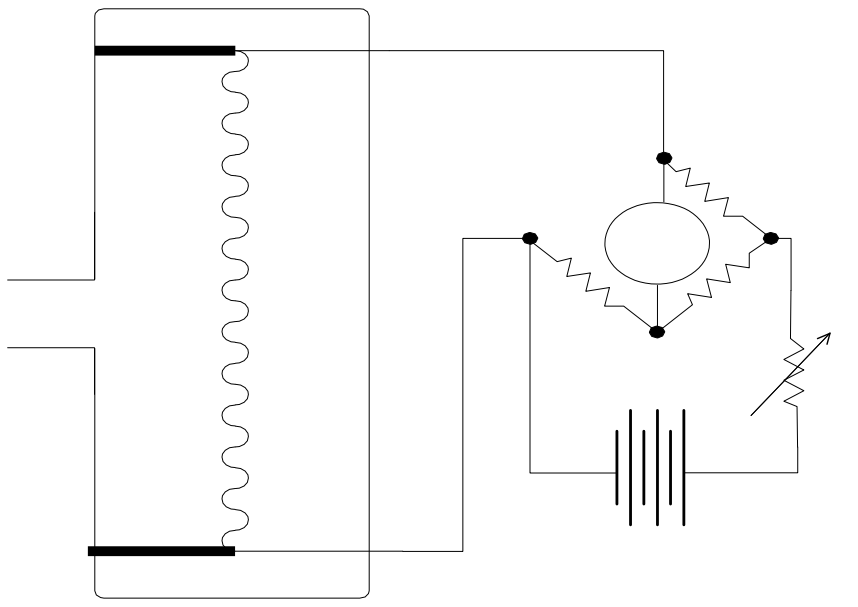

Figure 12 Pirani gauge. The filament is heated by a current and the resistance is measured with the bridge.

The thermocouple gauge is very similar to the Pirani gauge, but separates the roles of the heater and temperature measurement. The filament makes the heat, and a thermocouple sensor is spot-welded to it to measure the temperature. This gauge has very similar performance to the Pirani, but the circuit can be made somewhat simpler.

The thermistor gauge is one of the most commonly used gauges. It is extremely similar to a Pirani gauge except it uses a thermistor to measure the temperature. A thermistor is a semiconducting element which has a large negative temperature dependence on conductivity. This gauge has a more linear response to temperature than the regular Pirani gauge, which is its major advantage.

## 2.3.3 Ion Gauges

At very low pressures, neither force nor thermal gauges function effectively, and yet another type of gauge is required. High vacuum (low pressure) gauges are nearly always some form of ion gauge. As a general type, ion gauges function by ionizing the gas atoms in the vacuum volume and collecting and measuring the ion current produced. Because ionization is inherently a gas-type-dependent process, ion gauges are never gas-independent.

The standard hot-cathode gauge contains a heated filament (cathode) which produces electrons. These electrons are accelerated by a grid voltage of ~200V. Most electrons miss the wire grid and pass through to the other side. The anode is at a small negative voltage relative to the cathode (~ -10V) so the electrons come to a stop and return back towards the grid. They will oscillate across the grid several times before being captured by a wire. If they encounter a gas atom in their travels, they may ionize it. The ions will drift to the anode and the current observed here gives the pressure. A diagram of a standard gauge is shown in Figure 13a.

The Bayard-Alpert gauge is a modification of the standard hot-cathode gauge designed to improve on one of the limitations of the hot-cathode gauge. The excited electrons emit soft X rays which can photo eject electrons from the surface of the anode. This process produces an identical signal to that of positive ion collection, so the ultimate low pressure limit of the gauge is set by the background current of this X-ray process. For a standard gauge this is ~ 10^−6 Pa. The Bayard-Alpert gauge reverses the location of the filament and anode, which greatly reduces the area of the anode and hence the area for X-rays to hit. As a result, Bayard-Alpert gauges can be used to pressures of 10^−10 Pa, with an anode of 4μm.

A diagram of a Bayard-Alpert gauge is shown in Figure 13b.

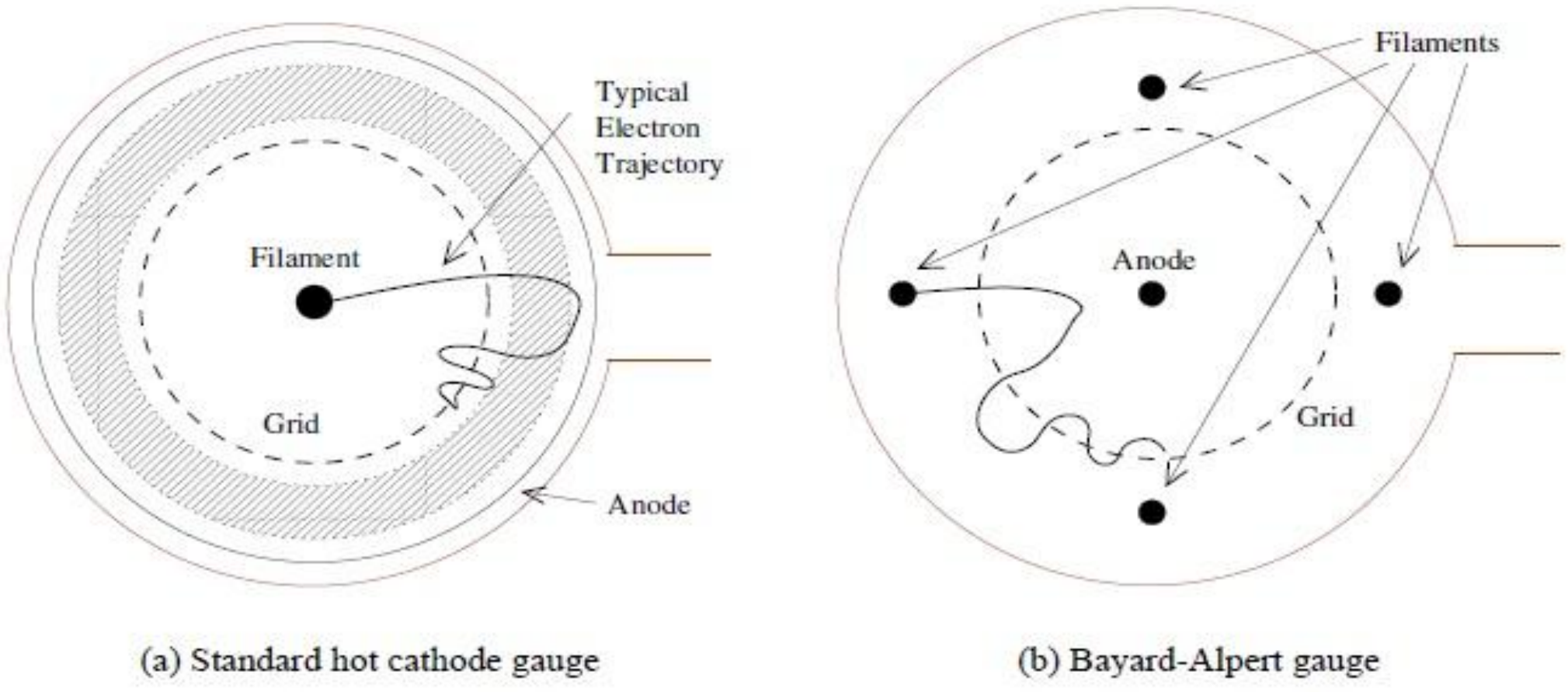

Figure 13 Ion gauges. The shaded region in (a) indicates the volume where produced ions will drift to the anode (rather than the cathode).

The lifetime of all hot-cathode gauges is limited by the lifetime of the filament. Filaments age rapidly when heated at pressures greater than 1 Pa, so a roughing gauge (such as a Pirani) must be used to monitor the pressure before turning on the hot-cathode gauge. Also, the presences of certain chemicals in vacuum system (such as chlorine) can very quickly destroy the hot filament.

Besides hot-cathode gauges, there are cold cathode gauges which avoid the filament destruction common to hot-type gauges. One common type of cold-cathode gauge is the Penning or Phillips gauge. In this gauge, a much higher voltage (∼2kV) is used than in hot-cathode gauges, but there is no heating to the cathode. The gauge is formed from two cathode plates with a wire loop anode halfway between the two. The electrons emitted from the cathode are accelerated towards the plane of the anode loop. The entire gauge is immersed in a magnetic field perpendicular to the plane of the plates. This field makes the electrons curve in a helix. The electrons curl down through the plane of the anode loop until they are pushed back by the other cathode and they oscillate a very large number of times in the field before encountering the ring of the anode. This long path and high kinetic energy give a very large chance of ionizing a gas molecule. The only disadvantage of the gauge is that it can be hard to start the discharge at low pressures. Sometimes a small filament or beta source is used to start the discharge. Once started, the process is self-sustaining.

The upper limit of this gauge's sensitivity is the pressure for glow discharges (~ 1 Pa) while its lower limit appears when the density becomes too low for the discharge to continue. Generally, Penning gauges are limited to ∼10−8 Pa. A diagram of a Penning gauge is shown in Figure 14.

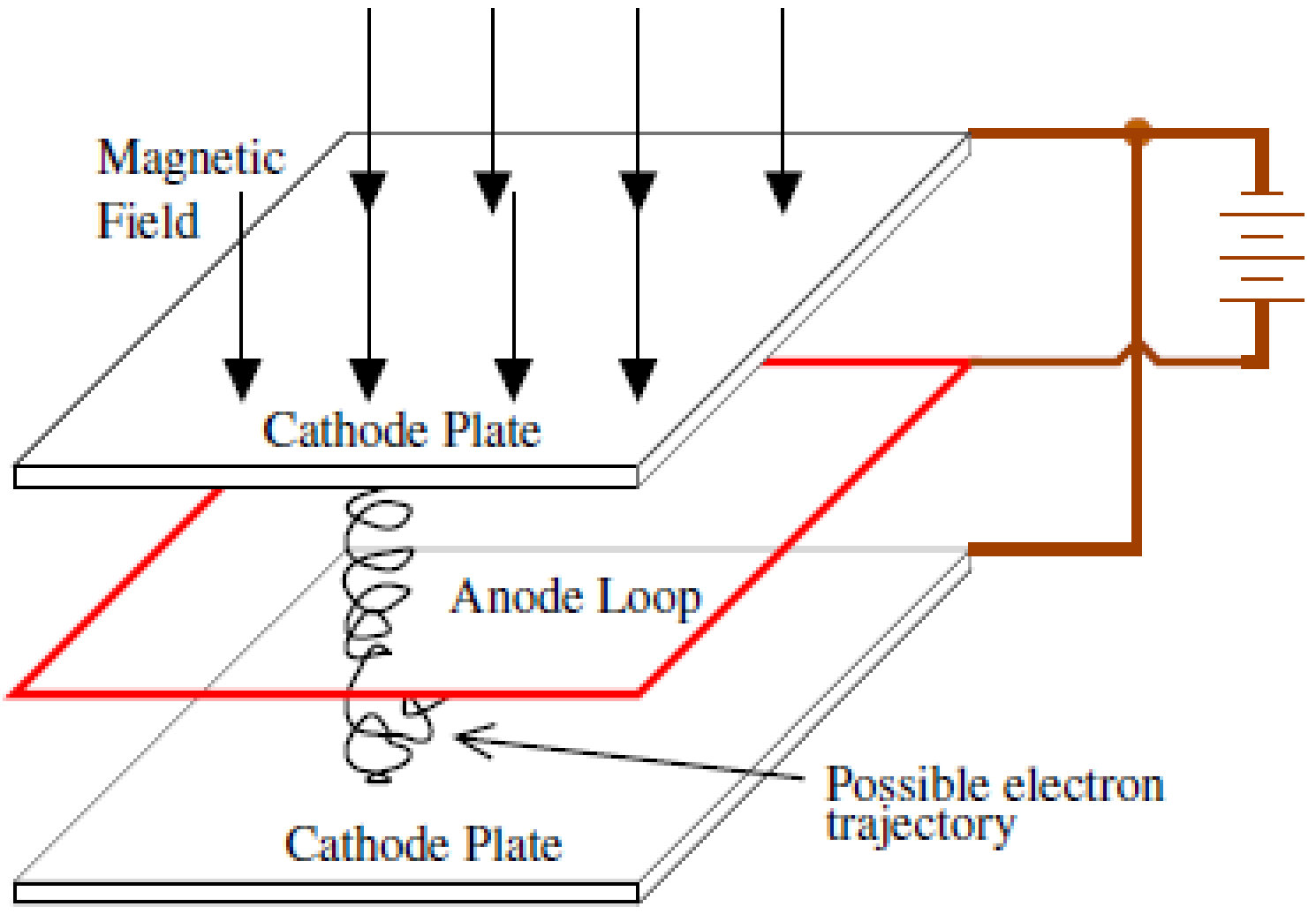

Figure 14 Penning Cold Cathode Gauge

Besides these common gauge types, many others have been produced for extremely high vacuum, some reaching down even to 10−16 Pa. The above are the most common types of ion gauges however, and most others are variations on the same theme. Every ion gauge also functions as an ion-pump (section 2.3.3) at some level, so the pressure measured by the gauge is lower than the pressure in the bulk volume.

# 3

# Data Acquisition & Control (DAQ)

Data acquisition is the process of sampling signals that measure real world physical conditions and converting the resulting samples into digital numeric values that can be manipulated by a computer. Data acquisition systems (abbreviated with the acronym DAS or DAQ) typically convert analog waveforms into digital values for processing.

The components of data acquisition systems include:

• Sensors that convert physical parameters to electrical signals.

• Signal conditioning circuitry to convert sensor signals into a form that can be converted to digital values.

• Analog-to-digital converters, which convert conditioned sensor signals to digital values.

Data acquisition applications are controlled by software programs developed using various general purpose programming languages such as BASIC, C, FORTRAN, Java, Lisp, and Pascal.

## 3.1 Definition

Although concepts like data acquisition and test and measurement can be surprisingly difficult to define completely, most computer users, engineers, and scientists agree there are several common elements:

• A personal computer (PC) is used to program test equipment and manipulate or store data. The term "PC" is used in a general sense to include any computer running any operating system and software that supports the desired result. The PC may also be used for supporting functions, such as real-time graphing or report generation. The PC may not necessarily be in constant control of the data acquisition equipment or even remain connected to the data acquisition equipment at all times.

• Test equipment can consist of data acquisition plug-in boards for PCs, external board chassis, or discrete instruments. External chassis and discrete instruments typically can be connected to a PC using either standard communication ports or a proprietary interface board in the PC.

• The test equipment can perform one or more measurement and control processes using various combinations of analog input, analog output, digital I/O, or other specialized functions.

The difficulty involved in differentiating between terms such as data acquisition, test and measurement, and measurement and control stems from the blurred boundaries that separate the different types of instrumentation in terms of operation, features, and performance. For example, some stand-alone instruments now contain card slots and microprocessors, use operating system software, and operate more like computers than like traditional instruments. Some external instruments now make it possible to construct test systems with high channel counts that gather data and log it to a controlling computer. Plug-in boards can transform computers into multi-range digital multimeters, oscilloscopes, or other instruments, complete with user friendly, on-screen virtual front panels.

For the sake of simplicity, this handbook uses the term “data acquisition and control” broadly to refer to a variety of hardware and software solutions capable of making measurements and controlling external processes. The term “computer” is also defined rather broadly; however, for most applications, a “computer” means an IBM compatible PC running Microsoft® Windows® 95 or later, unless otherwise noted.

## 3.2 Data Acquisition and Control Hardware

Data acquisition and control hardware is available in a number of forms, which offer varying levels of functionality, channel count, speed, resolution, accuracy, and cost. This section summarizes the features and benefits generally associated with the various categories, based on a broad cross-section of products.

### 3.2.1 Plug-in Data Acquisition Boards

Like display adapters, modems, and other types of expansion boards, plug-in data acquisition boards are designed for mounting in board slots on a computer motherboard. Today, most data acquisition boards are designed for the current PCI (Peripheral Component Interconnect) or earlier ISA (Industry Standard Architecture) buses. Data acquisition plug-in boards and interfaces have been developed for other buses (EISA, IBM Micro Channel®, and various Apple buses), but these are no longer considered mainstream products. As a category, plug-in boards offer a variety of test functions, high channel counts, high speed, and adequate sensitivity to measure moderately low signal levels, at relatively low cost.

**Features of plug-in data acquisition boards**

Least expensive method of computerized measurement and control.

High speed available (100 kHz to 1GHz and higher).

Available in multi-function versions that combine A/D, D/A, digital I/O, counting, timing, and specialized functions.

Good for tasks involving low-to-moderate channel counts.

Performance adequate to excellent for most tasks, but electrical noise inside the PC can limit ability to perform sensitive measurements.

Input voltage range is limited to approximately ±10V.

Use of PC expansion slots and internal resources can limit expansion potential and consume PC resources.

Making or changing connections to board’s I/O terminals can be inconvenient.

### 3.2.2 External Data Acquisition Systems

The original implementation of an external data acquisition system was a self-powered system that communicated with a computer through a standard or proprietary interface. As a boxed alternative to plug-in boards, this type of system usually offered more I/O channels, a quieter electrical environment, and greater versatility and speed in adapting to different applications. Today, external data acquisition systems often take the form of a stand-alone test and measurement solution oriented toward industrial applications. The applications for which they are used typically demand more than a system based on a PC with plug-in boards can provide or this type of architecture is simply inappropriate for the application. Modern external data acquisition systems offer:

• High sensitivity to low-level voltage signals, i.e., approximately 1mV or lower.

• Applications involving many types of sensors, high channel counts, or the need for stand-alone operation.

• Applications requiring tight, real-time process control.

Like the plug-in board based system, these external systems require the use of a computer for operation and data storage. However, the computer can be built up on boards, just as the instruments are, and incorporated into the board rack. There are several architectures for external industrial data acquisition systems, including VME, VXI, MXI, Compact PCI, and PXI. These systems use mechanically robust, standardized board racks and plug-in instrument modules that offer a full range of test and measurement functions. Some external system designs include microprocessor modules that support all the standard PC user interface elements, including keyboard, monitor, mouse, and standard communication ports. Frequently, these systems can also run Microsoft Windows and other PC applications. In this case, a conventional PC may only be needed to develop programs or off-load data for manipulation or analysis.

**Features of external data acquisition chassis**

Multiple board slots permit mixing-and-matching boards to support specialized acquisition and control tasks and higher channel counts.

Chassis offers an electrically quieter environment than a PC, allowing for more sensitive measurements.

Use of standard interfaces (IEEE-488, RS-232, USB, FireWire, and Ethernet) can facilitate daisy chaining, networking, long distance acquisition, and use with non-PC computers.

Dedicated processor and memory can support critical “real-time” control applications or stand-alone acquisition independent of a PC.

Standardized modular architectures are mechanically robust, easy to configure, and provide for a variety of measurement and control functions.

Required chassis, modules, and accessories are cost-effective for high channel counts.

Some architecture has minimal vendor support, limiting the sources of equipment and accessories available.

## 3.2.3 Real-Time Data Acquisition and Control

Critical real-time control is an important issue in data acquisition and control systems. Applications that demand real-time control are typically better suited to external systems than to systems based on PC plug-in boards. Although Microsoft Windows has become the standard operating system for PC applications, it is a non-deterministic operating system that can’t provide predictable response times in critical measurement and control applications. Therefore, the solution is to link the PC to a system that can operate autonomously and provide rapid, predictable responses to external stimuli.

## 3.2.3.1Discrete (Bench/Rack) Instruments

Originally, discrete electronic test instruments consisted mostly of single-channel meters, sources, and related instrumentation intended for general-purpose test applications. Over the years, the addition of communication interfaces and advances in instrument design, manufacturing, and measurement technology have extended the range and functionality of these instruments. New products such as scanners, multiplexers, SourceMeter ® instruments, counter/timers, nanovoltmeters, micro-ohmmeters, and other specialized instrumentation have made it possible to create computer-controlled test and measurement systems that offer exceptional sensitivity and resolution. Some systems of this type can service only one channel or just a few channels, so their cost per channel is high. However, the addition of switch matrices and multiplexers can lower the cost per channel by allowing one set of instruments to service many channels while preserving high signal integrity. These instruments can also be combined with computers that contain plug-in data acquisition boards.

## 3.2.4 Hybrid Data Acquisition Systems

Hybrid systems are a relatively recent development in external data acquisition systems. A typical hybrid system combines a DMM-type user interface with several standard data acquisition functions and expansion capabilities in a compact, instrument-like package. Typical functions include AC and DC voltage and current measurements, temperature and frequency measurements, event counting, timing, triggering, and process control. Keithley's Integra™ Series, which includes the Model 2700 and Model 2750 and their associated plug-in modules, provides multiple board slots for expanding the system's measurement capabilities and channel capacity

### Features of a hybrid data acquisition system

Delivers accuracy, measurement range, and sensitivity typical of bench
DMMs, and superior to standard data acquisition equipment.
DMM front end with digital display and front panel controls provides resolution equivalent to a DMM (18- to 22-bit A/D or better).
Built-in data and program storage memory for stand-alone data logging and process control.
Uses standard interfaces (IEEE-488) that support long-distance acquisition and provide compatibility with non-PC computers.
Cost-effective on a per-channel basis.
Limited expansion capacity (less critical because base test capability is already complete).
Generally slower than plug-in boards or external data acquisition systems.

# 4

# Communication Protocols

A communications protocol is a system of digital message formats and rules for exchanging those messages in or between computing systems and in telecommunications. A protocol may have a formal description. Protocols may include signaling, authentication and error detection and correction capabilities. Communicating systems use well-defined formats for exchanging messages. Each message has an exact meaning intended to provoke a particular response of the receiver. Thus, a protocol must define the syntax, semantics, and synchronization of communication; the specified behavior is typically independent of how it is to be implemented. A protocol can therefore be implemented as hardware, software, or both. Communications protocols have to be agreed upon by the parties involved. To reach agreement a protocol may be developed into a technical standard. A programming language describes the same for computations, so there is a close analogy between protocols and programming languages: protocols are to communications as programming languages are to computations.

## 4.1 Transmission Control Protocol (TCP)

The Transmission Control Protocol (TCP) is one of the core protocols of the Internet protocol suite. TCP is one of the two original components of the suite, complementing the Internet Protocol (IP), and therefore the entire suite is commonly referred to as TCP/IP. TCP provides reliable, ordered delivery of a stream of octets from a program on one computer to another program on another computer. TCP is the protocol used by major Internet applications such as the World Wide Web, email, remote administration and file transfer. Other applications, which do not require reliable data stream service, may use the User Datagram Protocol (UDP), which provides a datagram service that emphasizes reduced latency over reliability.

### 4.1.1 Network function

The protocol corresponds to the transport layer of TCP/IP suite. TCP provides a communication service at an intermediate level between an application program and the Internet Protocol (IP). That is, when an application program desires to send a large chunk of data across the Internet using IP, instead of breaking the data into IP-sized pieces and issuing a series of IP requests, the software can issue a single request to TCP and let TCP handle the IP details.

IP works by exchanging pieces of information called packets. A packet is a sequence of octets and consists of a header followed by a body. The header describes the packet's destination and, optionally, the routers to use for forwarding until it arrives at its destination. The body contains the data IP is transmitting. Due to network congestion, traffic load balancing, or other unpredictable network behavior, IP packets can be lost, duplicated, or delivered out of order. TCP detects these problems, requests retransmission of lost data, rearranges out-of-order data, and even helps minimize network congestion to reduce the occurrence of the other problems. Once the TCP receiver has reassembled the sequence of octets originally transmitted, it passes them to the application program. Thus, TCP abstracts the application's communication from the underlying networking details.

TCP is utilized extensively by many of the Internet's most popular applications, including the World Wide Web (WWW), E-mail, File Transfer Protocol, Secure Shell, peer-to-peer file sharing, and some streaming media applications.

TCP is optimized for accurate delivery rather than timely delivery, and therefore, TCP sometimes incurs relatively long delays (in the order of seconds) while waiting for out-of-order messages or retransmissions of lost messages. It is not particularly suitable for real-time applications such as Voice over IP. For such applications, protocols like the Real-time Transport Protocol (RTP) running over the User Datagram Protocol (UDP) are usually recommended instead.

TCP is a reliable stream delivery service that guarantees that all bytes received will be identical with bytes sent and in the correct order. Since packet transfer is not reliable, a technique known as positive acknowledgment with retransmission is used to guarantee reliability of packet transfers. This fundamental technique requires the receiver to respond with an acknowledgment message as it receives the data. The sender keeps a record of each packet it sends. The sender also keeps a timer from when the packet was sent, and retransmits a packet if the timer expires before the message has been acknowledged. The timer is needed in case a packet gets lost or corrupted.

TCP consists of a set of rules: for the protocol, that are used with the Internet Protocol, and for the IP, to send data "in a form of message units" between computers over the Internet. While IP handles actual delivery of the data, TCP keeps track of the individual units of data transmission, called segments that a message is divided into for efficient routing through the network. For example, when an HTML file is sent from a Web server, the TCP software layer of that server divides the sequence of octets of the file into segments and forwards them individually to the IP software layer (Internet Layer). The Internet Layer encapsulates each TCP segment into an IP packet by adding a header that includes (among other data) the destination IP address. Even though every packet has the same destination address, they can be routed on different paths through the network. When the client program on the destination computer receives them, the TCP layer (Transport Layer) reassembles the individual segments and ensures they are correctly ordered and error free as it streams them to an application.

## 4.2 Modbus

The Modbus protocol was developed in 1979 by Modicon, Incorporated, for industrial automation systems and Modicon programmable controllers. It has since become an industry standard method for the transfer of discrete/ analog I/O information and register data between industrial control and monitoring devices. Modbus is now a widely-accepted, open, public-domain protocol that requires a license, but does not require royalty payment to its owner.

Modbus devices communicate using a master-slave (client-server) technique in which only one device (the master/client) can initiate transactions (called queries). The other devices (slaves/servers) respond by supplying the requested data to the master, or by taking the action requested in the query. A slave is any peripheral device (I/O transducer, valve, network drive, or other measuring device) which processes information and sends its output to the master using Modbus. The Acromag I/O Modules form slave/server devices, while a typical master device is a host computer running appropriate application software. Other devices may function as both clients (masters) and servers (slaves).

Masters can address individual slaves, or can initiate a broadcast message to all slaves. Slaves return a response to all queries addressed to them individually, but do not respond to broadcast queries. Slaves do not initiate messages on their own, they only respond to queries from the master. A master's query will consist of a slave address (or broadcast address), a function code defining the requested action, any required data, and an error checking field. A slave's response consists of fields confirming the action taken, any data to be returned, and an error checking field. Note that the query and response both include a device address, a function code, plus applicable data, and an error checking field. If no error occurs, the slave's response contains the data as requested. If an error occurs in the query received, or if the slave is unable to perform the action requested, the slave will return an exception message as its response (see Modbus Exceptions). The error check field of the slave's message frame allows the master to confirm that the contents of the message are valid. Traditional Modbus messages are transmitted serially and parity checking is also applied to each transmitted character in its data frame.

At this point, it's important to make the distinction that Modbus itself is an application protocol, as it defines rules for organizing and interpreting data, but remains simply a messaging structure, independent of the underlying physical layer. As it happens to be easy to understand, freely available, and accessible to anyone, it is thus widely supported by many manufacturers.

## 4.3 What is ModbusTCP/IP?

Modbus TCP/IP (also Modbus-TCP) is simply the Modbus RTU protocol with a TCP interface that runs on Ethernet. The Modbus messaging structure is the application protocol that defines the rules for organizing and interpreting the data independent of the data transmission medium. TCP/IP refers to the Transmission Control Protocol and Internet Protocol, which provides the transmission medium for Modbus TCP/IP messaging. Simply stated, TCP/IP allows blocks of binary data to be exchanged between computers. It is also a world-wide standard that serves as the foundation for the World Wide Web. The primary function of TCP is to ensure that all packets of data are received correctly, while IP makes sure that messages are correctly addressed and routed. Note that the TCP/IP combination is merely a transport protocol, and does not define what the data means or how the data is to be interpreted (this is the job of the application protocol, Modbus in this case). So in summary, Modbus TCP/IP uses TCP/IP and Ethernet to carry the data of the Modbus message structure between compatible devices. That is, Modbus TCP/IP combines a physical network (Ethernet), with a networking standard (TCP/IP), and a standard method of representing data (Modbus as the application protocol). Essentially, the Modbus TCP/IP message is simply a Modbus communication encapsulated in an Ethernet TCP/IP wrapper.

In practice, Modbus TCP embeds a standard Modbus data frame into a TCP frame, without the Modbus checksum, as shown in the following diagram.

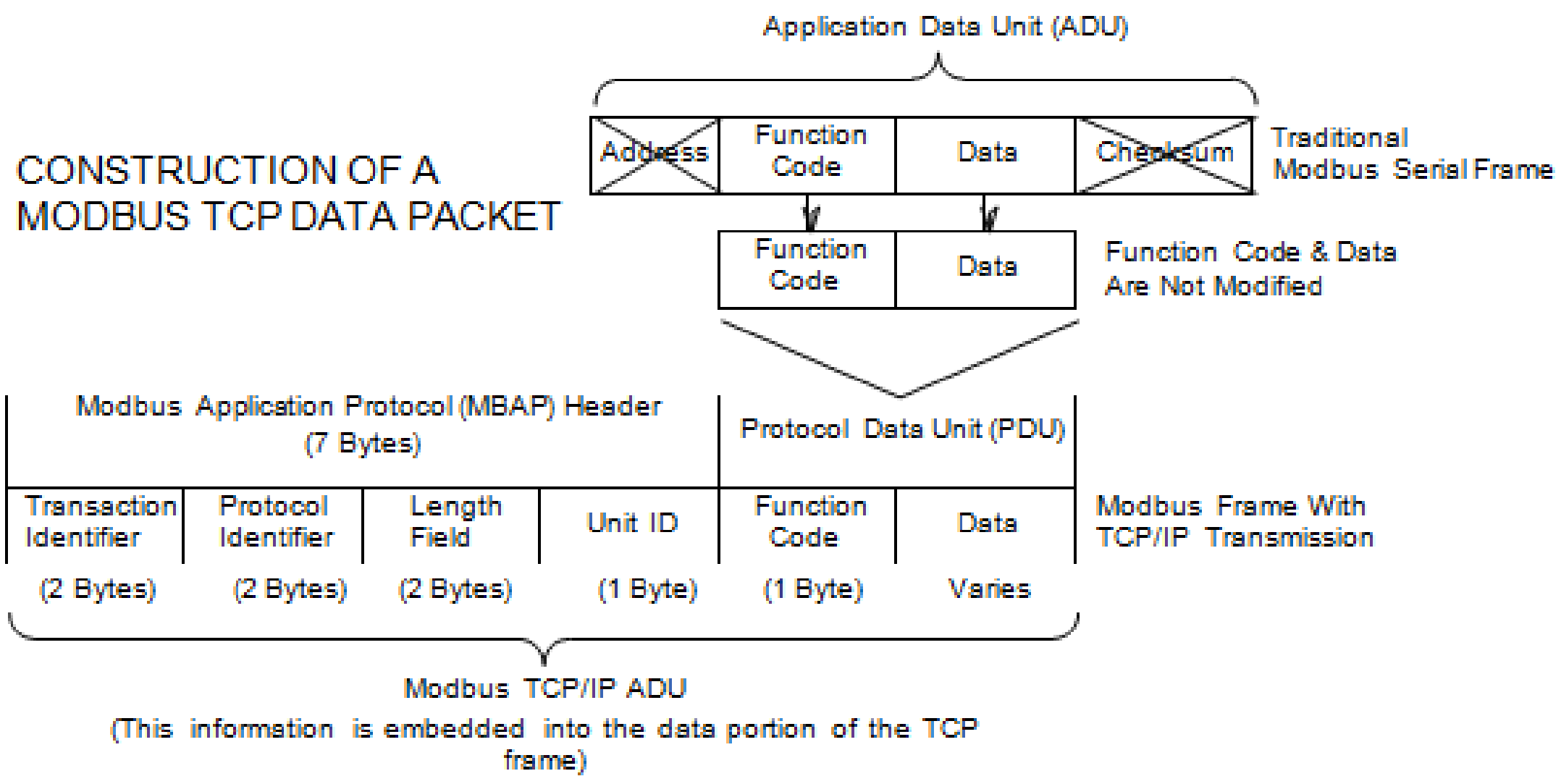

Figure 15 Construction of a Modbus TCP Data Packet

The Modbus commands and user data are themselves encapsulated into the data container of a TCP/IP telegram without being modified in any way. However, the Modbus error checking field (checksum) is not used, as the standard Ethernet TCP/IP link layer checksum methods are instead used to guaranty data integrity. Further, the Modbus frame address field is supplanted by the unit identifier in Modbus TCP/IP, and becomes part of the Modbus Application Protocol (MBAP) header (more on this later).

From the figure, we see that the function code and data fields are absorbed in their original form. Thus, a Modbus TCP/IP Application Data Unit (ADU) takes the form of a 7 byte header (transaction identifier + protocol identifier + length field + unit identifier), and the protocol data unit (function code + data). The MBAP header is 7 bytes long and includes the following fields:

• **Transaction/invocation Identifier (2 Bytes):** This identification field is used for transaction pairing when multiple messages are sent along the same TCP connection by a client without waiting for a prior response.

• **Protocol Identifier (2 bytes):** This field is always 0 for Modbus services and other values are reserved for future extensions.

• **Length (2 bytes):** This field is a byte count of the remaining fields and includes the unit identifier byte, function code byte, and the data fields.

• **Unit Identifier (1 byte):** This field is used to identify a remote server located on a non TCP/IP network (for serial bridging). In a typical Modbus TCP/IP server application, the unit ID is set to 00 or FF, ignored by the server, and simply echoed back in the response. The complete Modbus TCP/IP Application Data Unit is embedded into the data field of a standard TCP frame and sent via TCP to well-known system port 502, which is specifically reserved for Modbus applications. Modbus TCP/IP clients and servers listen and receive Modbus data via port 502.

We can see that the operation of Modbus over Ethernet is nearly transparent to the Modbus register/command structure..

## 4.4 Why Combine Modbus With Ethernet?

IEEE 802.3 Ethernet is a long-standing office networking protocol that has gained universal world-wide acceptance. It is also an open standard that is supported by many manufacturers and its infrastructure is widely available and largely installed. Consequently, its TCP/IP suite of protocols is used world-wide and even serves as the foundation for access to the World Wide Web. As many devices already support Ethernet, it is only natural to augment it for use in industrial applications.

Just as with Ethernet, Modbus is freely available, accessible to anyone, and widely supported by many manufacturers of industrial equipment. It is also easy to understand and a natural candidate for use in building other industrial communication standards. With so much in common, the marriage of the Modbus application protocol with traditional IEEE 802.3 Ethernet transmission forms a powerful industrial communication standard in Modbus TCP/IP. And because Modbus TCP/IP shares the same physical and data link layers of traditional IEEE 802.3 Ethernet and uses the same TCP/IP suite of protocols, it remains fully compatible with the already installed Ethernet infrastructure of cables, connectors, network interface cards, hubs, and switches.

## 4.5 What about Determinism?

Determinism is a term that is used here to describe the ability of the communication protocol to guaranty that a message is sent or received in a finite and predictable amount of time. We can surmise that, for critical control applications, determinism is very important. Historically, traditional Ethernet was not considered a viable field bus for industrial control and I/O networks because of two major shortcomings: inherent non-determinism, and low durability. However, new technology properly applied has mostly resolved these issues.

Originally, Ethernet equipment was designed for the office environment, not harsh industrial settings. Although, many factory Ethernet installations can use this standard hardware without a problem, new industrial-rated connectors, shielded cables, and hardened switches and hubs are now available to help resolve the durability issue. With respect to the non-deterministic behavior of Ethernet, this is largely a result of the arbitration protocol it uses for carrier transmission access on the network. That is, Carrier Sense Multiple Access with Collision Detect (CSMA/CD). Since any network device can try to send a data frame at any time, with CSMA/CD applied, each device will first sense whether the line is idle and available for use. If the line is available, the device will then begin to transmit its first frame. If another device also tries to send a frame at approximately the same time, then a collision occurs and both frames will be discarded. Each device then waits a random amount of time and retries its transmission until its frame is successfully sent. This channel-allocation method is inherently non-deterministic because a device may only transmit when the wire is free, resulting in unpredictable wait times before data may be transmitted. Additionally, because of cable signaling delay, collisions are still possible once the device begins to transmit the data, thus forcing additional retransmission/retry cycles.

As most control systems have a defined time requirement for packet transmission, typically less than 100ms, the potential for collisions and the CSMA/CD method of retransmission is not considered deterministic behavior and this is the reason that traditional Ethernet has had problems being accepted for use in critical control applications. However, CSMA/CD is naturally suppressed on a network of devices that are interconnected via Ethernet switches. Specifically, one device per switch port. This makeup is commonly referred to as switched Ethernet in an effort to distinguish itself from the non-deterministic behavior of traditional Ethernet.

Ethernet is made more deterministic via the use of fast Ethernet switches to interconnect devices. These switches increase the bandwidth of large networks by sub-dividing them into several smaller networks or separate "collision domains". The switch also minimizes network chatter by facilitating a direct connection from a sender to a receiver in such a way that only the receiver receives the data, not the entire network. So how does a switch (or switching hub) work to increase determinism? Each port of a switch forwards data to another port based on the MAC address contained in the received data packet/frame. The switch actually learns and stores the MAC addresses of every device it is connected to, along with the associated port number. Now the port of the switch does not require its own MAC address, and during retransmission of a received packet, the switch port will instead look like the originating device by having assumed its source address. In this way, the Ethernet collision domain is said to terminate at the switch port, and the switch effectively breaks the network into separate distinct data links or collision domains, one at each switch port. The ability of the switch to target a packet to a specific port, rather than forwarding it to all switch ports, also helps to eliminate the collisions that make Ethernet non-deterministic.

So, as switches have become less expensive, the current tendency in critical industrial control applications is to connect one Ethernet device per switch port, effectively treating the switch device as the hub of a star network. Since there is only one device connected to a port, there is no chance of collisions occurring. This effectively suppresses the CSMA/CD routine. In this manner, with only one network device connected per switch port, the switch can run full-duplex, with no chance of collisions. Thus, a 10/100 Ethernet switch effectively runs at 20/200 Mbps because it can transmit and receive at 10 or 100 Mbps simultaneously in both directions (full duplex). The higher transfer speed of full-duplex coupled without the need for invoking CSMA/CD produces a more deterministic mode of operation, helping critical control applications to remain predictable and on- time.

Unfortunately, broadcast traffic on a company network cannot be completely filtered by switches, and this may cause additional collisions reducing the determinism of a network connecting more than one device to a switch port. However, if the company network and the control & I/O network are instead separated, no traffic is added to the control network and its determinism is increased. Further, if a bridge is used to separate the two networks, then the bridge can usually be configured to filter unnecessary traffic.

So we see how combining good network design with fast switches and bridges where necessary raises the determinism of a network, making Ethernet more appealing. Other advances in Ethernet switches, such as, higher speeds, broadcast storm protection, virtual LAN support, SNMP, and priority messaging further help to increase the determinism of Ethernet networks. As Gigabit (Gbit), 10Gbit, and 100Gbit Ethernet enters the market, determinism will no longer be a concern.

## 4.6 THE OSI NETWORK MODEL

In order to better understand how Modbus TCP/IP is structured and the meaning of the term "open standard", we need to review the Open Systems Interconnect (OSI) Reference Model. This model was developed by the International Standards Organization and adopted in 1983 as a common reference for the development of data communication standards, like Modbus TCP/IP. It does not attempt to define an actual implementation, but rather it serves as a structural aide to understanding "what must be done" and "what goes where".

The traditional OSI model is presented below, along with the simplified 5- layer TCP/IP Standard (layers 5 & 6 suppressed). In the OSI model, the functions of communication are divided into seven (or five) layers, with every layer handling precisely defined tasks. For example, Layer 1 of this model is the physical layer and defines the physical transmission characteristics. Layer 2 is the data link layer and defines the bus access protocol. Layer 7 is the application layer and defines the application functions (this is the layer that defines how device data is to be interpreted).

By the OSI Model, we can infer that in order for two devices to be interoperable on the same network, they must have the same application- layer protocol. In the past, many network devices have used their own proprietary protocols and this has hindered their interoperability. This fact further drove the need for adoption of open network I/O solutions that would allow devices from a variety of vendors to seamlessly work together, and this drive for interoperability is a key reason Modbus TCP/IP was created.

Note that in the TCP/IP Standard Model, Ethernet handles the bottom 2 layers (1 & 2) of the seven layer OSI stack, while TCP/IP handles the next two layers (3 & 4). The application layer lies above TCP, IP, and Ethernet and is the layer of information that gives meaning to the transmitted data.

With Acromag 9xxEN-40xx Modbus TCP/IP modules, the application layer protocol is Modbus. That is, Modbus TCP/IP uses Ethernet media and TCP/IP to communicate using an application layer with the same register access method as Modbus RTU. Because many manufacturers happen to support Modbus RTU and TCP/IP, and since Modbus is also widely understood and freely distributed, Modbus TCP/IP is also considered an open standard.

| OSI 7-LAYER MODEL | | | TCP/IP Standard |
|---|---|---|---|
| 7 | Application | Used by software applications to prepare and interpret data for use by the other six OSI layers below it. Provides the application interface to the network. HTTP, FTP, email SMTP & POP3, CIP™, SNMP, are all found at this layer. | Application Layer |
| 6 | Presentation | Representation of data, coding type, and defines used characters. Performs data and protocol negotiation and conversion to ensure that data may be exchanged between hosts and transportable across the network. Also performs data compression and encryption. | |
| 5 | Session | Dialing control and synchronization of session connection. Responsible for establishing and managing sessions/ connections between applications & the network. Windows WinSock socket API is a common session layer manager. | |
| 4 | Transport | Sequencing of application data, controls start/end of transmission, provides error detection, correction, end-to-end recovery, and clearing. Provides software flow control of data between networks. TCP & UDP are found here. | Transport Layer |
| 3 | Network | Controls routing, prioritization, setup, release of connections, flow control. Establishes/maintains connections over a network & provides addressing, routing, and delivery of packets to hosts. IP, PPP, IPX, & X.25 are found here. | Internet, Network, or Internetwork Layer |
| 2 | Data Link | Responsible for ensuring reliable delivery at the lowest levels, including data frame, error detection and correction, sequence control, and flow control. Ethernet (IEEE 802.2) and MAC are defined at this level. | Network Access Layer<br><br>or<br><br>Host-to-Network Layer |
| 1 | Physical | Defines the electrical, mechanical, functional, and procedural attributes used to access and send a binary data stream over a physical medium (defines the RJ-45 connector & CAT5 cable of Ethernet). | |

Table 3: OSI 7-Layer Model (37)

So we see that Modbus TCP/IP is based on the TCP/IP protocol family and shares the same lower four layers of the OSI model common to all Ethernet devices. This makes it fully compatible with existing Ethernet hardware, such as cables, connectors, network interface cards, hubs, and switches.

### 4.6.1 The TCP/IP Stack

TCP/IP refers to the Transmission Control Protocol and Internet Protocol which were first introduced in 1974. TCP/IP is the foundation for the World Wide Web and forms the transport and network layer protocol of the internet that commonly links all Ethernet installations world-wide. Simply stated, TCP/IP allows blocks of binary data to be exchanged between computers. The primary function of TCP is to ensure that all packets of data are received correctly, while IP makes sure that messages are correctly addressed and routed. TCP/IP does not define what the data means or how the data is to be interpreted; it is merely a transport protocol.
To contrast, Modbus is an application protocol. It defines rules for organizing and interpreting data and is essentially a messaging structure that is independent of the underlying physical layer. It is freely available and accessible to anyone, easy to understand, and widely supported by many manufacturers.
Modbus TCP/IP uses TCP/IP and Ethernet to carry the data of the Modbus message structure between devices. That is, Modbus TCP/IP combines a physical network (Ethernet), with a networking standard (TCP/IP), and a standard method of representing data (Modbus).
TCP/IP is actually formed from a "suite" of protocols upon which all internet communication is based. This suite of protocols is also referred to as a protocol stack. Each host or router on the internet must run a protocol stack. The use of the word stack refers to the simplified TCP/IP layered Reference Model or "stack" that is used to design network software and outlined as follows:

**Table 4: TCP/IP Layered Reference Model**

| | | |
|---|---|---|
| 5 | Application | Specifies how an application uses a network. |
| 4 | Transport | Specifies how to ensure reliable data transport. |
| 3 | Internet/Network | Specifies packet format and routing. |
| 2 | Host-to-Network | Specifies frame organization and transmittal. |
| 1 | Physical | Specifies the basic network hardware. |

To better understand stack operation, the following table illustrates the flow of data from a sender to a receiver using the TCP/IP stack (we've renamed the Host-to-Network layer to the more commonly used Data Link Layer):

**TABLE 5: Flow of data from sender to a receiver using TCP/IP STACK**

| **SENDER** | → *Virtual Connection* → | **RECEIVER** |
|---|---|---|
| ↓ Application | ← *Equivalent Message* → | Application ↑ |
| ↓ Transport | ← *Equivalent Message* → | Transport ↑ |
| ↓ Internet/Network | ← *Equivalent Message* → | Internet/Network ↑ |
| ↓ Data Link Layer | ← *Equivalent Message* → | Data Link Layer ↑ |
| → Physical Hardware | →→→→→→→→→→→ | Physical Hardware ↑ |

Each layer on the sending stack communicates with the corresponding layer of the receiving stack through information stored in headers. As you move the data down the stack of the sender, each stack layer adds its own header to the front of the message that it receives from the next higher layer. That is, the higher layers are encapsulated by the lower layers. Conversely, this header information is removed by the corresponding layer at the Receiver. In this way, the headers are essentially peeled off as the data packet moves up the receiving stack to the receiver application.

**Table 6: Modbus TCP/IP Communication Stack**

| # | MODEL | IMPORTANT PROTOCOLS | Reference |
|---|---|---|---|
| 7 | Application | Modbus | |
| 6 | Presentation | | |
| 5 | Session | | |
| 4 | Transport | TCP | |
| 3 | Network | IP, ARP, RARP | |
| 2 | Data Link | Ethernet, CSMA/CD, MAC | IEEE 802.3 Ethernet |
| 1 | Physical | Ethernet Physical Layer | |

The following figure illustrates the construction of a TCP/IP-Ethernet packet for transmission. For Modbus TCP/IP, the application layer is Modbus and the Modbus Application Data Unit is embedded into the TCP data array. When an application sends its data over the network, the data is passed down through each layer--note how the upper layer information is wrapped into the data bytes of the next lowest layer (encapsulated). Each subsequent layer has a designated function and attaches its own protocol header to the front of its packet. The lowest layer is responsible for actually sending the data. This entire wrap-into procedure is then reversed for data received (the data received is unwrapped at each level and passed up thorough to the receiver's application layer).

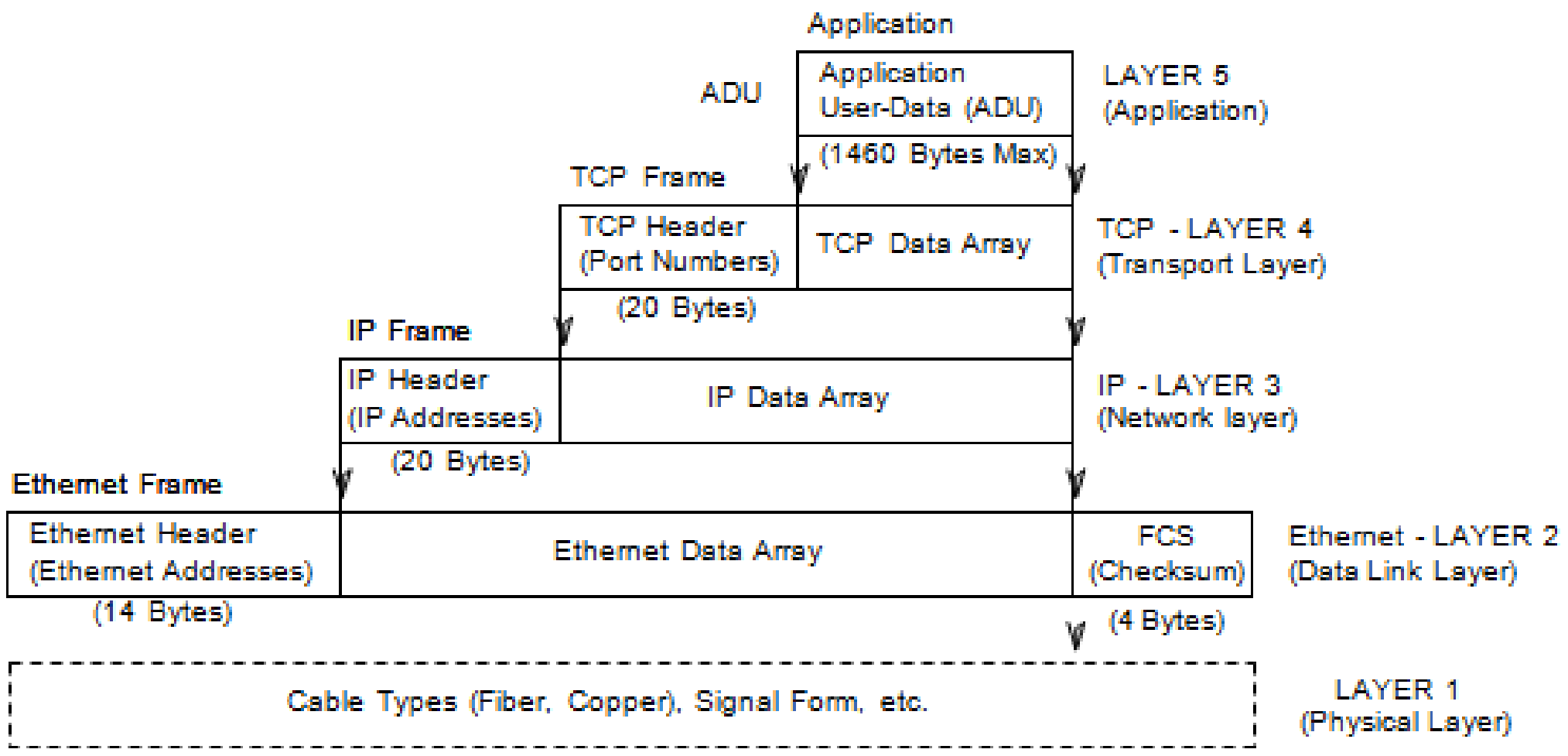

Figure 16 Communication protocol-I

To illustrate, with Modbus TCP/IP, the host (master/client) application forms its request, then passes its data down to the lower layers, which add their own control information to the packet in the form of protocol headers and sometimes footers. Finally the packet reaches the physical layer where it is electronically transmitted to the destination (slave/server). The packet then travels up through the different layers of its destination with each layer decoding its portion of the message and removing the header and footer that was attached by the same layer of the sending client computer. Finally the packet reaches the destination application. Although each layer only communicates with the layer just above or just below it, this process can be viewed as one layer at one end talking to its partner (peer) layer at the opposite end.

## 4.7 APPLICATION LAYER

The uppermost layer of the TCP/IP and OSI Reference Models is the Application Layer. There are many application layer protocols that may reside here, such as FTP, Telnet, HTTP, SMPT, DNS, and NNTP, among others. While each of these protocols has their own specific purpose, for Modbus TCP/IP, the primary application layer protocol of interest is Modbus.

### 4.7.1 Modbus Functions and Registers

The TCP/IP protocol suite (or stack of independent protocols) provides all the resources for two devices to communicate with each other over an Ethernet Local-Area Network (LAN), or global Wide-Area Network (WAN). But TCP/IP only guarantees that application messages will be transferred between these devices, it does not guaranty that these devices will actually understand or interoperate with one another. For Modbus TCP/IP, this capability is provided by the application layer protocol Modbus.

Modbus is an application protocol or messaging structure that defines rules for organizing and interpreting data independent of the data transmission medium. Traditional serial Modbus is a register-based protocol that defines message transactions that occur between masters and slaves. Slave devices listen for communication from the master and simply respond as instructed. The master always controls the communication and may communicate directly to one slave, or all connected slaves, but the slaves cannot communicate directly with each other.

Thus, we see that Modbus operates according to the common client/server (master/slave) model. That is, the client (master) sends a request telegram (service request) to the server (slave), and the server replies with a response telegram. If the server cannot process a request, it will instead return an error function code (exception response) that is the original function code plus 80H (i.e. with its most significant bit set to 1).

Modbus functions operate on memory registers to configure, monitor, and control device I/O. Modbus devices usually include a Register Map. You should refer to the register map for your device to gain a better understanding of its operation. You will also find it helpful to refer to the register map as you review the Modbus functions described later in this document.

The Modbus data model has a simple structure that only differentiates between four basic data types:

Discrete Inputs
Coils (Outputs)
Input Registers (Input Data)
Holding Registers (Output Data)

The service request (Modbus Protocol Data Unit) is comprised of a function code, and some number of additional data bytes, depending on the function. In most cases, the additional data is usually a variable reference, such as a register address, as most Modbus functions operate on registers.

The Modbus registers of a device are organized around the four basic data reference types noted above and this data type is further identified by the leading number of the reference address as follows:

**Table 7: Address Format Protocol-I**

| Reference | Description |
|---|---|
| 0xxxx | Read/Write Discrete Outputs or Coils. A 0x reference address is used to drive output data to a digital output channel. |
| 1xxxx | Read Discrete Inputs. The ON/OFF status of a 1x reference address is controlled by the corresponding digital input channel. |
| 3xxxx | Read Input Registers. A 3x reference register contains a 16-bit number received from an external source—e.g. an analog signal. |
| 4xxxx | Read/Write Output or Holding Registers. A 4x register is used to store 16-bits of numerical data (binary or decimal), or to send the data from the CPU to an output channel. |

IMPORTANT: The reference addresses noted in the memory map are not explicit hard-coded memory addresses. Internally, all Modbus devices use a zero-based memory offset computed from the reference address. However, the system interface of Modbus systems (software) will vary in this regard and may require you to enter the actual reference address, drop the leading number, or enter an absolute memory offset from 1, or a memory address offset from 0. This is system dependent and a common source of programming errors. Be wary of this when writing higher-level application programs to access these registers.

Note that not all Modbus functions operate on register map registers. All data addresses in Modbus messages are referenced to 0, with the first occurrence of a data item addressed as item number zero. Further, a function code field already specifies which register group it operates on (i.e. 0x, 1x, 3x, or 4x reference addresses). For example, holding register 40001 is addressed as register 0000 in the data address field of the message. The function code that operates on this register specifies a "holding register" operation and the "4xxxx" reference group is implied. Thus, holding register 40108 is actually addressed as register 006BH (107 decimal).

The function code field of the message (PDU) will contain one byte that tells the slave what kind of action to take. Valid function codes are from 1-255, but not all codes will apply to a module and some codes are reserved for future use. Additionally, the Modbus specification allocates function codes 65-72 and 100-110 for user-defined services.

The following table highlights a subset of standard Modbus functions commonly supported by Acromag modules (the reference register addresses that the function operates on are also indicated). The functions below are used to access the registers outlined in the register map of the module for sending and receiving data. The Report Slave ID command does not operate on a register map register.

**Table 8: Address Format Protocol-II**

| CODE | FUNCTION | REFERENCE |
|---|---|---|
| 01 (01H) | Read Coil (Output) Status | 0xxxx |
| 03 (03H) | Read Holding Registers | 4xxxx |
| 04 (04H) | Read Input Registers | 3xxxx |
| 05 (05H) | Force Single Coil (Output) | 0xxxx |
| 06 (06H) | Preset Single Register | 4xxxx |
| 15 (0FH) | Force Multiple Coils (Outputs) | 0xxxx |
| 16 (10H) | Preset Multiple Registers | 4xxxx |
| 17 (11H) | Report Slave ID | *Hidden* |

The client request data field provides the slave (server) with any additional information required by the slave to complete the action specified by the function code in the client's request. The data field typically includes register addresses, count values, and written data. For some messages, this field may not exist (has zero length), as not all messages will require data.

When the slave device responds to the master, it uses the function code field to indicate either a normal (error-free) response, or that some kind of error has occurred (an exception response). A normal response simply echoes the original function code of the query, while an exception response returns a code that is equivalent to the original function code with its most significant bit (MSB) set to logic 1.

For example, the Read Holding Registers command has the function code 0000 0011 (03H). If the slave device takes the requested action without error, it returns the same code in its response. However, if an exception occurs, it returns 1000 0011 (83H) in the function code field and appends a unique code in the data field of the response message that tells the master device what kind of error occurred, or the reason for the exception (see Modbus Exceptions).

The client application program must handle the exception response. It may choose to post subsequent retries of the original message, it may try sending a diagnostic query, or it may simply notify the operator of the exception error. The following paragraphs describe some of the Modbus functions commonly supported by Acromag 9xxEN modules. These examples are depicted from a Modbus TCP/IP perspective (the unit ID replaces the slave address; the traditional serial CRC/LRC error checking field is dropped). Only the first example will include the MBAP header information. You should refer to the Modbus specification for a complete description of all Modbus functions. To gain a better understanding of Modbus, please refer to your module's register map as you review this material.

When you review these examples and compare them to traditional serial Modbus commands, note that the slave address is supplanted by the unit identifier in Modbus TCP/IP (normally set to 00H or FFH). In addition, the error check field (CRC/LRC) is removed, as TCP/IP already applies its own error checking. For commands that support broadcast transmission, this applies to serial Modbus only, as Modbus TCP/IP is unicast only (except where an Ethernet-to-serial bridge is used).

The different fields of the of the Modbus TCP/IP ADU are encoded in Big- Endian format. This means that the most significant byte in the sequence is stored at the lowest storage address (i.e. it is first).The following example will include the format of the MBAP header information, but the header information will not be repeated in the successive examples.

#### 4.7.1.1 Read Coil Status (01)

This command will read the ON/OFF status of discrete outputs or coils (0x reference addresses) in the slave/server. For Acromag modules, its response is equivalent to reading the on/off status of solid-state output relays or switches. Broadcast transmission is not supported.

The Read Coil Status query specifies the starting coil (output channel) and quantity of coils to be read. Coils correspond to the discrete solid-state relays of this device and are addressed starting from 0 (up to 4 coils addressed as 0-3 for this model). The Read Coil Status in the response message is packed as one coil or channel per bit of the data field. For Acromag modules, the output status is indicated as 1 for ON (conducting current), and 0 for OFF (not conducting). The LSB of the first data byte corresponds to the status of the coil addressed in the query. The other coils follow sequentially, moving toward the high order end of the byte. Since this example has only 4 outputs, the remaining bits of the data byte will be set to zero toward the unused high order end of the byte.

**Table 9: Modbus Request ADU Example Header**

| MBAP Header Fields | Example Decimal (Hexadecimal) |
|---|---|
| Transaction ID High Order | 0 (00) *Client sets, unique value.* |
| Transaction ID Low Order | 1 (01) *Client sets, unique value.* |
| Protocol Identifier High Order | 0 (00) *Specifies Modbus service.* |
| Protocol Identifier Low Order | 0 (00) *Specifies Modbus service.* |
| Length High Order | 0 (00) *Client calculates.* |
| Length Low Order | 6 (06) *Client calculates.* |
| Unit Identifier | 255 (FF) or 0 (00) *Do not bridge.* |

The transaction identifier is used to match the response with the query when the client sends multiple queries without waiting for a prior response. It is typically a number from 1 to 16, but the maximum number of client transactions and the maximum number of server transactions will vary according to the device. The protocol identifier is always 0 for Modbus. The length is a count of the number of bytes contained in the data plus the function code (1 byte) and unit identifier (1 byte).

The unit identifier is 00H or FFH, as this module is Modbus TCP/IP. If this module was a traditional serial Modbus type (no Ethernet port), and it was being addressed via a bridge or gateway from an Ethernet client (an Ethernet-to-serial bridge), then the unit identifier is equivalent to the traditional serial Modbus slave address (1-247). Using 00H or FFH as shown here will cause the any serial bridge/gateway device to block the passage of this client message across the bridge. This is why some text will show the unit identifier as part of the query itself since it supplants the traditional slave address (note that the length includes the unit identifier byte), while others show it as part of the MBAP header as is done here.

**Table 10: Modbus Request ADU Example - Read Coil Status Query**

| Field Name | Example Decimal (Hexadecimal) |
|---|---|
| Function Code | 1 (01) |
| Starting Address High Order | 0 (00) |
| Starting Address Low Order | 0 (00) |
| Number Of Points High Order | 0 (00) |
| Number Of Points Low Order | 4 (04) |

Note that the leading character of the 0x reference address is implied by the function code and omitted from the address specified. In this example, the first address is 00001, referenced via 0000H, and corresponding to coil 0.

**Table 11: Modbus Response ADU Example Header**

| MBAP Header Fields | Example Decimal (Hexadecimal) |
|---|---|
| Transaction ID High Order | 0 (00) *Echoed back, no change* |
| Transaction ID Low Order | 1 (01) *Echoed back, no change* |
| Protocol Identifier High Order | 0 (00) *Echoed back, no change* |
| Protocol Identifier Low Order | 0 (00) *Echoed back, no change* |
| Length High Order | 0 (00) *Server calculates* |
| Length Low Order | 4 (04) *Server calculates.* |
| Unit Identifier | 255 (FF) or 0 (00) *No change* |

**Table 12: Modbus Response ADU Example - Read Coil Status Response**

| Field Name | Example Decimal (Hexadecimal) |
|---|---|
| Function Code | 1 (01) |
| Byte Count | 1 (01) |
| Data (Coils 3-0) | 10 (0A) |

Note that the response function code is the same as the request function code. The transaction identifier is preserved by the server and returned. The protocol identifier remains 0 for Modbus. The length of the response is calculated by the server and is the size of the Modbus server's PDU, plus the unit identifier (1 byte). The unit identifier is the same as what was received from the client.

If an error had occurred, the response function code is modified and set equal to the request function code plus 80H. The transaction ID, protocol ID, and unit identifier stay the same. The length becomes 0002H (2 bytes). The PDU then becomes the exception code value itself (1 byte). Refer to Modbus Exceptions for information on exception codes. To summarize, the status of coils 3-0 is shown as the byte value 0A hex, or 00001010 binary. Coil 3 is the fifth bit from the left of this byte, and coil 0 is the LSB. The four remaining bits (toward the high-order end) are zero. Reading left to right; the output status of coils 3.0 is ON-OFF-ON-OFF.

| Bin | 0 | 0 | 0 | 0 | 1 | 0 | 1 | 0 |
|---|---|---|---|---|---|---|---|---|
| Hex | 0 | | | | A | | | |
| Coil | NA | NA | NA | NA | 3 | 2 | 1 | 0 |

The following examples do not repeat the information contained in the MBAP request and response headers, only the Modbus PDU information is provided. Refer to the previous example for MBAP header format.

#### 4.7.1.2 Read Holding Registers (03)

Reads the binary contents of holding registers (4x reference addresses) in the slave device. Broadcast transmission is not supported. The Read Holding Registers query specifies the starting register and quantity of registers to be read. Note that registers are addressed starting at 0 (registers 1-16 addressed as 0-15). The Read Holding Registers response message is packed as two bytes per register, with the binary contents right-justified in each byte. For each register, the first byte contains the high order bits and the second byte the low order bits.

**Table 13: Modbus PDU Example - Read Holding Register Query**

| Field Name | Example Decimal (Hexadecimal) |
|---|---|
| Function Code | 3 (03) |
| Starting Address High Order | 0 (00) |
| Starting Address Low Order | 5 (05) |
| Number Of Points High Order | 0 (00) |
| Number Of Points Low Order | 3 (03) |

**Table 14: Modbus PDU Example - Read Holding Register Response**

| Field Name | Example Decimal (Hexadecimal) |
|---|---|
| Function Code | 3 (03) |
| Byte Count | 6 (06) |
| Data High (Register 40006) | (3A) |
| Data Low (Register 40006) | 75%=15000 (98) |
| Data High (Register 40007) | (13) |
| Data Low (Register 40007) | 25%=5000 (88) |
| Data High (Register 40008) | (00) |
| Data Low (Register 40008) | 1%=200 (C8) |

To summarize our example, the contents of register 40006 (2 bytes) is the channel 0 high limit of 75% (15000=3A98H). The contents of register 40007 (2 bytes) is the channel 0 low limit of 25% (5000=1388H). The contents of register 40008 is the channel 0 deadband value (2 bytes) of 1% (200=00C8H).

#### 4.7.1.3 Read Input Registers (04)

This command will read the binary contents of input registers (3x reference addresses) in the slave device. Broadcast transmission is not supported. The Read Input Registers query specifies the starting register and quantity of registers to be read. Note that registers are addressed starting at 0. That is, registers 1-16 are addressed as 0-15. The Read Input Registers response message is packed as two bytes per register, with the binary contents right-justified in each byte. For each register, the first byte contains the high order bits and the second byte the low order bits.

**Table 15: Modbus PDU Example - Read Input Registers Query**

| Field Name | Example Decimal (Hexadecimal) |
|---|---|
| Function Code | 4 (04) |
| Starting Address High Order | 0 (00) |
| Starting Address Low Order | 2 (02) |
| Number Of Points High Order | 0 (00) |
| Number Of Points Low Order | 2 (02) |

To summarize our example, the contents of register 30003 (2 bytes) is the channel 1 input value of 80% (16000=3E80H). The contents of register 30004 (2 bytes) is the channel 0 status flags of 136 (0088H)—i.e. flagging high limit exceeded.

#### 4.7.1.4 Force Single Coil (05)

Forces a single coil/output (0x reference address) ON or OFF. With broadcast transmission (address 0), it forces the same coil in all networked slaves (serial Modbus only). The Force Single Coil query specifies the coil reference address to be forced, and the state to force it to. The ON/OFF state is indicated via a constant in the query data field. A value of FF00H forces the coil to be turned ON (i.e. the corresponding solid-state relay is turned ON or closed), and 0000H forces the coil to be turned OFF (i.e. the solid-state output relay is turned OFF or opened). All other values are invalid and will not affect the coil.

Coils are referenced starting at 0—up to 4 coils are addressed as 0-3 for our example and this corresponds to the discrete output channel number.

**Table 16: Modbus PDU Example - Force Single Coil Query and Response**

| Field Name | Example Decimal (Hexadecimal) |
|---|---|
| Function Code | 5 (05) |
| Coil Address High Order | 0 (00) |
| Coil Address Low Order | 3 (03) |
| Force Data High Order | 255 (FF) |
| Force Data Low Order | 0 (00) |

The Force Single Coil response message is simply an echo (copy) of the query as shown above, but returned after executing the force coil command. No response is returned to broadcast queries from a master device (serial Modbus only).

#### 4.7.1.5 Preset Single Register (06)

This command will preset a single holding register (4x reference address) to a specific value. Broadcast transmission is supported by this command (serial Modbus only) and will act to preset the same register in all networked slaves.

The Preset Single Register query specifies the register reference address to be preset, and the preset value. Note that registers are addressed starting at 0--registers 1-16 are addressed as 0-15. The Preset Single Registers response message is an echo of the query, returned after the register contents have been preset.

**Table 17: Modbus PDU Example - Preset Holding Register Query and Response**

| Field Name | Example Decimal (Hexadecimal) |
|---|---|
| Function Code | 6 (06) |
| Register Address High Order | 0 (00) |
| Register Address Low Order | 1 (01) |
| Preset Data High Order | 0 (00) |
| Preset Data Low Order | 2 (02) |

The response message is simply an echo (copy) of the query as shown above, but returned after the register contents have been preset. No response is returned to broadcast queries from a master (serial Modbus only).

#### 4.7.1.6 Force Multiple Coils (15)

Simultaneously forces a series of coils (0x reference address) either ON or OFF. Broadcast transmission is supported by this command (serial Modbus only) and will act to force the same block of coils in all networked slaves.

The Force Multiple Coils query specifies the starting coil reference address to be forced, the number of coils, and the force data to be written in ascending order. The ON/OFF states are specified by the contents in the query data field. A logic 1 in a bit position of this field requests that the coil turn ON, while a logic 0 requests that the corresponding coil be turned OFF. Unused bits in a data byte should be set to zero. Note that coils are referenced starting at 0—up to 4 coils are addressed as 0-3 for this example and this also corresponds to the discrete output channel number.

**Table 18: Modbus PDU Example - Force Multiple Coils Query**

| Field Name | Example Decimal (Hexadecimal) |
|---|---|
| Function Code | 15 (0F) |
| Coil Address High Order | 0 (00) |
| Coil Address Low Order | 0 (00) |
| Number Of Coils High Order | 0 (00) |
| Number Of Coils Low Order | 4 (04) |
| Byte Count | 01 |
| Force Data High (First Byte) | 5 (05) |

Note that the leading character of the 0x reference address is implied by the function code and omitted from the address specified. In this example, the first address is 00001 corresponding to coil 0 and referenced via 0000H. Thus, in this example the data byte transmitted will address coils 3...0, with the least significant bit addressing the lowest coil in this set as follows (note that the four unused upper bits of the data byte are set to zero):

| Bin | 0 | 0 | 0 | 0 | 0 | 1 | 0 | 1 |
|---|---|---|---|---|---|---|---|---|
| Hex | 0 | | | | 5 | | | |
| Coil | NA | NA | NA | NA | 3 | 2 | 1 | 0 |

**Table 19: Modbus PDU Example - Force Multiple Coils Response**

| Field Name | Example Decimal (Hexadecimal) |
|---|---|
| Function Code | 15 (0F) |
| Coil Address High Order | 0 (00) |
| Coil Address Low Order | 0 (00) |
| Number Of Coils High Order | 0 (00) |
| Number Of Coils Low Order | 4 (04) |

The Force Multiple Coils normal response message returns the slave address, function code, starting address, and the number of coils forced, after executing the force instruction. Note that it does not return the byte count or force value. No response is returned to broadcast queries from a master device (serial Modbus).

### 4.7.1.7 Preset Multiple Registers (16)

Presets a block of holding registers (4x reference addresses) to specific values. Broadcast transmission is supported by this command and will act to preset the same block of registers in all networked slaves (serial Modbus only).

**Table 20: Modbus PDU Example - Preset Multiple Registers Query**

| Field Name | Example Decimal (Hexadecimal) |
|---|---|
| Function Code | 16 (10) |
| Starting Register High Order | 0 (00) |
| Starting Register Low Order | 0 (00) |
| Number Of Registers High Order | 0 (00) |
| Number Of Registers Low Order | 3 (03) |
| Preset Data High (First Register) | 0 (00) |
| Preset Data Low (First Register) | 200 (C8) |
| Preset Data High (Second Reg) | 0 (00) |
| Preset Data Low (Second Reg) | 5 (05) |
| Preset Data High (Third Reg) | 0 (00) |
| Preset Data Low (Third Reg) | 2 (02) |

**Table 21: Modbus PDU Example - Preset Multiple Registers Response**

| Field Name | Example Decimal (Hexadecimal) |
|---|---|
| Function Code | 16 (10) |
| Starting Register High Order | 0 (00) |
| Starting Register Low Order | 0 (00) |
| Number Of Registers High Order | 0 (00) |
| Number Of Registers Low Order | 3 (03) |

The Preset Multiple Registers query specifies the starting register reference address, the number of registers, and the data to be written in ascending order. Note that registers are addressed starting at 0--registers 1-16 are addressed as 0-15.

The Preset Multiple Registers normal response message returns the slave address, function code, starting register reference, and the number of registers preset, after the register contents have been preset. Note that it does not echo the preset values. No response is returned to broadcast queries from a master device (serial Modbus only).

### 4.7.1.8 Report Slave ID (17)

This command returns the model, serial, and firmware number for an Acromag slave/server device (97xEN for this example), the status of the Run indicator, and any other information specific to the device. This command does not address Register Map registers and broadcast transmission is not supported (serial Modbus).

**Table 22: Modbus PDU Example - Report Slave ID Query**

| Field Name | Example Decimal (Hexadecimal) |
|---|---|
| Function Code | 17 (11) |

**Table 23: Modbus PDU Example - Report Slave ID Response (Acromag 97xEN)**

| FIELD | DESCRIPTION |
|---|---|
| Unit ID | Echo Unit ID Sent In Query |
| Function Code | 11 |
| Byte Count | 42 |
| Slave ID (Model No.) | 08=972EN-4004 (4 Current Outputs)<br>09=972EN-4006 (6 Current Outputs)<br>0A=973EN-4004 (4 Voltage Outputs)<br>0B=973EN-4006 (6 Voltage Outputs) |
| Run Indicator Status | FFH (ON) |
| Firmware Number String (Additional Data Field) | 41 43 52 4F 4D 41 47 2C 39 33 30 30 2D **31 32 37** 2C 39 **37 32** 45 4E 2D **34 30 30 36** 2C<br>*30 31 32 33 34 35 41* 2C *30 31 32 33 34 35*<br>("ACROMAG,9300-**127**,9**72**EN-**4006**,*serial number&rev*,*six-byteMACID*") |

### 4.7.2 Supported Data Types

**Table 24: Summary Of Data Types Used By Acromag 900MB/900EN Modules**

| Data Types | Description |
|---|---|
| Count Value | A 16-bit signed integer value representing an A/D count, a DAC count, time value, or frequency with a range of –32768 to +32767. |
| Count Value | A 16-bit unsigned integer value representing an A/D count, a DAC count, time value, or frequency with a range of 0 to 65535. |
| Percentage | A 16-bit signed integer value with resolution of 0.005%/lsb. ±20000 is used to represent ±100%. For example, -100%, 0% and +100% are represented by decimal values –20000, 0, and 20000, respectively. The full range is –163.84% (-32768 decimal) to +163.835% (+32767 decimal). |
| Temperature | A 16-bit signed integer value with resolution of 0.1°C/lsb. For example, a value of 12059 is equivalent to 1205.9°C, a value of –187 equals –18.7°C. The maximum possible temperature range is –3276.8°C to +3276.7°C. |
| Discrete | A discrete value is generally indicated by a single bit of a 16-bit word. The bit number/position typically corresponds to the discrete channel number for this model. Unless otherwise defined for outputs, a 1 bit means the corresponding output is closed or ON, a 0 bit means the output is open or OFF. For inputs, a value of 1 means the input is in its high state (usually >> 0V), while a value of 0 specifies the input is in its low state (near 0V). |

All I/O values are accessed via 16-bit Input Registers or 16-bit Holding Registers (see Register Map). Input registers contain information that is read-only. For example, the current input value read from a channel, or the states of a group of digital inputs. Holding registers contain read/write information that may be configuration data or output data. For example, the high limit value of an alarm function operating at an input, or an output value for an output channel.

I/O values of Acromag modules take the following common forms of data to represent temperature, percentage, and discrete on/off, as required. This is not a Modbus standard and will vary between devices. With Modbus TCP/IP, error checking of the data is handled by the underlying TCP protocol and traditional serial Modbus error checking will not be reviewed here.

### 4.7.3 Modbus Exceptions

Recall that a server may generate an exception response to a client request and this is normally flagged by returning the original function code plus 80H (the original code with its most significant bit set). Additionally, it may also return an exception code in the data field of the response that can be used to trouble-shoot the problem.

For example, if a client requests an unsupported service (specifies an invalid function code), then the server may return exception code 01 (Illegal Function) in the data field of the response message. Likewise, if a holding register is written with an invalid value, then exception code 03 (Illegal Data Value) will be returned in the data field of the response message. The following table gives some common error codes:

**Table 25: Modbus Exception Codes**

| Code | Exception | Description |
|---|---|---|
| 01 | Illegal Function | The function code received in the query is not allowed or invalid. |
| 02 | Illegal Data Address | The data address received in the query is not an allowable address for the slave or is invalid. |
| 03 | Illegal Data Value | A value contained in the query data field is not an allowable value for the slave or is invalid. |
| 04 | Slave/Server Device Failure | The server failed during execution. An unrecoverable error occurred while the slave/server was attempting to perform the requested action. |
| 05 | Acknowledge | The slave/server has accepted the request and is processing it, but a long duration of time is required to do so. This response is returned to prevent a timeout error from occurring in the master. |
| 06 | Slave/Server Device Busy | The slave is engaged in processing a long-duration program command. The master should retransmit the message later when the slave is free. |
| 07 | Negative Acknowledge | The slave cannot perform the program function received in the query. This code is returned for an unsuccessful programming request using function code 13 or 14 (codes not supported by this model). The master should request diagnostic information from the slave. |
| 08 | Memory Parity Error | The slave attempted to read extended memory, but detected a parity error in memory. The master can retry the request, but service may be required at the slave device. |
| 0A | Gateway Problem | Gateway path(s) not available. |
| 0B | Gateway Problem | The target device failed to respond (the gateway generates this exception). |
| FF | Extended Exception Response | The exception response PDU contains extended exception information. A subsequent 2 byte length field indicates the size in bytes of this function-code specific exception information. |

In a normal response, the slave simply echoes the function code of the original query in the function field of the response. All function codes have their most-significant bit (msb) set to 0 (their values are below 80H). In an exception response, the slave sets the msb of the function code to 1 in the returned response (i.e. exactly 80H higher than normal) and returns the exception code in the data field. This is used by the client/master application to actually recognize an exception response and to direct an examination of the data field for the applicable exception code.

TCP is a data-stream based protocol, it may send almost any length IP packet it chooses, and it can parse this message as required. For example, it may encapsulate two back-to-back encapsulation messages in a single TCP/IP/MAC packet, or it may divide an encapsulation message across two separate TCP/IP/MAC packets.

### 4.7.4 Modbus TCP/IP ADU Format

In the introduction, we talked about how a traditional Modbus message (Modbus Application Data Unit) was stripped of its checksum and device address field, then combined with an MBAP header (Modbus Application Protocol), to build a Modbus TCP/IP Application Data Unit. This information is then nested into the data/payload field of a standard TCP frame, the total of which is then nested into the IP frame, which is then nested into the Ethernet/MAC frame for transmission over Ethernet. This nesting is the message encapsulation process that is commonly referred to. The following sections will attempt to describe the encapsulation that occurs at each layer as we move down the stack to the connection media, starting from the application layer, Modbus.

We know that the application layer is said to ride on top of TCP. Prior to passing the application message via TCP, a Modbus TCP/IP Application Data Unit is formed from a 7-byte Modbus Application Protocol (MBAP) header and the Protocol Data Unit (Modbus function code and data). This packet takes the following form (Refer to Modbus TCP/IP Application Data Unit (ADU)).

The 7-byte MBAP header includes the following fields:

- **Transaction/Invocation Identifier (2 Bytes):** This identification field is used for transaction pairing when several Modbus transactions are sent along the same TCP connection without waiting for completion of the prior transaction.
- **Protocol Identifier (2 bytes):** This field is always 0 for Modbus services and other values are reserved for future extensions.
- **Length (2 bytes):** This field is a byte count of the remaining fields and includes the destination identification and data fields.
- **Unit Identifier (1 byte):** This field is used to identify a remote server located on a non TCP/IP network (for bridging Ethernet to a serial sub-network). In a typical slave application, the unit ID is ignored and just echoed back in the response. It is recommended that a unit ID of FF be used to keep this value insignificant to a serial bridge or gateway (see below).

# Modbus TCP/IP Application Data Unit (ADU)

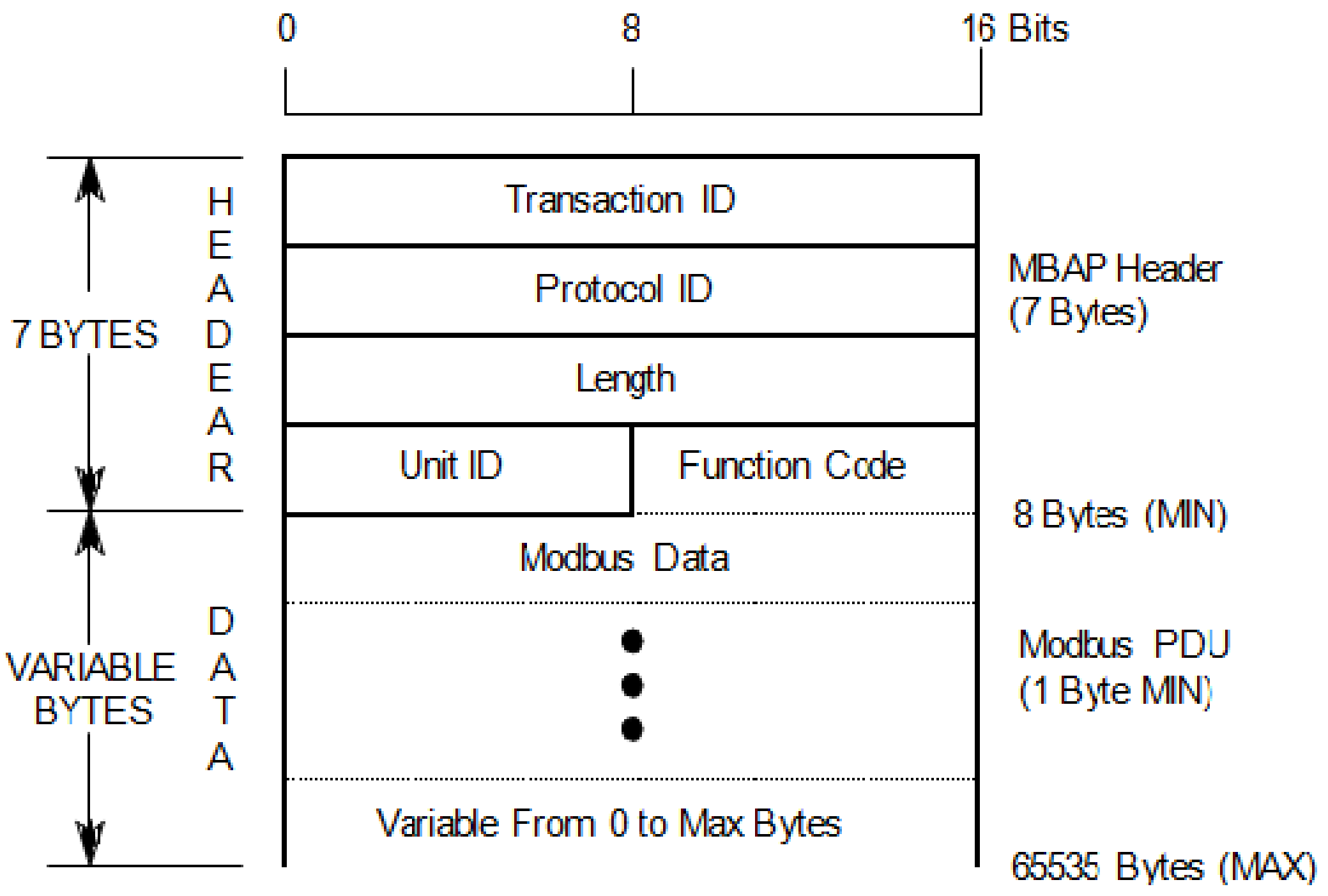

Figure 17 Modbus TCP/IP Application Data Unit

The Protocol Data Unit (PDU) is the Modbus function code and data field in their original form. The original Modbus error checking field (checksum) is not used, as the standard ethernet TCP/IP link layer checksum methods are instead used to guaranty data integrity. Further, the original Modbus device address field is supplanted by the unit identifier in Modbus TCP/IP and becomes part of the Modbus Application Protocol (MBAP) header. The original device address is not needed because Ethernet devices already contain their own unique MAC addresses. However, it is used if a serial bridge or gateway is being used to bridge Ethernet to a serial sub-network of Modbus devices.

With traditional serial Modbus, a client can only send one request at a time and must wait for an answer before sending a second request. However, Modbus TCP/IP devices may send several requests to the same server without waiting for the prior response. In this instance, the transaction identifier is use to match a future response with its originating request and must be unique per transaction. It is commonly a TCP sequence number driven by a counter that is incremented by each request. The maximum number of client transactions will vary from device to device, but is generally a number from 1 to 16. Likewise the maximum number of server transactions also varies. For Acromag 9xxEN-4 modules, this number is 10.

The unit identifier field was intended to facilitate communication between Modbus TCP/IP Ethernet devices and traditional Modbus serial devices by using a bridge or gateway to route traffic from one network to a serial line sub-network. In this case, the destination IP address identifies the bridge or gateway device to send the message to, while the bridge device itself uses the Modbus Unit Identifier to forward the request to the specific slave device of the sub network. Recall that serial Modbus uses addresses 1 to 247 decimal, and reserves 0 as a broadcast address. Thus, the unit identifier assumes the same assignment for these applications. Further, this is the only way that broadcast messages are supported with Modbus TCP/IP, as TCP alone only sends unicast (point-to-point) messages.

With TCP/IP devices, a Modbus server is addressed using its IP address, rendering the unit identifier non-functional and FFH is used in its place. The hex address FF remains non-significant to a gateway or bridge and will continue to be ignored if the network is later expanded or augmented with serial bridge or gateway devices.

The Modbus TCP/IP ADU is then inserted into the data field of a standard TCP frame and sent via TCP on well-known system port 502, which is specifically reserved for Modbus applications. Thus, this packet is encapsulated by the data frames imposed by the TCP/IP stack of protocols (TCP/IP/MAC) before being transmitted onto the network. The term encapsulation refers to the action of packing (embedding) this message into the TCP container, the IP container, and the MAC container. This lower level encapsulation is illustrated as follows:

| | | | | |
|---|---|---|---|---|
| Ethernet Header (14 Bytes) | IP Header (20 Bytes) | TCP Header (20 Bytes) | Encapsulation Message | CRC |

Figure 18 TCP/IP/MAC Encapsulation (Explicit Message)

Because TCP is a connection-oriented protocol, a TCP connection must first be established before a message can be sent via Modbus TCP/IP.

Following the client-server principle, this connection is established by the client (master). This connection can be handled explicitly by the client user- application software, or automatically by the client TCP connection manager. More commonly, this is handled automatically by the client protocol software via the TCP socket interface and this operation remains transparent to the application.

All Modbus TCP/IP message connections are point-to-point communication paths between two devices, which require a source address, a destination address, and a connection ID in each direction. Thus Modbus TCP/IP communication is restricted to unicast messages only.

Well-known port 502 has been specifically reserved for Modbus applications. A Modbus server will listen for communication on port 502. When a Modbus client wants to send a message to a remote Modbus server, it opens a connection with remote port 502. As soon as a connection is established, the same connection can be used to transfer user data in either direction between a client and server. A client and server may also establish several TCP/IP connections simultaneously. When a connection is established, all the transmissions that are part of that connection are associated with a Connection ID (CID). If this connection involves transmission in both directions, then two Connection ID's are assigned. The maximum number of connections allowed is dependent on the specifications of the particular TCP/IP interface (for Acromag 9xxEN modules, this number is 10). In the case of cyclic transmission between a client and server, a permanent connection can also be employed. If it is only necessary to transfer parameter or diagnostic information when a special event occurs, then this connection can be closed after each data transmission and reopened as needed.

## 4.8 TRANSPORT LAYER

The Transport Layer resides just below the Application Layer and is responsible for the transmission, reception, and error checking of the data. There are a number of Transport Layer protocols that may operate at this layer, but the primary one of interest for Modbus TCP/IP is the Transport Control Protocol (TCP).

### 4.8.1 TCP- Transport Control Protocol

The Transport Control Protocol (TCP) resides one layer above the Internet Protocol (IP) and is responsible for transporting the application data and making it secure, while IP is responsible for the actual addressing and delivery of the data. The TCP packet is inserted into the data portion of the IP packet below it. IP itself is an unsecured, connectionless protocol and must work together with the overlaying TCP in order to operate. In this way, TCP is generally considered the upper layer of the IP platform that serves to guaranty secure data transfer. The use of the label Modbus-TCP (versus Modbus TCP/IP) does not imply that IP is not used or not important.

If data is lost, it must be retransmitted. This type of data exchange refers to explicit messaging and is commonly used for exchanging information that is not time-critical, but still necessary. TCP uses explicit messaging and will work to ensure that a message is received, but not necessarily on time.

TCP is a connection-oriented protocol. TCP establishes a connection between two network stations for the duration of the data transmission. While establishing this connection, conditions such as the size of the data packets are specified (which apply to the entire connection session).

TCP also follows the Client-Server communication model. That is, whichever network station takes the initiative and establishes the connection is referred to as the TCP Client. The station to which the connection is made is called the TCP Server. In Modbus TCP/IP, the communication is always controlled by the master (client) and the master/client will establish the connection. The server (slave) cannot initiate communication on its own, but just waits for the client (master) to make contact with it. The client then makes use of the service offered by the server (note that depending on the service, one server may accommodate several clients at one time).

TCP verifies the sent user data with a checksum and assigns a sequential number to each packet sent. The receiver of a TCP packet uses the checksum to verify having received the data correctly. Once the TCP server has correctly received the packet, it uses a predetermined algorithm to calculate an acknowledgement number from the sequential number. The acknowledgement number is returned to the client with the next packet it sends as an acknowledgement. The server also assigns a sequential number to the packet it sends, which is then subsequently acknowledged by the client with an acknowledgement number. This process helps to ensure that any loss of TCP packets will be noticed, and that if needed, they can then be re-sent in the correct sequence.

TCP also directs the user data on the destination computer to the correct application program by accessing various application services using various well-known port numbers. For example, Telnet can be reached through Port 23, FTP through port 21, and Modbus through port 502. In this way, the port number is analogous to the room number in a large office building—if you address a letter to the public relations office in room 312, you are indicating that you wish to utilize the services of the public relations office.

A port is the address that is used locally at the transport layer (on one node) and identifies the source and destination of the packet inside the same node. Port numbers are divided between well-known port numbers (0- 1023), registered user port numbers (1024-49151), and private/dynamic port numbers (49152-65535). Ports allow TCP/IP to multiplex and demultiplex a sequence of IP datagrams that need to go to many different (simultaneous) application processes.

To reiterate, with TCP, the transmitter expects the receiver to acknowledge receipt of the data packets. Failure to acknowledge receipt of the packet will cause the transmitter to send the packet again, or the communication link to be broken. Because each packet is numbered, the receiver can also determine if a data packet is missing, or it can reorder packets not received into the correct order. If any data is detected as missing, all subsequent received data will be buffered. The data will then be passed up the protocol stack to the application, but only when it is complete and in the correct order. Of course this error checking mechanism of the connection-oriented TCP protocol takes time and will operate more slowly than a connection-less protocol like UDP (UDP is not used in Modbus TCP/IP). Thus, sending a message via TCP makes the most sense where continuous data streams or large quantities of data must be exchanged, or where a high degree of data integrity is required (with the emphasis being on secure data).

The following figure illustrates the construction of a TCP Packet:

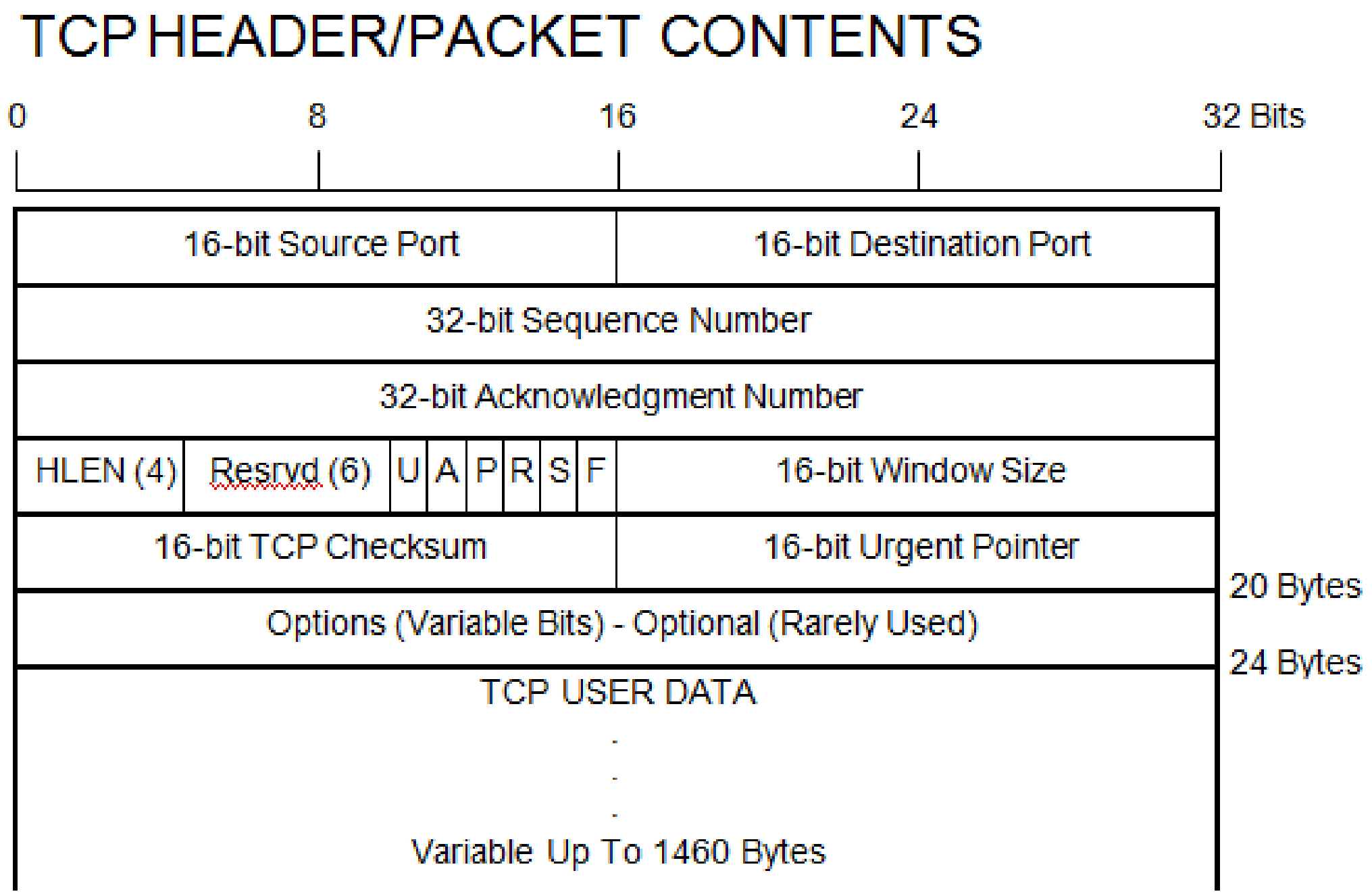

Figure 18 TCP Header Packet Contents

**TCP Header Field Definitions (Left-to-Right and Top-to-Bottom):**

**Source Port (SP)** – Port of sender's application (the port the sender is waiting to listen for a response from the destination machine).
**Destination Port (DP)** – Port number of the receiver's application (the port of the remote machine the sent packet will be received at).
**Sequence Number (SN)** – Offset from the first data byte relative to the start of the TCP flow which is used to guaranty that a sequence is maintained when a large message requires more than one transmission. **Acknowledgment Number (AN)** – This is the sequence number expected in the next TCP packet to be sent and works by acknowledging the sequence number as sent by the remote host. That is, the local host's AN is a reference to the remote machine's SN, and the local machine's SN is related to the remote machine's AN.
**Header Length (HLEN)** – A measure of the length of the header in increments of 32-bit sized words.
**Reserved** – These 6 bits are reserved for possible future use.
**UARPSF Flags (URG, ACK, PSH, RST, SYN, FIN)** – U=Urgent flag which specifies that the urgent point included in this packet is valid; A=Acknowledgement flag specifies that the portion of the header that has the acknowledgement number is valid; P=Push flag which tells the TCP/IP stack that this should be pushed up to the application layer program that needs it or requires it as soon as time allows; R=Reset flag used to reset the connection; S=Synthesis flag used to synchronize sequence numbers with acknowledgement numbers for both hosts (synthesis of the connection); F=Finish flag used to specify that a connection is finished according to the side that sent the packet with the F flag set.
**Window Size (WS)** – This indicates how many bytes may be received on the receiving side before being halted from sliding any further and receiving more bytes as a result of a packet at the beginning of the sliding window not having been acknowledged or received.
**TCP Checksum (TCPCS)** – This is a checksum that covers the header and data portion of a TCP packet to allow the receiving host to verify the integrity of an incoming TCP packet.

**Urgent Pointer (UP)** – This allows for a section of data as specified by the urgent pointer to be passed up by the receiving host quickly.
**IP Options** – These bits are optional and rarely used.
**TCP User Data** – This portion of the packet may contain any number of application layer protocols (CIP™, HTTP, SSH, FTP, Telnet, etc.).

### 4.8.2 TCP Example

The following simplified example illustrates a typical TCP transaction. In this example a network client (web browser) initiates data transfer with a web server. The client is a PC running Internet Explorer and connected to the network via a Network Interface Card (NIC).

Earlier we talked about how ports are used to send and receive messages via TCP. For example, local port 502 was reserved for listening/receiving Modbus messages. Likewise, port 80 is another well-known port that is reserved for web applications.

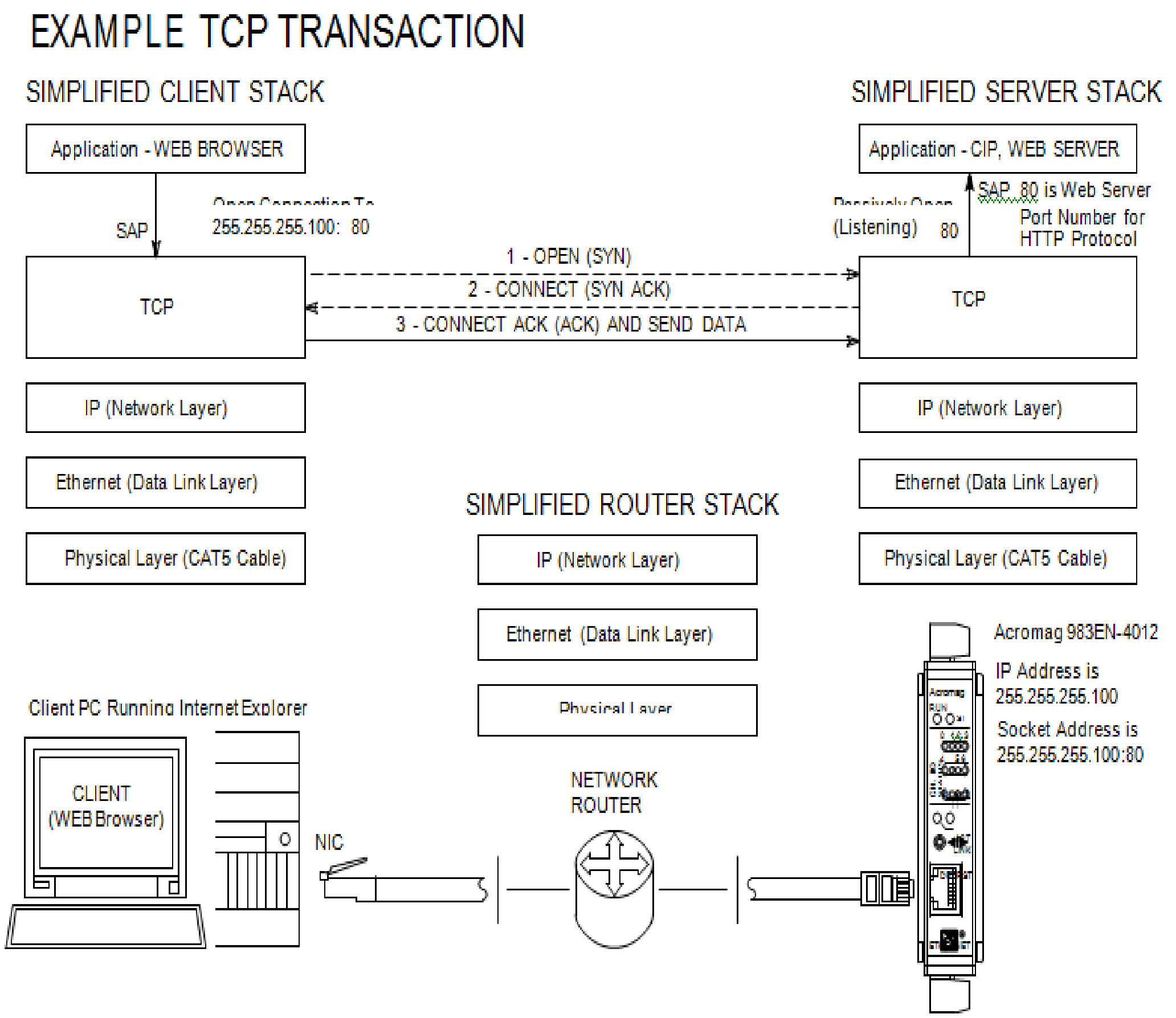

Figure 19 Example of TCP Transaction

Note that a web browser always uses TCP for communication with a web server. The web browser (client application) starts by making a service request to TCP of its transport layer for the opening of a connection for reliable data transport. It uses the IP address of the remote server combined with the well-known port number 80 (HTTP Protocol) as its socket address. TCP opens the connection to its peer entity at the web server by initiating a three-way handshake. If this handshake can complete and the connections successfully open, then data can flow between the web browser (client) and the web server (Acromag module).

Once the connection is made, the web browser and remote server assume that a reliable open data pipe has formed between them and they begin transporting their data in sequence, and without errors, as long as TCP does not close the connection. TCP will monitor the transaction for missing packets and retransmit them as necessary to ensure reliability.

Note that in the figure above, an observer in the data paths at either side of the router would actually see the beginning of the message from the client to the web server begin only in the third data frame exchanged (the client's request message is combined with the connection acknowledge of the third exchange).

## 4.9 NETWORK LAYER

### 4.9.1 IP – Internet Protocol

Although Modbus TCP/IP are named together, they are really complimentary protocols. The Internet Protocol (IP) manages the actual addressing and delivery of the data packets. IP provides a connectionless and unacknowledged method for sending data packets between two devices on a network. IP does not guaranty delivery of the data packet, it relies on a transport layer protocol (like TCP) or application layer protocol (like Modbus) to do that. IP also makes it possible to assemble an indefinite number of individual networks into a larger overall network, without regard to physical implementation of the sub networks. That is, the data is sent from one network station to another, transparent to these differences.

An IP packet is a chunk of data transferred over the Internet using standard Internet Protocol (IP). Each packet begins with a header containing addressing and system control information. Unlike uniform ATM "cells", IP packets vary in length depending on the data being transmitted.

The following illustrates the contents of the IP header. The first 5 rows are commonly used (20 Bytes), while the 6th (or more) rows will depend on how many 32-bit option words are required by special options. The data is the encapsulated packet of the upper Transport Layer, a TCP or UDP packet (Modbus TCP/IP does not utilize UDP).

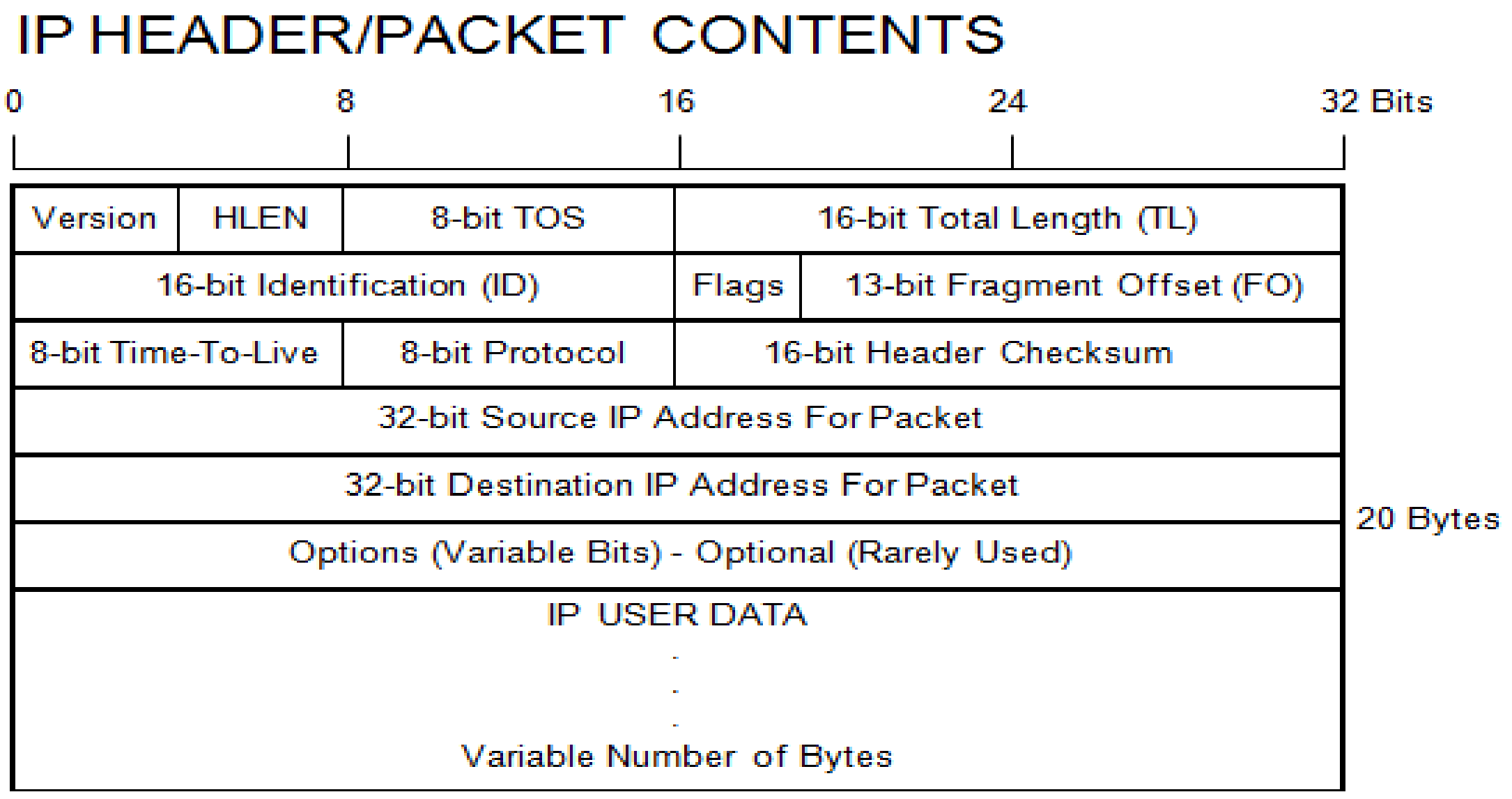

Figure 20 IP Header/Packet Contents

**IP Header Field Definitions (Left-to-Right and Top-to-Bottom):**

**Version** – A 4-bit field that specifies what version of the Internet Protocol is being used (currently IP version 4). IP version 6 has many advantages over IP version 4, but is not in widespread use yet.
**Header Length (HLEN)** – A 4-bit number that specifies the increments of 32-bit words that tell the machine decoding the IP packet where the IP header is supposed to end (this dictates the beginning of the data). For example, "0101" (5) would specify an IP packet having only the first 5 rows as header information and its data thus beginning with the 6th row.
**Type-of-Service (TOS)** – Used for special IP packet information that may be accessed by routers passing along the packet, or by receiving interfaces. The first 3 bits are reserved, the fourth bit is set to 0, and the remaining 4 bits are used to flag the following (respectively): minimize delay for this packet, maximize throughput for this packet, maximize reliability for this packet, and minimize monetary costs. Many application layer protocols have recommended values for each of these bits based on the kind of service they are using. For example, NNTP (Net News Transfer Protocol) is not a very time critical operation. NNTP is used for USENET posts, and group synchronization between servers, or to a client from a server. If it happens to take a long time to transfer all this data, that's OK. Since it's not time sensitive, the "minimize monetary costs" bit may be set for a server synchronizing itself with another server under these conditions. Thus, it is left to the router to determine the paths which are the cheapest and then route a packet based on the flags that are set. If a router has only two routes (one to/from the internet, a second to/from a Local Area Network), then these 4 bits are often ignored since there are not multiple routes to its destination. These bits may be useful where a router may have four routes to a distant network, each route using a medium that has specific costs related to bandwidth, reliability, or latency (fiber, satellite, LAN line, or VPN for example). With each of these links up through a router, an incoming packet may be routed via any of these paths, but a properly configured router may be able to take advantage of how these packet bits are set by the sender in determining which route to take. These bits are sometimes known as the Differentiated Services Code Point (DSCP) which defines one of a set of classes of service. This value is usually set to 0, but may be used to indicate a particular Quality of Service request from the network.
**Total Length (TL)** – The total length of the IP packet in 8-bit (byte) increments. By subtracting header length (HL) from total length (TL), you can determine how many bytes long the data portion of the IP packet is. As a 16-bit value, valid ranges would be from 20 (minimum size of IP header) to 65535 bytes. A TL of only 20 is unlikely (no data), but could happen if something was broken. Very large IP packets (greater than 1500 bytes) are also uncommon, since they must typically be fragmented onto some networks.
**Identification (ID)** – A 16-bit number used to distinguish one sent IP packet from another by having each IP packet sent increment the ID by 1 over the previous IP packet sent.
**Flags** – A sequence of 3 fragmentation flags used to control where routers are allowed to fragment a packet (via Don't Fragment flag) and to indicate the parts of a packet to the receiver: 001=More, 010=Don't Fragment, 100= Unused.
**Fragmentation Offset (FO)** – A byte count from the start of the original packet sent and set by any router that performs IP router fragmentation.
**Time-to-Live (TTL)** – An 8-bit value that is used to limit the number of routers through which a packet may travel before reaching its destination (the number of hops/links which the packet may be routed over). This number is decremented by 1 by most routers and is used to prevent accidental routing loops. If the TTL drops to zero, the packet is discarded by either the server that has last decremented it, or the next server that receives it.
**Protocol** – An 8-bit value that is used to allow the networking layer to know what kind of transport layer protocol is in the data segment of the IP packet. For example, 1=ICMP, 2=IGMP, 6=TCP, 17=UDP.
**Header Checksum** – A 16-bit checksum (1's complement value) for the header data that allows a packet to offer verification data to help ensure that the IP header is valid. This checksum is originally inserted by the sender and then updated whenever the packet header is modified by a router. This is used to detect processing errors on the part of the router or bridge where the packet is not already protected by a link layer Cyclic Redundancy Check (CRC). Packets with an invalid checksum are discarded by all nodes in an IP network.

**<u>Source IP Address (32 bits)</u>** – The IP address of the source machine sending the data onto the network. This address is commonly represented by 4 octets representing decimal values and separated by periods (255.255.255.10 for example).

**<u>Destination IP Address (32 bits)</u>** – The IP address of the destination machine to which the packet is being routed for delivery. This address is commonly represented by 4 octets representing decimal values and separated by periods (255.255.255.10 for example).

**<u>Options (Variable Number of Bits/Words)</u>** – These bits are reserved for special features and are rarely used, but when they are used, the IP header length will be greater than 5 (five 32-bit words) to indicate the relative size of
the option field.

**<u>IP Data (Variable Number of Bits/Words)</u>** - This portion of the packet may contain any number of nested protocols (TCP, UDP, ICMP, etc.).

### 4.9.2 Ethernet (MAC) Address

The Ethernet Address or MAC Address refers to the Media Access Control Address that uniquely identifies the hardware of any network device. This is a unique, 48-bit, fixed address assigned and hard-coded into an Ethernet device at the factory. This is usually expressed in hexadecimal form as 12 hex characters (6 bytes), with the first 3 bytes (6 leftmost hex characters) representing the device manufacturer, and the last 3 bytes (6 rightmost hex characters) uniquely assigned by the manufacturer. All six bytes taken together uniquely identify the network device.

Do not confuse the Ethernet Address (MAC address) with the Internet Protocol (IP) Address, which is a 32-bit number assigned to your computer (see below) that can change each time you connect to a network.

### 4.9.3 Internet (IP) Address

IP addresses are 32-bit numbers that are administered by an independent authority (InterNIC) and are unique for any device on the network. The IP address is a 32-bit value made up of four octets (8 bits), with each octet having a value between 0-255 (00H-FFH). It is commonly expressed as four decimal numbers (8-bit values) separated by a decimal point. This provides about 4.3 billion possible combinations.

Large networks of corporations, communications companies, and research institutions will obtain large blocks of IP addresses, then divide them into sub networks within their own organization and distribute these addresses as they see fit. The smaller networks of universities and companies will acquire smaller blocks of IP addresses to distribute among their users. Because these numbers are ultimately assigned by the Internet Assigned Number's Authority, an IP address can be used to approximate a machine's location.

Similar to the Ethernet Address, the IP address is comprised of two parts: the network address or Net ID (first part), and the host address or Host ID (last part). This last part refers to a specific machine on the given sub- network identified by the first part. The number of octets of the four total that belong to the network address depend on the Class definition (Class A, B, or C) and this refers to the size of the network.

A Subnet is a contiguous string of IP addresses. The first IP address in a subnet is used to identify the subnet and usually addresses the server for the subnet. The last IP address in a subnet is always used as a broadcast address and anything sent to the last IP address of a subnet is sent to every host on that subnet.

Subnets are further broken down into three size classes based on the 4 octets that make up the IP address. A Class A subnet is any subnet that shares the first octet of the IP address. The remaining 3 octets of a Class A subnet will define up to 16,777,214 possible IP addresses (224 – 2). A Class B subnet shares the first two octets of an IP address (providing 216 – 2, or 65534 possible IP addresses). Class C subnets share the first 3 octets of an IP address, giving 254 possible IP addresses. Recall that the first and last IP addresses are always used as a network number and broadcast address respectively, and this is why we subtract 2 from the total possible unique addresses that are defined via the remaining octet(s).

A Subnet Mask is used to determine which subnet an IP address belongs to. The use of a subnet mask allows the network administrator to further divide the host part of this address into two or more subnets. The subnet mask flags the network address part of the IP address, plus the bits of the host part, that are used for identifying the sub-network. By mask convention, the bits of the mask that correspond to the sub-network address are all set to 1's (it would also work if the bits were set exactly as in the network address). It's called a mask because it can be used to identify the unique subnet to which an IP address belongs to by performing a bitwise AND operation between the mask itself, and the IP address, with the result being the sub- network address, and the remaining bits the host or node address.

For our example, the default IP address of this module is 128.1.1.100. If we assume that this is a Class C network address (based on the default Class C subnet mask of 255.255.255.0), then the first three numbers represent this Class C network at address 128.1.1.0, the last number identifies aunique host/node on this network (node 100) at address 128.1.1.100.

To make sure that the broadcast message is recognized by all connected network stations, the IP driver uses "FF FF FF FF FF FF" as the Ethernet address. The station that recognizes its own IP address in the ARP request will confirm this with an ARP reply. The ARP reply is a data packet addressed to the ARP request sender with an ARP identifier indicated in the protocol field of the IP header.

The IP driver then extracts the Ethernet address obtained from the ARP reply and enters it into the ARP table. Normally, these dynamic entries do not remain in the ARP table and are aged out, if the network station is not subsequently contacted within a few minutes (typically 2 minutes under Windows).

The ARP table may also support static address entries, which are fixed addresses manually written into the ARP table and not subject to aging. Static entries are sometimes used for passing the desired IP address to new network devices which do not yet have an IP address.

Recall that ARP allowed a station to recover the destination hardware (MAC) address when it knows the IP address. The Reverse Address Resolution Protocol (RARP) is a complimentary protocol used to resolve an IP address from a given hardware address (such as an Ethernet address). A station will use RARP to determine its own IP address (it already knows its MAC address).

For our example, if we wish to further divide this network into 14 subnets, then the first 4 bits of the host address will be required to identify the sub network (0110), then we would use "11111111.11111111.11111111.11110000" as our subnet mask. This would effectively subdivide our Class C network into 14 sub networks of up to 14 possible nodes each. With respect to the default settings of Acromag 9xxEN modules:

Subnet Mask 255.255.255.0 (11111111.11111111.11111111.00000000)
IP Address: 128.1.1.100 (10000000.00000001.00000001.01100100)
Subnet Address: 128.1.1.0 (1000000.00000001.00000001.00000000)

Subnetwork address 128.1.1.0 has 254 possible unique node addresses. We are using node 100 of 254 possible for our module.

At this point, we see that each layer (application, transport, network, and data link layer) uses its own address method. The application layer uses socket numbers, which combine the IP address with the port number. The transport layer uses port numbers to differentiate simultaneous applications. The network layer uses the IP address, and the Data Link layer uses the MAC address.

### 4.9.3.1 ARP –Address Resolution Protocol

The Address Resolution Protocol (ARP) is a TCP/IP function that resides at the network layer (layer 3) with the Internet Protocol (IP), and its function is to map Ethernet addresses (the MAC ID) to IP addresses, and maintain a mapping table within the network device itself. This protocol allows a sending station to gather address information used to form a layer 2 frame complete with the IP address and hardware (MAC) address. Every TCP/IP- based device contains an ARP Table (or ARP cache) that is referred to by a router when it is looking up the hardware address of a device for which it knows the IP address and needs to forward a datagram to. If this device wants to transmit an IP packet to another device, it first attempts to look-up the Ethernet address of that device in its ARP table. If it finds a match, it will pass the IP packet and Ethernet address to the Ethernet driver (physical layer). If no hardware address is found in the ARP table, then an ARP broadcast is sent onto the network. The ARP protocol will query the network via a local broadcast message to ask the device with the corresponding IP address to return its Ethernet address. This broadcast is read by every connected station, including the destination station. The destination station sends back an ARP reply with its hardware address attached so that the IP datagram can now be forwarded to it by the router. The hardware address of the ARP response is then placed in an internal table and used for subsequent communication.

It is important to note that when we used the term "local broadcast message" above, that Ethernet broadcast messages will pass through hubs, switches, and bridges, but will not pass through routers. As such, broadcast messages are confined to the subnet on which they originate and will not propagate out onto the worldwide web.

To make sure that the broadcast message is recognized by all connected network stations, the IP driver uses "FF FF FF FF FF FF" as the Ethernet address. The station that recognizes its own IP address in the ARP request will confirm this with an ARP reply. The ARP reply is a data packet addressed to the ARP request sender with an ARP identifier indicated in the protocol field of the IP header.

The IP driver then extracts the Ethernet address obtained from the ARP reply and enters it into the ARP table. Normally, these dynamic entries do not remain in the ARP table and are aged out, if the network station is not subsequently contacted within a few minutes (typically 2 minutes under Windows).

The ARP table may also support static address entries, which are fixed addresses manually written into the ARP table and not subject to aging. Static entries are sometimes used for passing the desired IP address to new network devices which do not yet have an IP address.

Recall that ARP allowed a station to recover the destination hardware (MAC) address when it knows the IP address. The Reverse Address Resolution Protocol (RARP) is a complimentary protocol used to resolve an IP address from a given hardware address (such as an Ethernet address). A station will use RARP to determine its own IP address (it already knows its MAC address).

#### 4.9.3.2 RARP – Reverse Address Resolution Protocol

RARP essentially allows a node in a local area network to request its IP address from a gateway server's ARP table. The ARP table normally resides in the LAN's gateway router and maps physical machine addresses (MAC addresses) to their corresponding Internet Protocol (IP) addresses. When a new machine is added to a LAN, its RARP client program requests its IP address from the RARP server on the router. Assuming that an entry has been set up in the router table, the RARP server will return the IP address to the machine which will then store it for future use.

ARP and its variant RARP are needed because IP uses logical host addresses (the IP address), while media access control protocols (Ethernet, Token-Ring, FDDI, etc.) need MAC addresses. The IP addresses are assigned by network managers to IP hosts and this is usually accomplished by configuration file options and driver software. That is why these are sometimes referred to as software addresses. LAN topologies cannot use these software addresses and they require that the IP addresses be mapped to their corresponding MAC addresses.

For example, a diskless workstation cannot read its own IP address from configuration files. They will send an RARP request (or BOOTP request) to a RARP server (or BOOTP server). The RARP server will find the corresponding IP address in its configuration files using the requesting station's MAC address as a lookup, and then send this IP address back in a RARP reply packet.

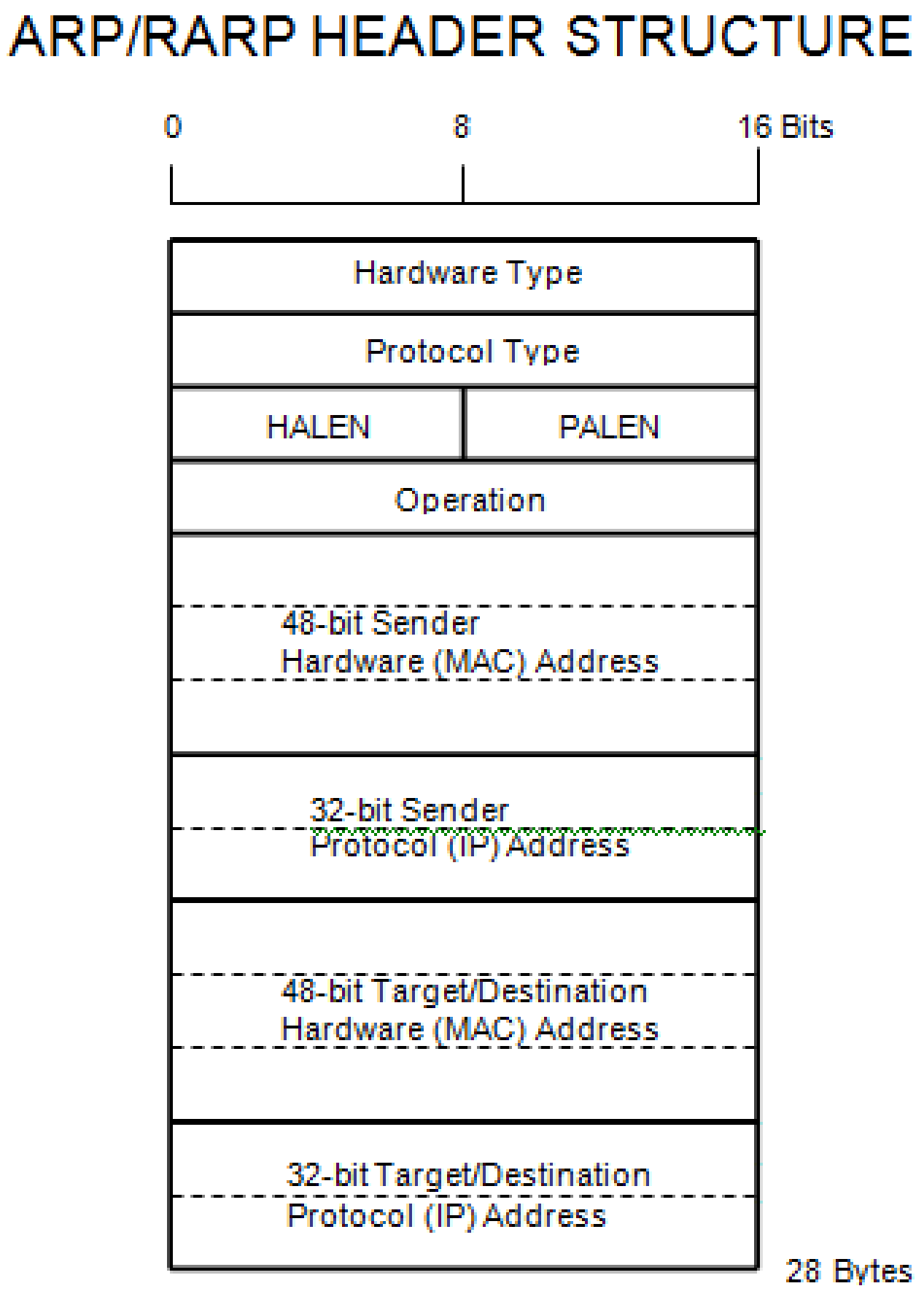

Figure 21: ARP/Header Structure

**ARP/RARP Header Field Definitions (Left-to-Right, Top-to-Bottom):**

**Hardware Type** – Specifies a hardware interface type from which the sender requires a response. This is "1" for Ethernet.
**Protocol Type** – This is the protocol type used at the network layer and is used to specify the type of high-level protocol address the sender has supplied.
**Hardware Address Length (HALEN)** – This is the hardware address length in bytes which is "6" for an Ethernet (MAC Address).
**Protocol Address Length (PALEN)** – This is the protocol address length in bytes which is "4" for a TCP/IP (IP Address).
**Operation Code** – This code is commonly used to indicate whether the packet is an ARP request (1), or an ARP response (2). Valid codes are as follows:
1 = ARP Request; 2 = ARP Response; 3 = RARP Request; 4 =
RARP Response; 5 = Dynamic RARP Request; 6 = Dynamic RARP Reply;
7 = Dynamic RARP Error; 8 = InARP Request; 9 = InARP Reply.
**Sender Hardware (MAC) Address** – This is the 48-bit hardware (MAC) address of the source node.
**Sender Protocol (Software) Address** – This is the 32-bit senders protocol address (the layer 3/network layer address—the IP address). **Target/Destination Hardware (MAC) Address** – Used in an RARP request. The RARP request response carries both the target hardware address (MAC address) and layer 3 address (IP address). **Target/Destination Protocol (Software) Address** – Used in an ARP request. The ARP response carries both the target hardware address (MAC address) and layer 3 address (IP address).

Most network stations will send out a gratuitous ARP request when they are initializing their own IP stack. This is really an ARP request for their own IP address and is used to check for a duplicate IP address. If there is a duplicate IP address, then the stack does not complete its initialization.

## 4.10 DATA LINK (MAC) LAYER

The Data Link Layer or Host-to-Network Layer provides the protocol for connecting the host to the physical network. Specifically, this layer interfaces the TCP/IP protocol stack to the physical network for transmitting IP packets.

Recall from Figure 1 of the TCP/IP Stack section, how the various protocols at each of the different layers are encapsulated (nested) into the data frame of the next lowest layer. That is, packets generally carry other packet types inside them and their function is to often contain or encapsulate other packets. In this section, we will look at the lowest encapsulation layer (often referred to as the data link layer or the MAC layer) where Ethernet resides.

Note that bits are transmitted serially with the least significant bit of each byte transmitted first at the physical layer. However, when the frame is stored on most computers, the bits are ordered with the least significant bit of each byte stored in the rightmost position (bits are generally transmitted right-to-left within the octets, and the octets are transmitted left-to-right).

### 4.11 CSMA/CD

The data link layer also uses the CSMA/CD protocol (Carrier Sense Multiple Access w/ Collision Detection) to arbitrate access to the shared Ethernet medium.

Recall that with CSMA/CD, any network device can try to send a data frame at any time, but each device will first try to sense whether the line is idle and available for use. If the line is available, the device will begin to transmit its first frame. If another device also tries to send a frame at approximately the same time (perhaps because of cable signaling delay), then a collision occurs and both frames are subsequently discarded. Each device then waits a random amount of time and retries its transmission until it is successfully sent.

With switched Ethernet, a deliberate effort has been made to suppress CSMA/CD in order to increase determinism. This is done by using Ethernet switches to interconnect devices, connecting only one device per switch port. With only one device per switch port, there is no chance of collisions and devices will communicate full-duplex, at effectively double the base data rate.

## 4.12 Medium Access Control (MAC) Protocol

The Medium Access Control (MAC) protocol provides the services required to form the data link layer of Ethernet. This protocol encapsulates its data by adding a 14-byte header containing the protocol control information before the data, and appending a 4 byte CRC value after the data. The entire frame is preceded by a short idle period (the minimum inter-frame gap), and an 8-byte preamble.

**Ethernet (MAC) Frame Definitions (Left-to-Right & Top-to-Bottom):**

**Preamble & SFD (56-bits+8-bits SFD)** – Technically, the preamble is not part of the Ethernet frame, but is used to synchronize signals between stations. It is comprised of a pattern of 62 alternating 1’s & 0’s, followed by two set bits “11” (the last 8-bits of this pattern are actually referred to as the Start-of-Frame Delimiter or SFD byte). This 8-byte preamble/SFD is also preceded by a small idle period that corresponds to the minimum inter-frame gap period of 9.6us (at 10Mbps). After transmitting a frame, the transmitter must wait for this period to allow the signal to propagate through the receiver of its destination. The preamble allows the receiver at each station to lock into the Digital Phase Lock Loop used to synchronize the receive data clock to the transmit data clock. When the first bit of the preamble arrives, the receiver may be in an arbitrary state (out of phase) with respect to its local clock. It corrects this phase during the preamble and may miss/gain a number of bits in the process. That is why the special pattern of two set bits (11) is used to mark the last two bits of the preamble. When it receives these two bits, it starts assembling the bits into bytes for processing by the MAC layer. When using Manchester encoding at 10Mbps, the 62 alternating 1/0 bits of the preamble will resemble a 5MHz square wave.

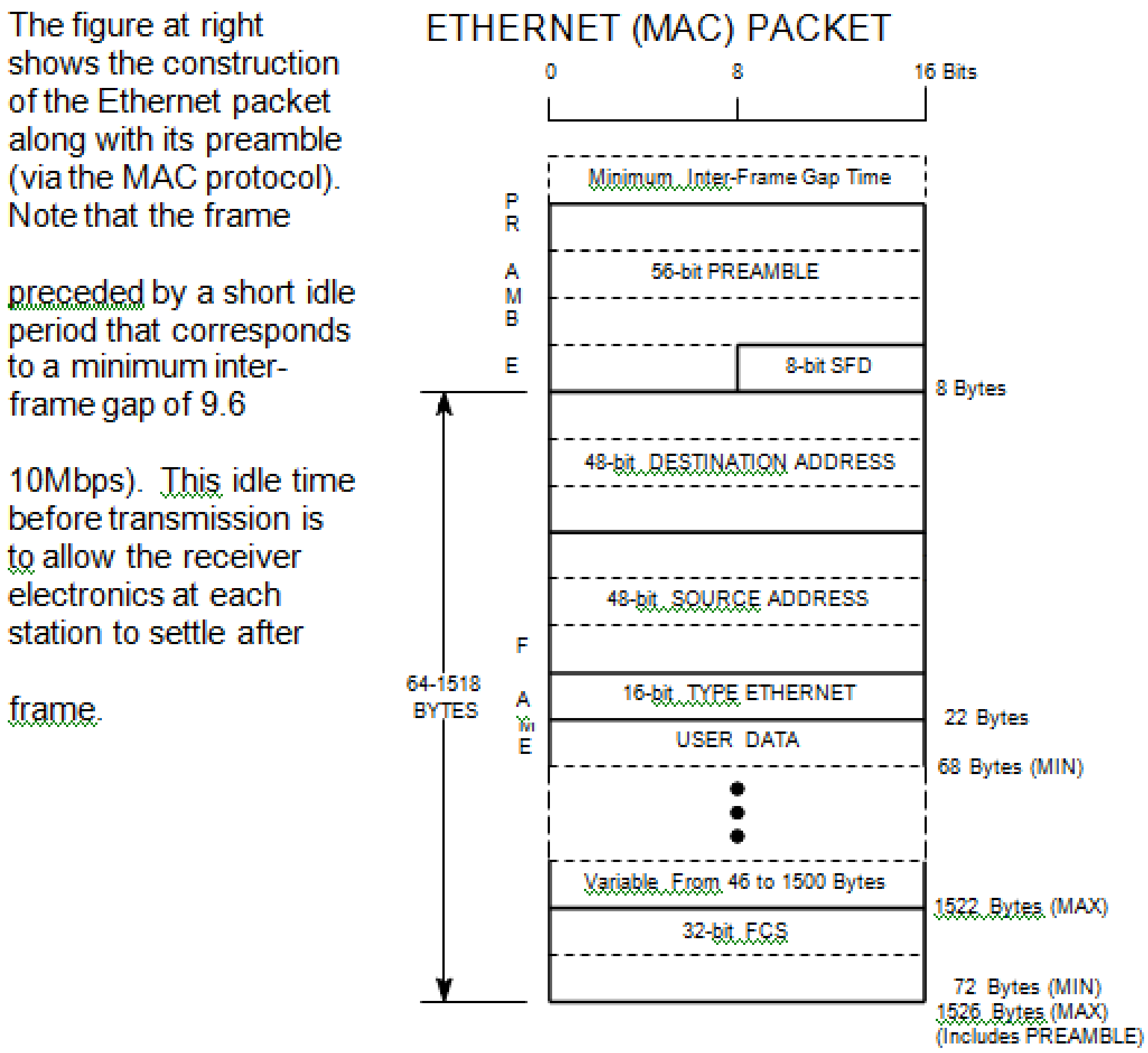

Figure 22 Ethernet (MAC) Packet

## 4.13 RARP – Reverse Address Resolution Protocol

**Destination Address (48-bits)** – This 6-byte address is the destination Ethernet address (MAC Address). It may address a single receiver node (unicast), a group of nodes (multicast), or all receiving nodes (broadcast). **Source Address (48-bits)** – This 6-byte address is the sender's unique node address and is used by the network layer protocol to identify the sender and also to permit switches and bridges to perform address learning.

**Type (16-bits)** – This 2-byte field provides a Service Access Point (SAP) and is used to identify the type of network layer protocol being carried. The value 0800H would be used to indicate an IP network protocol, other values indicate other network layer protocols. For example, 0806H would indicate an ARP request, 0835H would indicate a RARP request. For IEEE 802.3 LLC (Logical Link Control), this field may alternately be used to indicate the length of the data portion of the packet.

**CRC Cyclic Redundancy Check (32-bits)** – The CRC is added at the end of a frame to support error detection for cases where line errors or transmission collisions result in a corrupted MAC frame. Any frame with an invalid CRC is discarded by a MAC receiver without further processing and the MAC protocol does not provide any other indication that a frame has been discarded due to an invalid CRC.

The Ethernet standard dictates a minimum frame size which requires at least 46 data bytes in a MAC

frame. If a network layer tries to send less than 46 bytes of data, the MAC protocol adds the requisite number of 0 bytes (null padding characters) to satisfy this requirement. The maximum data size which may be carried in a MAC frame over Ethernet is 1500 bytes.

Any received frame less than 64 bytes is illegal and referred to as a “runt”. Runts may result from a collision and a receiver will discard all runt frames. Any received frame which does not contain an integral multiple of octets (bytes) is also illegal (misaligned frame), as the receiver cannot compute the CRC for the frame and these will also be discarded by the receiver. Any received frame greater than the maximum frame size is referred to as a “giant” and these frames are also discarded by an Ethernet receiver.

# 5

# Hardware Overview

In this section, we give an overview of all the hardware components which were used by us in setting up the Vacuum Measurement system. We primarily required 6 Cold Cathode Pirani Gauges for measuring vacuum, 1 Maxiguage Controller to control and monitor the 6 gauges, a DAQ Module ADAM 5000 to condition the signal to be transmitted over the ethernet to the controlling system i.e. a computer. The DAQ module is also used when remote monitoring of the vacuum measurement is required; lastly we require a computer and a power supply.

## 5.1 Cold cathode Pirani Gauge

The Pirani measurement circuit is always on. The cold cathode measurement circuit is controlled by the Pirani circuit and is activated only at pressures <1×10-2 mbar.

As long as the cold cathode measurement circuit has not ignited, the measurement value of the Pirani is output as measuring signal (if p < 5×10-4 mbar, "Pirani underrange" is displayed). The measuring signal depends on the type of gas being measured. The curves are accurate for N2, O2, dry air and CO. They can be mathematically converted for other gases. When cold cathode measurement systems are activated, an ignition delay occurs. The delay time increases at low pressures and is typically:

10-5 mbar ≈ 1 second
10-7 mbar ≈ 20 seconds
5×10-9 mbar ≈ 2 minutes

As long as the cold cathode measurement circuit has not yet ignited, the measurement value of the Pirani is output as measuring signal ("Pirani under-range" is displayed for pressures <5×10-4 mbar).

### 5.1.1 Technical Specifications

Storage: -40 °C … +65 °C
Operation: + 5 °C … +55 °C
Bakeout: +150 °C (without electronics unit and magnetic shielding)
Use: Indoors only altitude up to 2000 m (6600 ft)
Measurement range (air, N2): 5×10-9 … 1000 mbar
Accuracy: ≈±30% in the range 1×10-8 … 100 mbar
Reproducibility: ≈±5% in the range 1×10-8 … 100 mbar

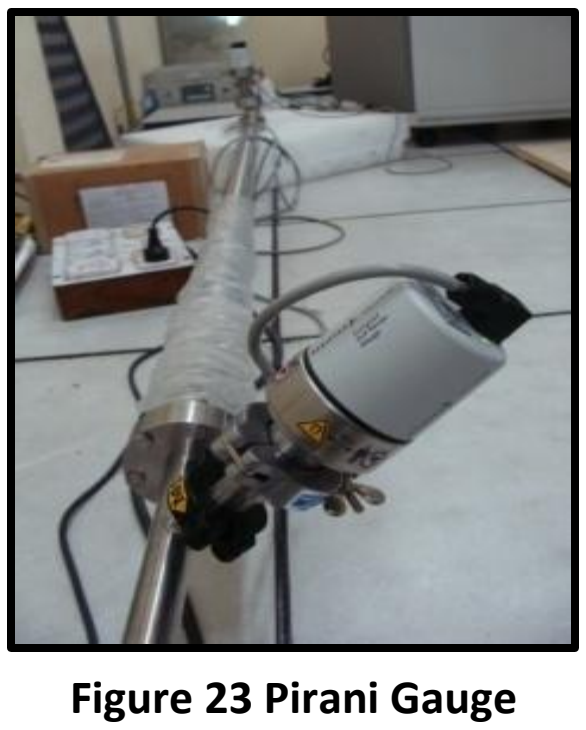

**Figure 23 Pirani Gauge**

Type of protection: IP 40
Relative humidity: Max. 80% at temperatures ≤+31°C decreasing to 50% at +40°C
Maximum pressure (absolute): 10 bar only for inert gases <55 °C
Voltage at the gauge: 15.0 … 30.0 VDC (max. ripple 1 Vpp)
Power consumption: ≤2 W
Voltage at the supply unit with maximum cable length 16.0 … 30.0 VDC (max. ripple 1 Vpp)
Electrical connection Hirschmann compact connector type GO 6, 6 pins, male
Maximum cable length: 75 m (0.25 mm² conductor)
100 m (0.34 mm² conductor)
300 m (1.00 mm² conductor)

This gauge consists of two separate measurement systems (Pirani and cold cathode system according to the inverted magnetron principle). They are combined in such a way that for the user, they behave as one single measurement system. The optimum measurement configuration for the particular pressure range, in which measurement is performed, is used is from 10-4 mbar to 5 × 10-9 mbar.

### 5.1.2 Measuring Signal versus Pressure

p = 10^ (1.667×U-d) ⇔ U = c + 0.6×log10 p

| P | U | c | d |
|---|---|---|---|
| [mbar] | [V] | 6.8 | 11.33 |
| [μbar] | [V] | 5.0 | 8.333 |
| [Torr] | [V] | 6.875 | 11.46 |
| [mTorr] | [V] | 5.075 | 8.458 |
| [micron] | [V] | 5.075 | 8.458 |
| [Pa] | [V] | 5.6 | 9.333 |
| [kPa] | [V] | 7.4 | 12.33 |

Where, U= measuring signal; p= pressure; c, d= constants (pressure unit dependent)

## 5.2 Maxigauge Controller

The Maxigauge is a versatile instrument that is used in between the compact full range gauge and the ADAM 5000 module. It has an arrangement for accommodating a maximum of 6 pressure gauges and the corresponding values may be displayed either simultaneously or singly (depending upon the user requirements). At the back side it has a 15 pin port for the control purpose along with a separate port for serial communication using the RS-232. It also has another port for the connection with the relay.

### 5.2.1 Technical Specifications

Connections for transmitter: 6 (max. 3 IMR 265/ PBR 260 / CMR27x)
Display rate: 4.01/s
Error signal: Switching current max. 3A
Error signal: Switching

Voltage max. 60 V DC
Error signal: Working contact, potential-free 1 piece
Filter time constant 2.1/0.32/0.1 s
Interface RS-232-C, RS-422, RS-422 isolated, RS-485 isolated
Mains requirement: frequency (range) 50-60 Hz
Mains requirement: power consumption 60 VA
Mains requirement: voltage (range) 90-250 V
Measurement range (max.): 55000 hPa
Measurement range (min.): 5 · 10-11 hPa
Measurement rate: 100 1/s
Protection category: IP 30
Safety: EN61010-1 / IEC1010, EN60950, EN 50081-1&2
Set point: Changeover contact, potential-free 6, 0 pieces
Set point: Current max. 3 A
Set point: Voltage max. 60 V DC
Signal output: Measuring value, analog 0-10 V DC

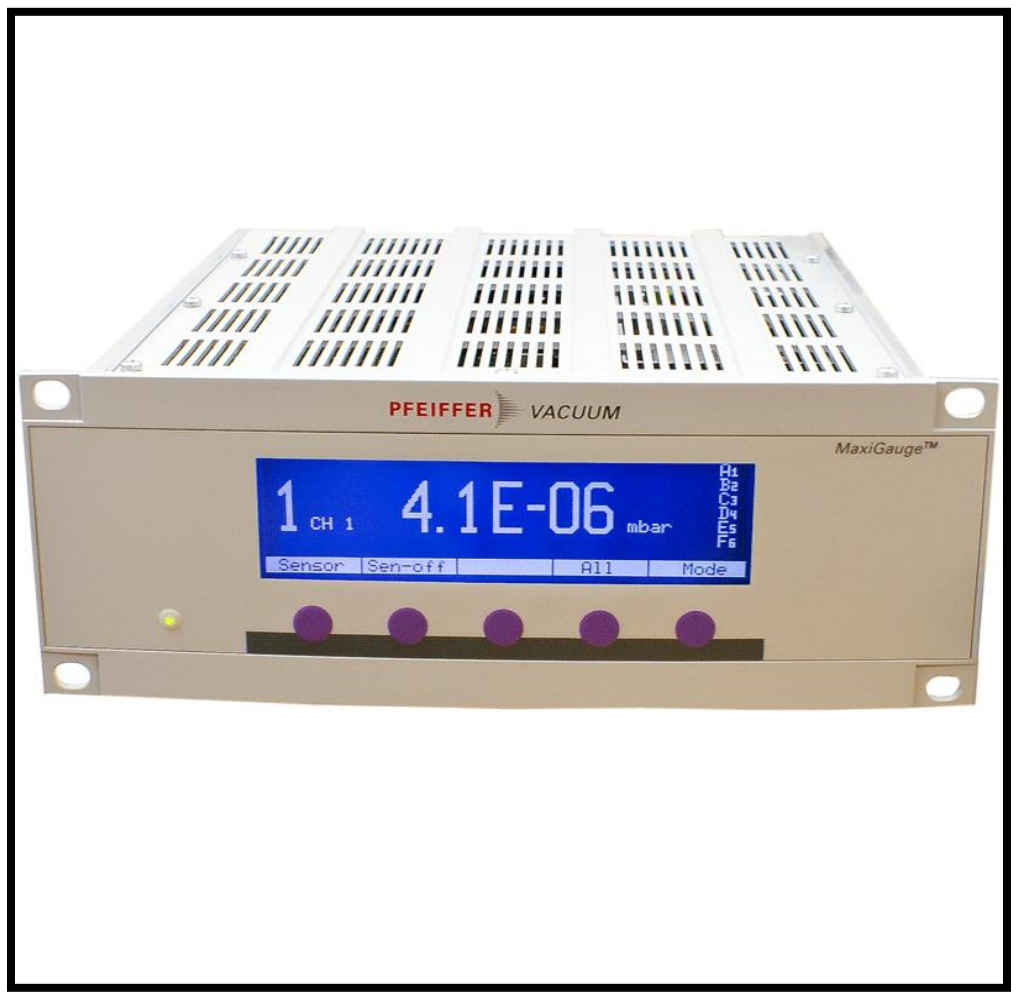

**Figure 24 Maxiguage Controller**

Signal output: Output resistance 660
Switching voltage 240 V with RI 256
Temperature: Operating 5-40 °C
Temperature: Storage -20-+60 °C
Weight 2.1 kg

# 5.3 ADAM Module 5000/TCP

ADAM module is a distributed input-output device that can accept or transmit any kind of signal whether it is analog or digital. It is a generalized signal acceptor and transmitter. ADAM-5000L/TCP is an Ethernet-based I/O system. Without a repeater, ADAM-5000L/TCP can cover a communication distance up to 100 m. This allows remote configuration via Ethernet and eight PCs can simultaneously access the data. The ADAM-5000L/TCP provides the solutions for easy configuration and efficient management. It is an ideal and cost-effective solution for e-automation architecture.

## 5.3.1 Major Features

### 5.3.1.1 Communication Network

By adopting a 32-bit RISC CPU, the ADAM-5000/TCP has greatly advanced data processing abilities for the user, especially for network communications (response time < 5ms). There is a standard RJ-45 modular jack Ethernet port on the ADAM-5000/TCP'S CPU board, and I/O modules field signals would be able to link with the Ethernet directly without assistance from other hardware devices such as converters or data gateways. The communication speeds can be auto-switched between 10 M and 100 Mbps data transfer rate depending upon the network environment. Through an Ethernet network, your DA&C systems, computer workstations, and higher-level enterprise MIS servers can access plant floor data. Such data can be used in system supervising, product scheduling, statistical quality control, and more.

#### 5.3.1.2 Modbus/TCP Protocol

Modbus/TCP is one of the most popular standards for industrial Ethernet networks. Following this communication protocol, the ADAM-5000/TCP is easy to integrate with any HMI software packages or user-developed applications that support Modbus. Users do not have to prepare a specific driver for the ADAM-5000/TCP when they install the DA&C system with their own operating application. It will certainly reduce engineer effort. Moreover, the ADAM-5000/TCP works as a Modbus data server. It allows eight PCs or tasks to access its current data simultaneously from anywhere: LAN, Intranet, or Internet.

#### 5.3.1.3 Hardware Capacity & Diagnostic

Advantech's ADAM-5000/TCP is designed with a high I/O capacity and supports all types of ADAM-5000 I/O modules. Providing eight slots for any mixed modules, this DA&C system handles up to 128 I/O points (four ADAM-5024s allowed). Different from other main units, the ADAM- 5000/TCP not only has a higher I/O capacity, but it also has a smarter diagnostic ability. There are eight indicators on the front case of the CPU module. Users can read the system status clearly, including power, CPU, Ethernet link, Communication active, communication rate, and more. In addition, there are also Tx and Rx LEDs on the Ethernet port, indicating data transfer and reception.

#### 5.3.1.4 Communicating Isolation

High-speed transient suppressors isolate ADAM-5000/TCP Ethernet port from dangerous voltage up to 1500VDC power spikes and avoid surge damage to whole system.

#### 5.3.1.5 Completed set of I/O modules for total solutions

The ADAM-5000/TCP uses a convenient backplane system common to the ADAM-5000 series. Advantech's complete line of ADAM-5000 modules integrates with the ADAM-5000/TCP to support your applications (not include ADAM-5090). Full ranges of digital module supports 10 to 30 VDC input and outputs. A set of analog modules provide 16-bit resolution and programmable input and output (including bipolar) signal ranges.

#### 5.3.1.6 Built-in real-time OS and watchdog timer

The microprocessor also includes a real-time OS and watchdog timer. The real-time OS is available to handle several tasks at the same time. The watchdog timer is designed to automatically reset the microprocessor if the system fails. This feature greatly reduces the level of maintenance required and makes the ADAM-5000/TCP ideal for use in applications which require a high level of system performance and stability.

#### 5.3.1.7 Software Support

Based on the Modbus standard, the ADAM-5000/TCP firmware is a built-in Modbus/TCP server. Therefore, Advantech provides the necessary DLL drivers, OCX component OPC Server, and Windows Utility for users for client data for the ADAM-5000/TCP. Users can configure this DA&C system via Windows Utility; integrate with HMI software package via Modbus/TCP driver or Modbus/TCP OPC Server. Even more, you can use the DLL driver or OCX component to develop your own applications.

#### 5.3.1.8 Security Setting

Though Ethernet technology comes with great benefits in speed and integration, there also exist risks about network invasion from outside. For this reason, a security protection design was built into the ADAM- 5000/TCP. Once the user has set the password into the ADAM-5000/
TCP firmware, important system configurations (Network, Firmware, and Password) can only be changed through password verification.

#### 5.3.1.9 UDP Data Stream

Most of time, each host PC in a DA&C system needs to regularly request the I/O devices via TCP/IP packs to update current data. It may cause to data collision and lower performance on the network, especially when there are frequent communication between multi-servers and I/O devices. To reduce the communication loading of the host computer on your Ethernet network, the ADAM-5000/TCP also supports UDP (User Datagram Protocol) protocol to broadcast the data packs to specific IPs without requesting commands. Users can apply this great feature to implement Data Stream, Event Trigger, and other advanced functions.

#### 5.3.1.10 Modbus Ethernet Data Gateway

Much more than an I/O system, ADAM-5000/TCP provides an RS-485 network interface for other Modbus devices integration. It works as Ethernet Data Gateway, upgrading Modbus serial network devices up to Ethernet layer. Maximum 32 nodes of ADAM-5511 or 3’rd party products supported Modbus protocol are allowed to integrate with an ADAM- 5000/TCP. This great feature enlarges our system scope, as opposed to other general dummy I/O system.

### 5.3.2 Technical Specifications of ADAM-5000/TCP System

#### 5.3.2.1 System

• CPU: ARM 32-bit RISC CPU
• Memory: 4 MB Flash RAM
• Operating System: Real-time O/S
• Timer BIOS: Yes
• I/O Capacity: 8 slots (four ADAM-5024 allowed)
• Status Indicator: Power (3.3V, 5V), CPU, Communication (Link, Collide, 10/100 Mbps, Tx, Rx)
• CPU Power Consumption: 5.0W
• Reset Push Bottom: Yes

#### 5.3.2.2 Ethernet Communication

• Ethernet: 10 BASE-T IEEE 802.3,100 BASE-TX IEEE 802.3u
• Wiring: UTP, category 5 or greater
• Bus Connection: RJ45 modular jack
• Comm. Protocol: Modbus/TCP
• Data Transfer Rate: Up to 100 Mbps
• Max Communication Distance: 100 meters
• Even Response Time: < 5 ms
• Data Stream Rate: 50 ms to 7 days

#### 5.3.2.3 Serial Communication
• RS-485 signals: DATA +, DATA-
• Mode: Half duplex, multi-drop
• Connector: Screw terminal
• Transmission Speed: Up to 115.2 Kbps
• Max. Transmission Distance: 4000 feet (1220 m)

#### 5.3.2.4 Power
• Unregulated 10 to 30VDC
• Protection: Over-voltage and power reversal

#### 5.3.2.5 Isolation
• Ethernet Communication: 1500 V DC
• I/O Module: 3000 V DC

#### 5.3.2.6 Mechanical
• Case: KJW with captive mounting hardware
• Plug-in Screw Terminal Block: Accepts 0.5 mm 2 to 2.5 mm 2, 1 - #12 or 2 - #14 to #22 AWG

#### 5.3.2.7 Environment
• Operating Temperature: - 10 to 70º C (14 to 158º F)
• Storage Temperature: - 25 to 85º C (-13 to 185º F)
• Humidity: 5 to 95%, non-condensing
• Atmosphere: No corrosive gases

## 5.3.3 Dimensions
### 5.3.3.1 Basic functional block diagram

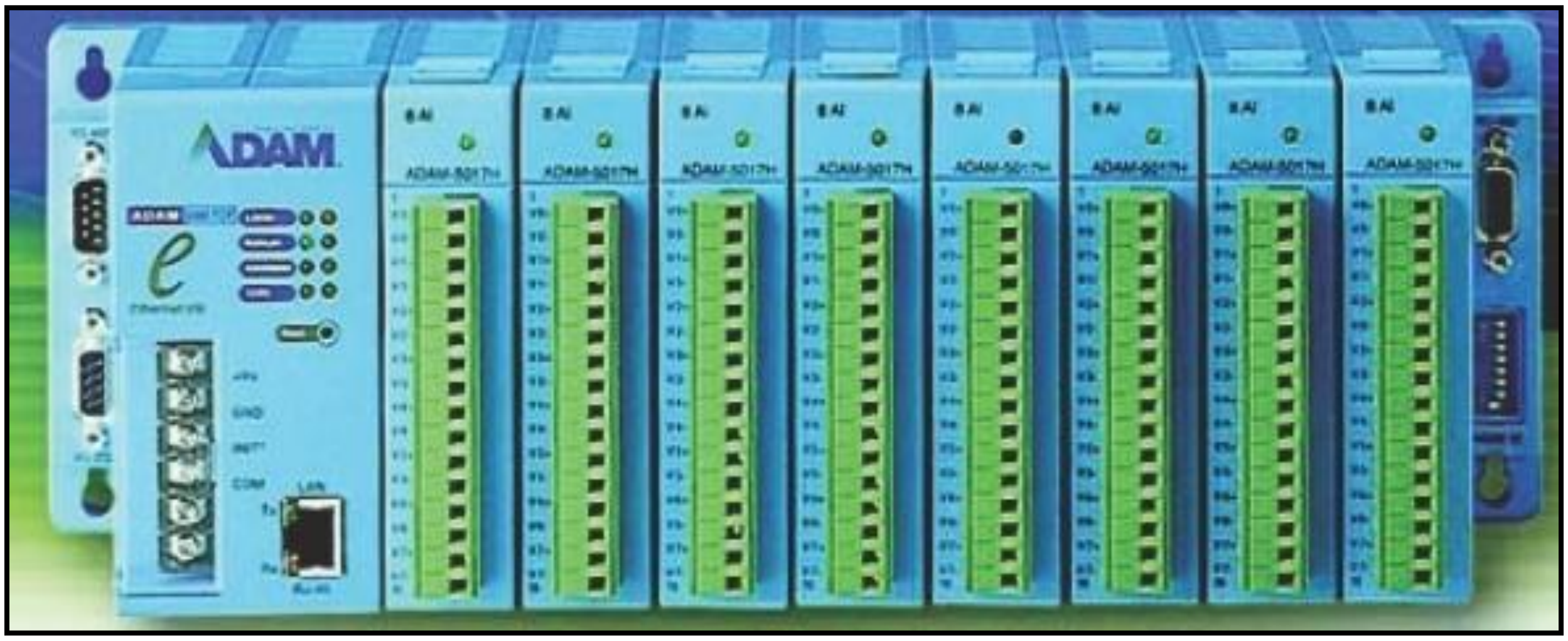

**Figure 25 Functional Block Diagram**

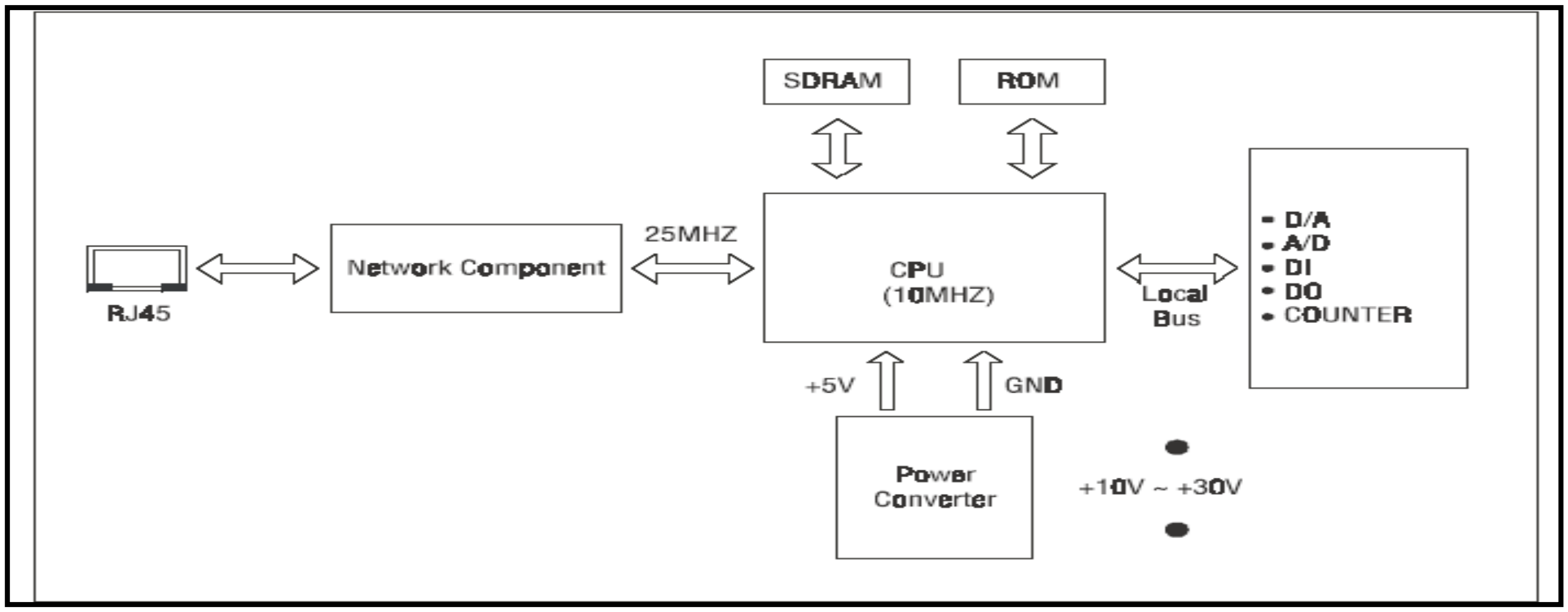

**Figure 26 Block Diagram**

# 5.3 Required Adam Module for the Proposed System

## 8 Channel Analog Input Module ADAM-5017

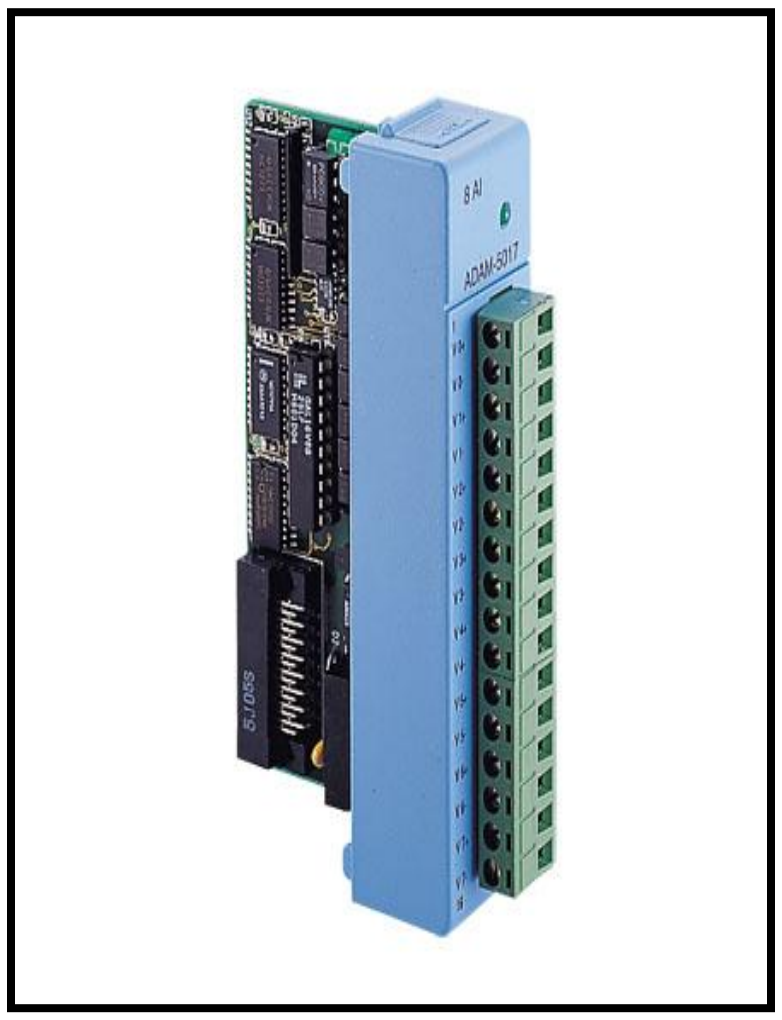

**Figure 27 ADAM-5017 Module**

The ADAM-5017 is an 8-channel analog differential input module that provides programmable input ranges on all channels. It accepts millivolt inputs (±150mV, ±500 mV), voltage inputs (±1V, ±5V, ±10V) and current input (20mA, requires 125Ω resistors).

Specifications:

| | |
|---|---|
| Input Range | ±150mV,±500mV,±1V,±5V,±10V,±20mA |
| Input Impedance | 2MΩ |
| Accuracy | ±0.1% or better |
| Power Requirement | +10 to +30 VDC (non regulated) |
| Power Consumption | 1.2 W |
| Sampling Rate | 10 samples/sec (total) |

# 5.4 Computer/PC

In our project we need to deal with only one process variable (pressure) i.e. one analog input and one digital output. So use of PLC (Programmable Logic Controller) is discarded. Moreover it is relatively expensive. Another option of implementing microprocessor based control was eliminated as it employs long and tedious programming with the complex circuit components. Henceforth, we adopted computer based control algorithm. This was not only cost effective but also optimized the process by enabling easy programming using LABVIEW software. Below we list the requisite specifications of the computer which we required for the process. Any regular computers which are used in small offices can be used. But as the saying goes, the more the better so a computer with heavier configuration than provided would always be better in terms of the number of clock cycles per second, better processing speed and higher data storage capacity etc.

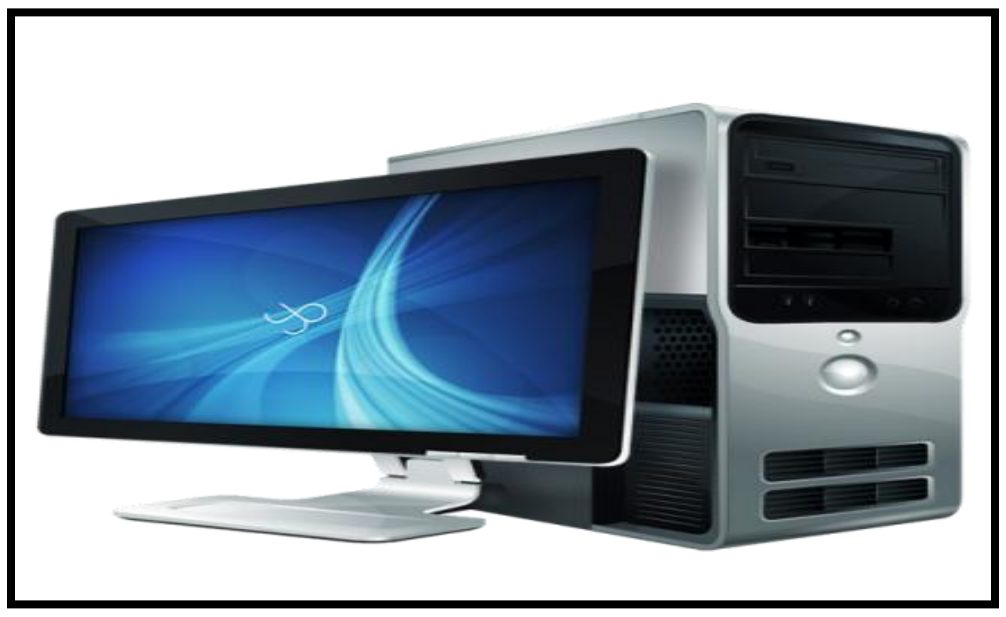

Figure 28 PC

Specifications:

| | |
|---|---|
| Processor | Intel Core i5 |
| Motherboard | Intel Desktop Motherboard DG41WV |
| Hard Drive | 500GB |
| RAM | 4GB |
| Cache Memory | 4MB |
| Operating System | Windows XP/7 |

## 5.5 Traco Power Supply

This power supply is a special kind of switching power supply available in different ranges of 12V DC, 24V DC. It is used here to give power to ADAM module.

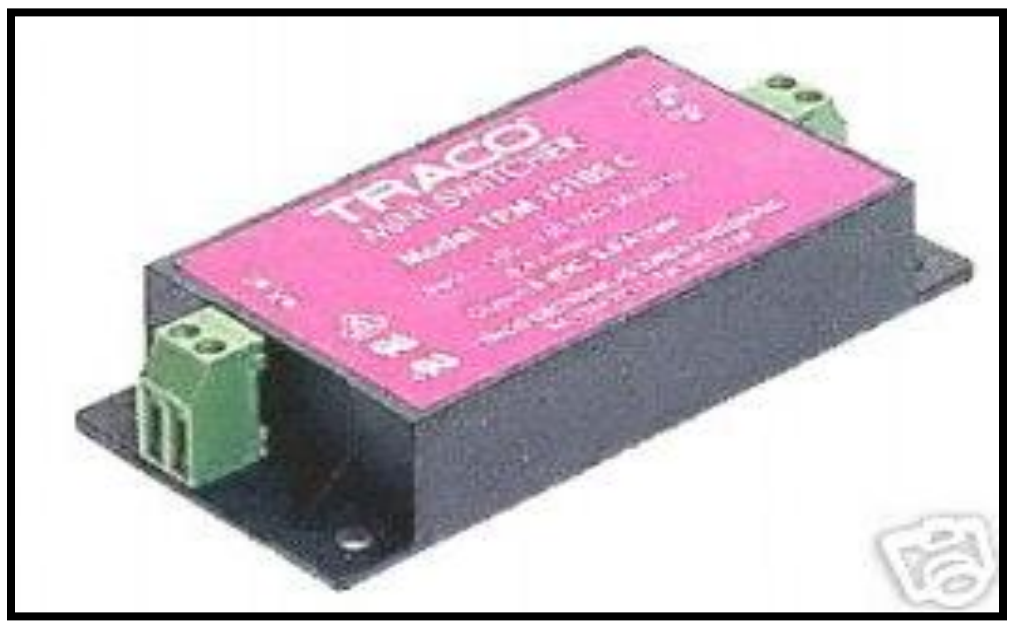

Figure 29 Power Supply

Specifications:

| | |
|---|---|
| Input voltage range | 100-240 VAC |
| Input frequency | 50/60 Hz |
| DC output voltage | 24 V |
| Max Current output | 3A |

# 6

# Software Overview

Without the software, any hardware is just a piece of junk. So in this section we are going to give a brief overview of the software we had used in this project i.e.NI LabVIEW.

## 6.1 Dataflow Programming

In computer programming, dataflow programming is a programming paradigm that models a program as a directed graph of the data flowing between operations, thus implementing dataflow principles and architecture. Dataflow programming languages share some features of functional languages, and were generally developed in order to bring some functional concepts to a language more suitable for numeric processing.

### 6.1.1 Properties of dataflow programming languages

Dataflow programming focuses on how things connect, unlike imperative programming, which focuses on how things happen. In imperative programming a program is modeled as a series of operations (things that "happen"), the flow of data between these operations is of secondary concern to the behavior of the operations themselves. However, dataflow programming models programs as a series of (sometimes interdependent) connections, with the operations between these connections being of secondary importance.

One of the key concepts in computer programming is the idea of "state", essentially a snapshot of the measure of various conditions in the system. Most programming languages require a considerable amount of state information in order to operate properly, information which is generally hidden from the programmer. For a real world example, consider a three-way light switch. Typically a switch turns on a light by moving it to the "on" position, but in a three-way case that may turn the light back off — the result is based on the state of the other switch, which is likely out of view.

In fact, the state is often hidden from the computer itself as well, which normally has no idea that this piece of information encodes state, while that is temporary and will soon be discarded. This is a serious problem, as the state information needs to be shared across multiple processors in parallel processing machines. Without knowing which state is important and which isn't, most languages force the programmer to add a considerable amount of extra code to indicate which data and parts of the code are important in this respect.

This code tends to be both expensive in terms of performance, as well as difficult to debug and often downright ugly; most programmers simply ignore the problem. Those that cannot must pay a heavy performance cost, which is paid even in the most common case when the program runs on one processor. Explicit parallelism is one of the main reasons for the poor performance of Enterprise Java Beans when building data-intensive, non-OLTP applications. Dataflow languages promote the data to become the main concept behind any program. It may be considered odd that this is not always the case, as programs generally take in data, process it, and then feed it back out. This was especially true of older programs, and is well represented in the Unix operating system which pipes the data between small single-purpose tools. Programs in a dataflow language start with an input, perhaps the command line parameters, and illustrate how that data is used and modified. The data is now explicit, often illustrated physically on the screen as a line or pipe showing where the information flows.

Operations consist of "black boxes" with inputs and outputs, all of which are always explicitly defined. They run as soon as all of their inputs become valid, as opposed to when the program encounters them. Whereas a traditional program essentially consists of a series of statements saying "do this, now do this", a dataflow program is more like a series of workers on an assembly line, who will do their assigned task as soon as the materials arrive. This is why dataflow languages are inherently parallel; the operations have no hidden state to keep track of, and the operations are all "ready" at the same time.

Dataflow programs are generally represented very differently inside the computer as well. A traditional program is just what it seems, a series of instructions that run one after the other. A dataflow program might be implemented as a hash table instead, with uniquely identified inputs as the keys, used to look up pointers to the instructions. When any operation completes, the program scans down the list of operations until it finds the first operation where all of the inputs are currently valid, and runs it. When that operation finishes it will typically put data into one or more outputs, thereby making some other operation become valid.

For parallel operation only the list needs to be shared; the list itself is the state of the entire program. Thus the task of maintaining state is removed from the programmer and given to the language's runtime instead. On machines with a single processor core where an implementation designed for parallel operation would simply introduce overhead, this overhead can be removed completely by using a different runtime. There are many hardware architectures oriented toward the efficient implementation of dataflow programming models. MIT's tagged token dataflow architecture was designed by Greg Papadopoulos. Data flow has also been proposed as an abstraction for specifying the global behavior of distributed system components: in the live distributed objects programming model, distributed data flows are used to store and communicate state, and as such, they play the role analogous to variables, fields, and parameters in Java-like programming languages.

## 6.2 Visual Programming Language

In computing, a visual programming language (VPL) is any programming language that lets users create programs by manipulating program elements graphically rather than by specifying them textually. A VPL allows programming with visual expressions, spatial arrangements of text and graphic symbols used either as elements of syntax or secondary notation. For example, many VPLs (known as dataflow or diagrammatic programming) are based on the idea of "boxes and arrows", where boxes or other screen objects are treated as entities, connected by arrows, lines or arcs which represent relations.

VPLs may be further classified, according to the type and extent of visual expression used, into icon-based languages, form-based languages, and diagram languages. Visual programming environments provide graphical or iconic elements which can be manipulated by users in an interactive way according to some specific spatial grammar for program construction.
A visually transformed language is a non-visual language with a superimposed visual representation. Naturally visual languages have an inherent visual expression for which there is no obvious textual equivalent. Current developments try to integrate the visual programming approach with dataflow programming languages to either have immediate access to the program state resulting in online debugging or automatic program generation and documentation (i.e. visual paradigm). Dataflow languages also allow automatic parallelization, which is likely to become one of the greatest programming challenges of the future.
An instructive counterexample for visual programming languages is the Microsoft Visual Studio. The languages it encompasses (Visual Basic, Visual C#, Visual J#, etc.) are commonly confused to be but are not visual programming languages. All of these languages are textual and not graphical. MS Visual Studio is a visual programming environment, but not a visual programming language, hence the confusion.

## 6.3 NI LabVIEW

LabVIEW (short for Laboratory Virtual Instrument Engineering Workbench) is a system-design platform and development environment for a visual programming language from National Instruments.
The graphical language is named "G" (not to be confused with G-code). Originally released for the Apple Macintosh in 1986, LabVIEW is commonly used for data acquisition, instrument control, and industrial automation on a variety of platforms including Microsoft Windows, various versions of UNIX, Linux, and Mac OS X. The latest version of LabVIEW is LabVIEW 2012, released in August 2012.

### 6.3.1 Dataflow programming

The programming language used in LabVIEW, also referred to as G, is a dataflow programming language. Execution is determined by the structure of a graphical block diagram (the LV-source code) on which the programmer connects different function-nodes by drawing wires. These wires propagate variables and any node can execute as soon as all its input data become available. Since this might be the case for multiple nodes simultaneously, G is inherently capable of parallel execution. Multi-processing and multi-threading hardware is automatically exploited by the built-in scheduler, which multiplexes multiple OS threads over the nodes ready for executions.

### 6.3.2 Graphical programming

LabVIEW ties the creation of user interfaces (called front panels) into the development cycle. LabVIEW programs/subroutines are called virtual instruments (VIs). Each VI has three components: a block diagram, a front panel and a connector panel. The last is used to represent the VI in the block diagrams of other, calling VIs. The front panel is built using controls and indicators. Controls are inputs – they allow a user to supply information to the VI. Indicators are outputs – they indicate, or display, the results based on the inputs given to the VI. The back panel, which is a block diagram, contains the graphical source code. All of the objects placed on the front panel will appear on the back panel as terminals. The back panel also contains structures and functions which perform operations on controls and supply data to indicators. The structures and functions are found on the Functions palette and can be placed on the back panel.

Collectively controls, indicators, structures and functions will be referred to as nodes. Nodes are connected to one another using wires – e.g. two controls and an indicator can be wired to the addition function so that the indicator displays the sum of the two controls. Thus a virtual instrument can either be run as a program, with the front panel serving as a user interface, or, when dropped as a node onto the block diagram, the front panel defines the inputs and outputs for the given node through the connector panel. This implies each VI can be easily tested before being embedded as a subroutine into a larger program.
The graphical approach also allows non-programmers to build programs by dragging and dropping virtual representations of lab equipment with which they are already familiar. The LabVIEW programming environment, with the included examples and documentation, makes it simple to create small applications. This is a benefit on one side, but there is also a certain danger of underestimating the expertise needed for high-quality G programming. For complex algorithms or large-scale code, it is important that the programmer possesses an extensive knowledge of the special LabVIEW syntax and the topology of its memory management. The most advanced LabVIEW development systems offer the possibility of building stand-alone applications. Furthermore, it is possible to create distributed applications, which communicate by a client/server scheme, and are therefore easier to implement due to the inherently parallel nature of G. The image above is an illustration of a simple LabVIEW program showing the dataflow source code in the form of the block diagram in the lower left frame and the input and output variables as graphical objects in the upper right frame. The two are the essential components of a LabVIEW program referred to as a Virtual Instrument.

## 6.3.3 Benefits

### 6.3.3.1 Interfacing

A key benefit of LabVIEW over other development environments is the extensive support for accessing instrumentation hardware. Drivers and abstraction layers for many different types of instruments and buses are included or are available for inclusion. These present themselves as graphical nodes. The abstraction layers offer standard software interfaces to communicate with hardware devices. The provided driver interfaces save program development time. The sales pitch of National Instruments is, therefore, that even people with limited coding experience can write programs and deploy test solutions in a reduced time frame when compared to more conventional or competing systems. A new hardware driver topology (DAQmxBase), which consists mainly of G-coded components with only a few register calls through NI Measurement Hardware DDK (Driver Development Kit) functions, provides platform independent hardware access to numerous data acquisition and instrumentation devices. The DAQmxBase driver is available for LabVIEW on Windows, Mac OS X and Linux platforms. Although not a .NET language, LabVIEW also offers an interface to .NET Framework assemblies, which makes it possible to use, for instance, databases and XML files in automation projects.

### 6.3.3.2 Code compilation

In terms of performance, LabVIEW includes a compiler that produces native code for the CPU platform. The graphical code is translated into executable machine code by interpreting the syntax and by compilation. The LabVIEW syntax is strictly enforced during the editing process and compiled into the executable machine code when requested to run or upon saving. In the latter case, the executable and the source code are merged into a single file. The executable runs with the help of the LabVIEW run-time engine, which contains some precompiled code to perform common tasks that are defined by the G language. The run-time engine reduces compile time and also provides a consistent interface to various operating systems, graphic systems, hardware components, etc. The run-time environment makes the code portable across platforms. Generally, LabVIEW code can be slower than equivalent compiled C code, although the differences often lie more with program optimization than inherent execution speed.

#### 6.3.3.3 Large libraries

Many libraries with a large number of functions for data acquisition, signal generation, mathematics, statistics, signal conditioning, analysis, etc., along with numerous graphical interface elements are provided in several LabVIEW package options. The number of advanced mathematic blocks for functions such as integration, filters, and other specialized capabilities usually associated with data capture from hardware sensors is immense. In addition, LabVIEW includes a text-based programming component called MathScript with additional functionality for signal processing, analysis and mathematics. MathScript can be integrated with graphical programming using "script nodes" and uses a syntax that is generally compatible with MATLAB [citation needed].

#### 6.3.3.4 Code Re-use

The fully modular character of LabVIEW code allows code reuse without modifications: as long as the data types of input and output are consistent, two sub VIs are interchangeable. The LabVIEW Professional Development System allows creating stand-alone executables and the resultant executable can be distributed an unlimited number of times. The run-time engine and its libraries can be provided freely along with the executable. A benefit of the LabVIEW environment is the platform independent nature of the G code, which is (with the exception of a few platform-specific functions) portable between the different LabVIEW systems for different operating systems (Windows, Mac OS X and Linux). National Instruments is increasingly focusing on the capability of deploying LabVIEW code onto an increasing number of targets including devices like Phar Lap or VxWorks OS based LabVIEW Real-Time controllers, FPGAs, PocketPCs, PDAs, Wireless sensor network nodes, and even Lego Mindstorms NXT.

#### 6.3.3.5 Parallel programming

With LabVIEW it is very easy to program different tasks that are performed in parallel by means of multithreading. This is, for instance, easily done by drawing two or more parallel while loops. This is a great benefit for test system automation, where it is common practice to run processes like test sequencing, data recording, and hardware interfacing in parallel.

#### 6.3.3.6 Ecosystem

Due to the longevity and popularity of the LabVIEW language, and the ability for users to extend the functionality, a large ecosystem of 3rd party add-ons has developed through contributions from the community. This ecosystem is available on the LabVIEW Tools Network, and is a marketplace for both free and paid LabVIEW add-ons.

#### 6.3.3.7 User community

There is a low-cost LabVIEW Student Edition aimed at educational institutions for learning purposes. There is also an active community of LabVIEW users who communicate through several e-mail groups and Internet forums.

# 7

# Hardware Design

In this chapter we are going to discuss the hardware setup of our vacuum measurement system in full detail. We will first state our aim and objectives of the project and then we would introduce our proposed model along with the necessary details required to set up the measurement system.

## 7.1 Objectives of the Project:

- Measuring pressure of a multiple-gauge vacuum system.
- Study & develop a suitable scheme to transmit obtained signals from field instruments to control room.
- Study and develop a suitable graphical user interface (GUI) to display transmitted signals as a process variable and record the data obtained after specified time intervals.
- Study and develop a generalized test set up for pressure measurement with our proposed scheme.
- Date acquisition, handling and preservation.
- Develop a system which can be controlled from a remote location.
- Study and develop automation of the whole process.

## 7.2 Proposed Model:

Any basic process control loop looks like:-

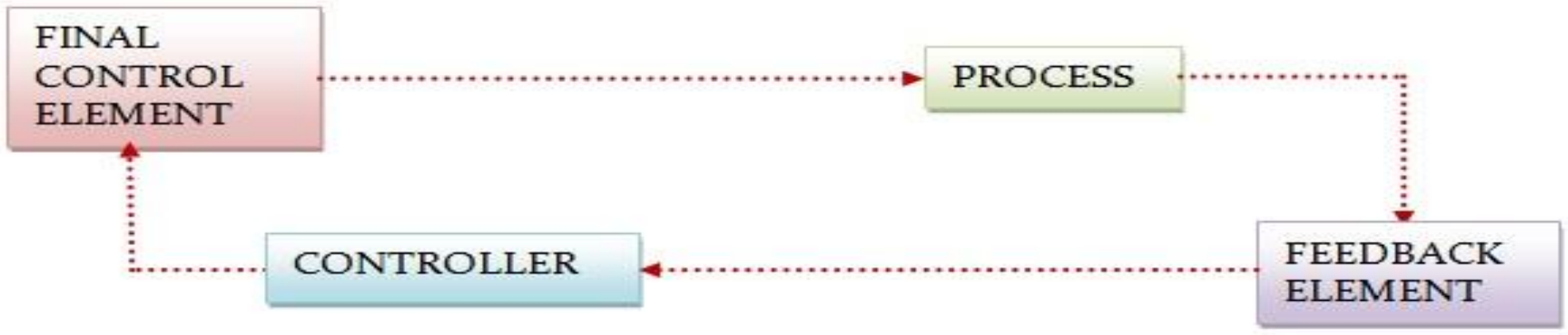

**Figure 30 Feedback control System**

In this project, vacuum measurement part was taken as the main objective of the scheme. Though the project is actually a part of a vacuum control system, the control part was also studied but could not be fully executed due to the time constraint. Finally the project emerged as a generalized setup for testing of various vacuum systems from a remote location. Here, we firstly require a pressure measuring element to obtain the instantaneous value of the pressure. The output of the measuring element is fed into the controller. The controller conditions this electrical signal to a level of 0-10 volts. This signal is then taken up by the ADAM module. It then transmits this data to the control room via the TCP/Modbus protocol. In this project, computer was used for the graphical display of the data. The programming of the control algorithm was done using LABVIEW and control action was generated.

The following diagram illustrates this:

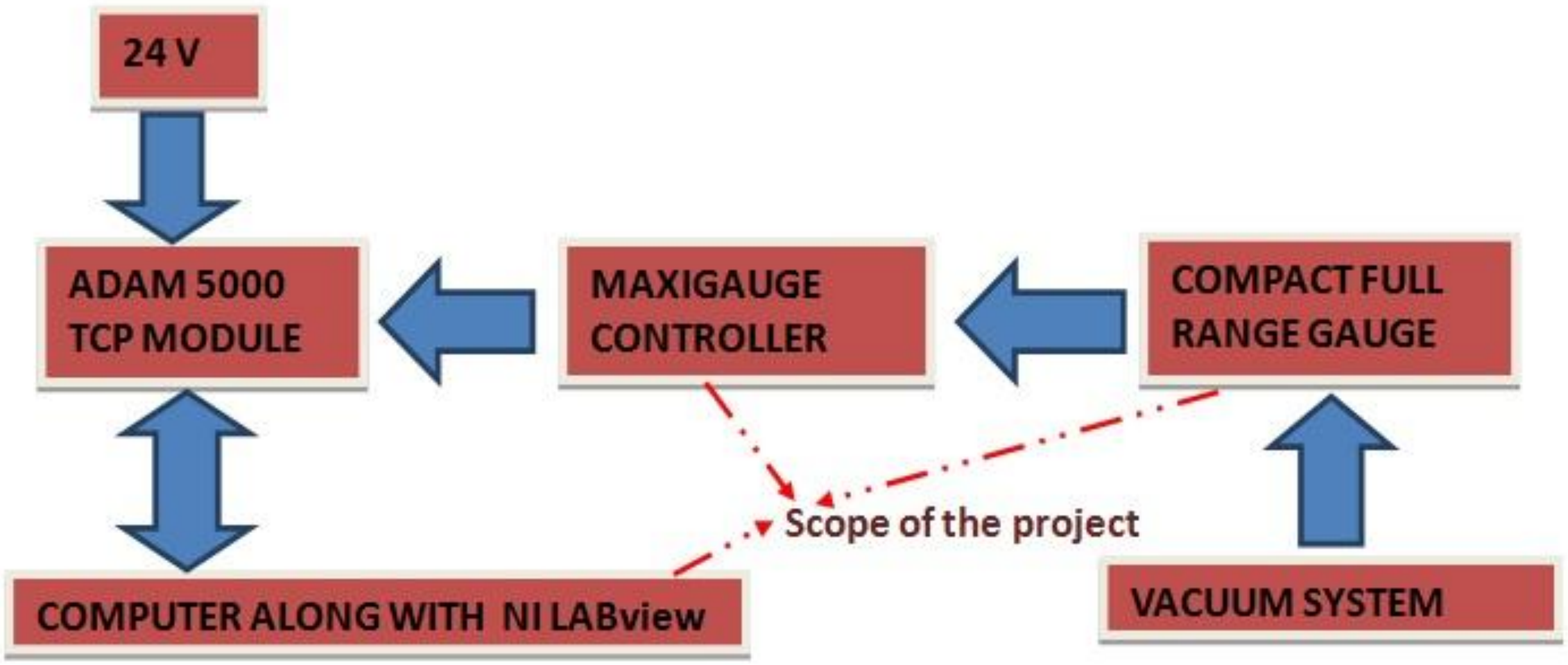

## 7.3 Setup

Pirani gauges were used for measurement of vacuum in our vacuum measurement system. A figure illustrating the setup is given below:

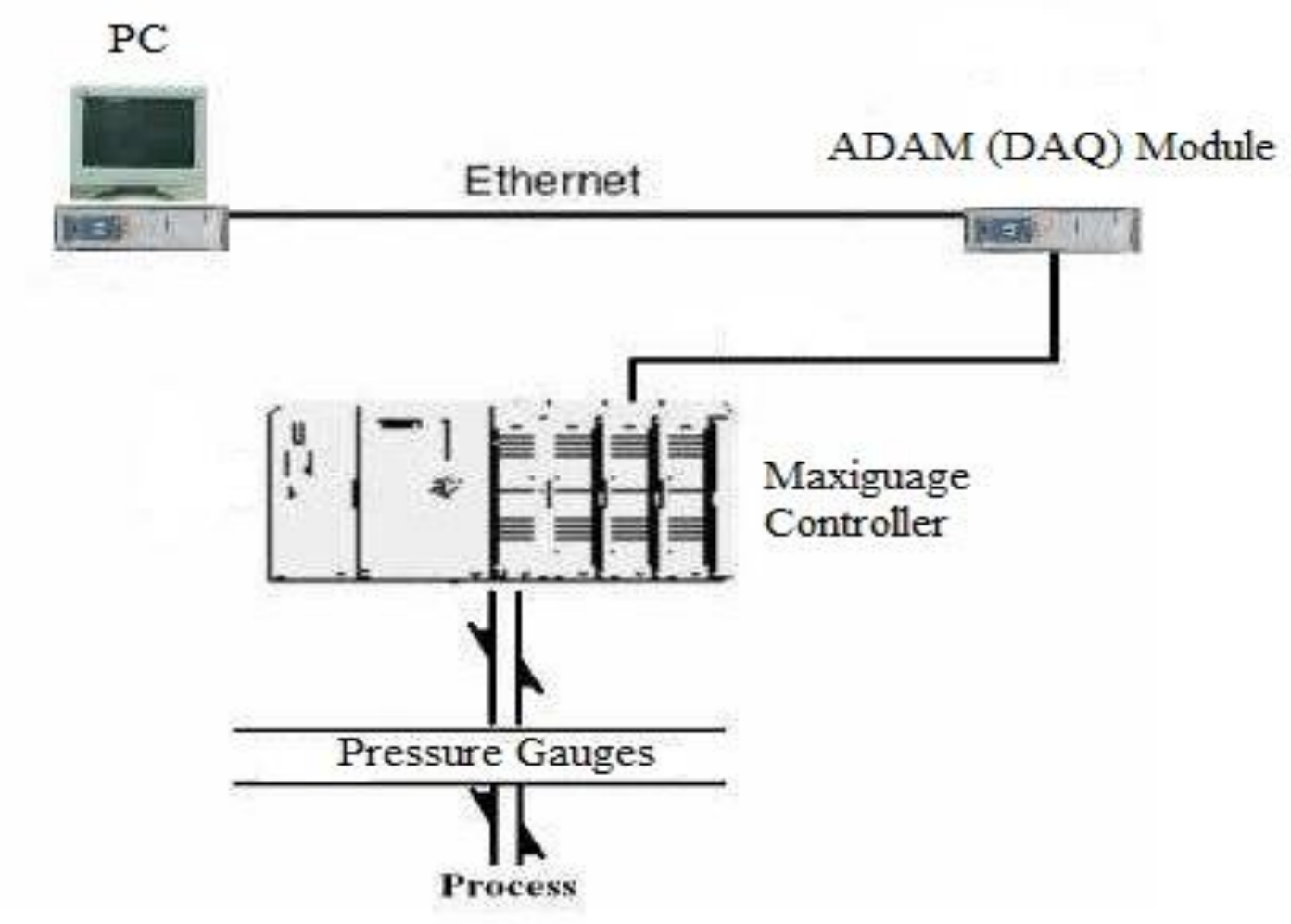

**Figure 31 Functional Setup Scheme**

As seen here, we feed the output of the measuring element i.e. the Pirani gauges into the Maxigauge controller. The controller then takes up the analog signal and conditions this electrical signal to a level of 0-10 volts. This output conditioned signal is then taken up by the ADAM module which converts this analog signal into a suitable digital signal which is suitable for transmission over the ethernet lines to the control room via the TCP/Modbus protocol. From the communication point of view; the Pirani gauge, Maxigauge controller, ADAM Module would resemble the transmitter section. The transmitted signal is then received at the receiving end by a computer which is essentially the receiver section. In this project, computer was used for the graphical display of the data as well as data logging purposes. The programming of the control algorithm was done using LABVIEW and control action was generated.

## ▪ Interfacing of the Circuit

Here in the following diagram we would see how the Pirani gauge is actually connected to the Maxiguage controller which in turn is connected to the ADAM Module from where the ethernet lines originate:

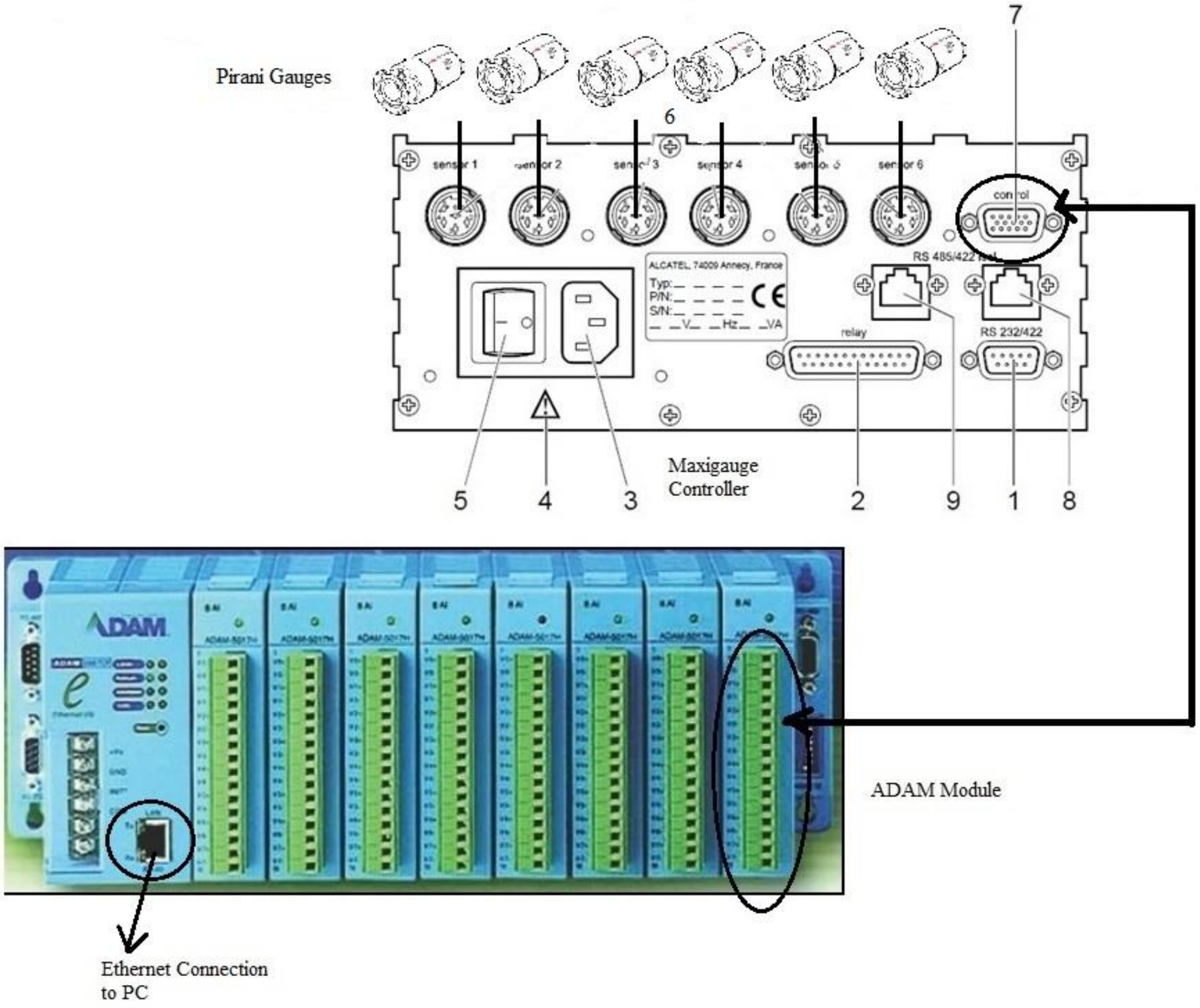

**Figure 32 Interfacing with different Module**

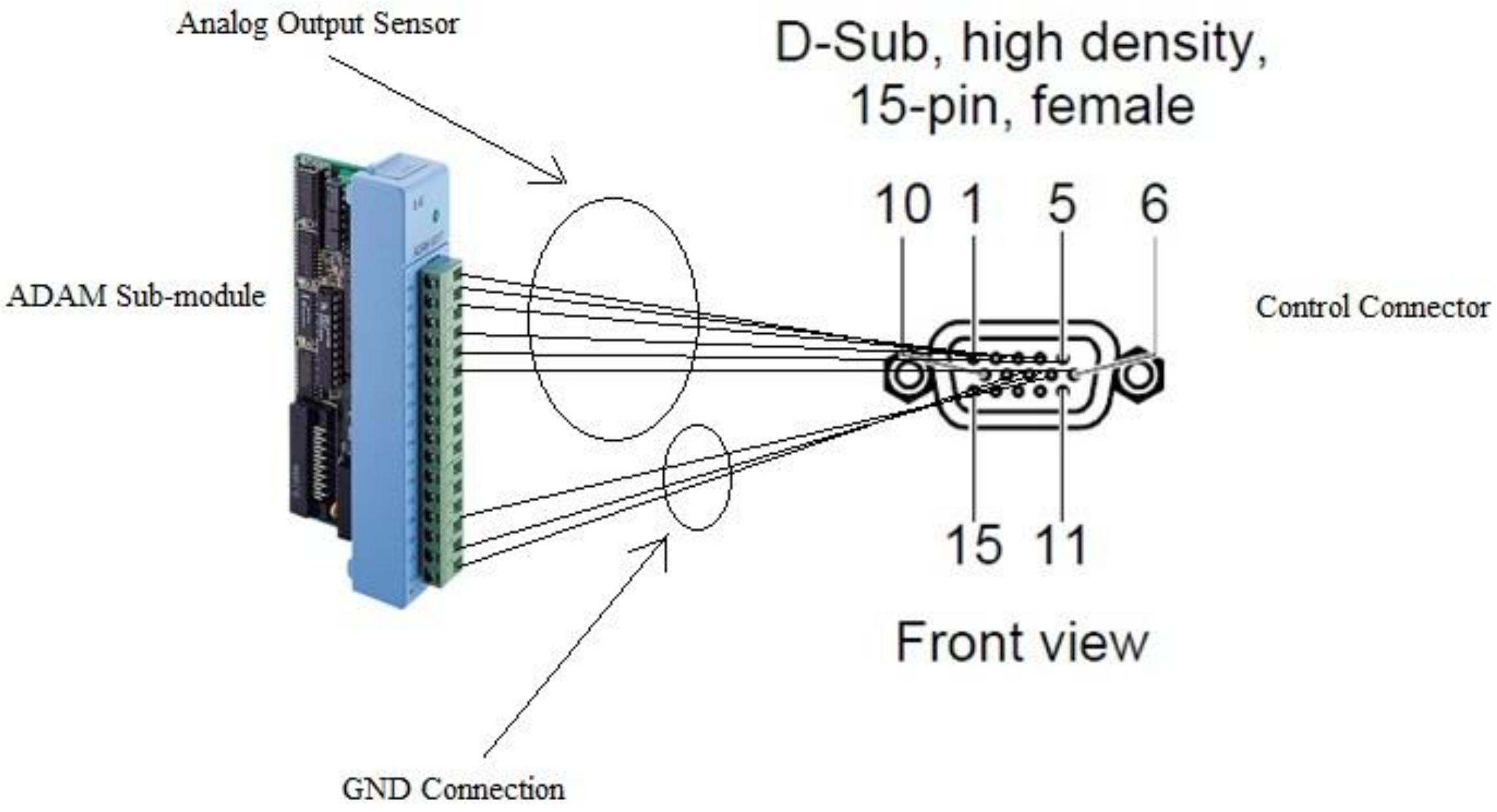

**Figure 33 Connection between ADAM 5017 and Maxiguage Controller**

## 7.3.1 Pin Specifications of Maxiguage Controller

TPG 256 A part number PT G28 760:

| | | |
|---|---|---|
| 1 | *RS 232/422* | Pinout for serial interface RS232C or RS422 (not isolated) |
| 2 | *relay* | Connector for relay switch contacts |
| 3 | | Power inlet 3-pin |
| 4 | ⚠ | Reference for fuses inside the unit (replacement only by Pfeiffer Vacuum Service) |
| 5 | | Power switch |
| 6 | *sensor 1 ... ... sensor 6* | Connectors for gauges |
| 7 | *control* | Connector for control functions |

TPG 256 A part number PT G28 761, additional features:

| | | |
|---|---|---|
| 8,9 | *RS 485/422 isol.* | Port for serial interface RS485 (addressable, isolated) and RS422 (isolated) |

### 7.3.2 Pin Specifications of Control Connector

Pin assignment

| Pin | Assignment |
|---|---|
| 1 | Analog output sensor 1 |
| 2 | Analog output sensor 2 |
| 3 | Analog output sensor 3 |
| 4 | Analog output sensor 4 |
| 5 | Analog output sensor 5 |
| 6 | Analog output sensor 6 |
| 7 | GND |
| 8 | GND |
| 9 | GND |
| 10 | External control sensor 1 |
| 11 | External control sensor 2 |
| 12 | External control sensor 3 |
| 13 | External control sensor 4 |
| 14 | External control sensor 5 |
| 15 | External control sensor 6 |

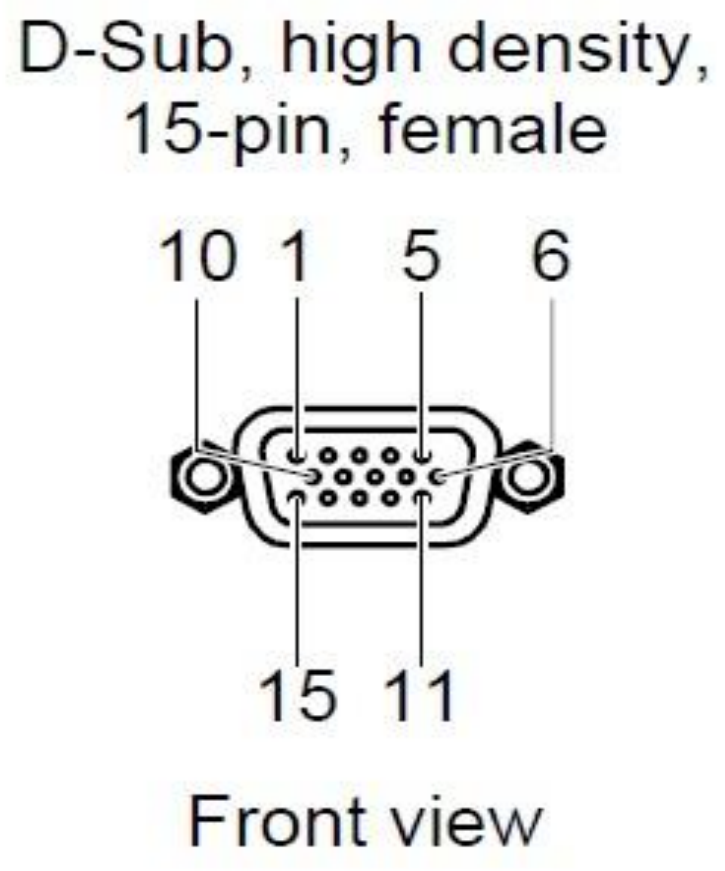

## 7.4 Performance

The system has been tested in the Laboratory. The critical response time was 150-200ms when the Client runs on the same Server machine. However, the performance degrades to 0.5 to 1 Sec, depending on the network traffic over the LAN.

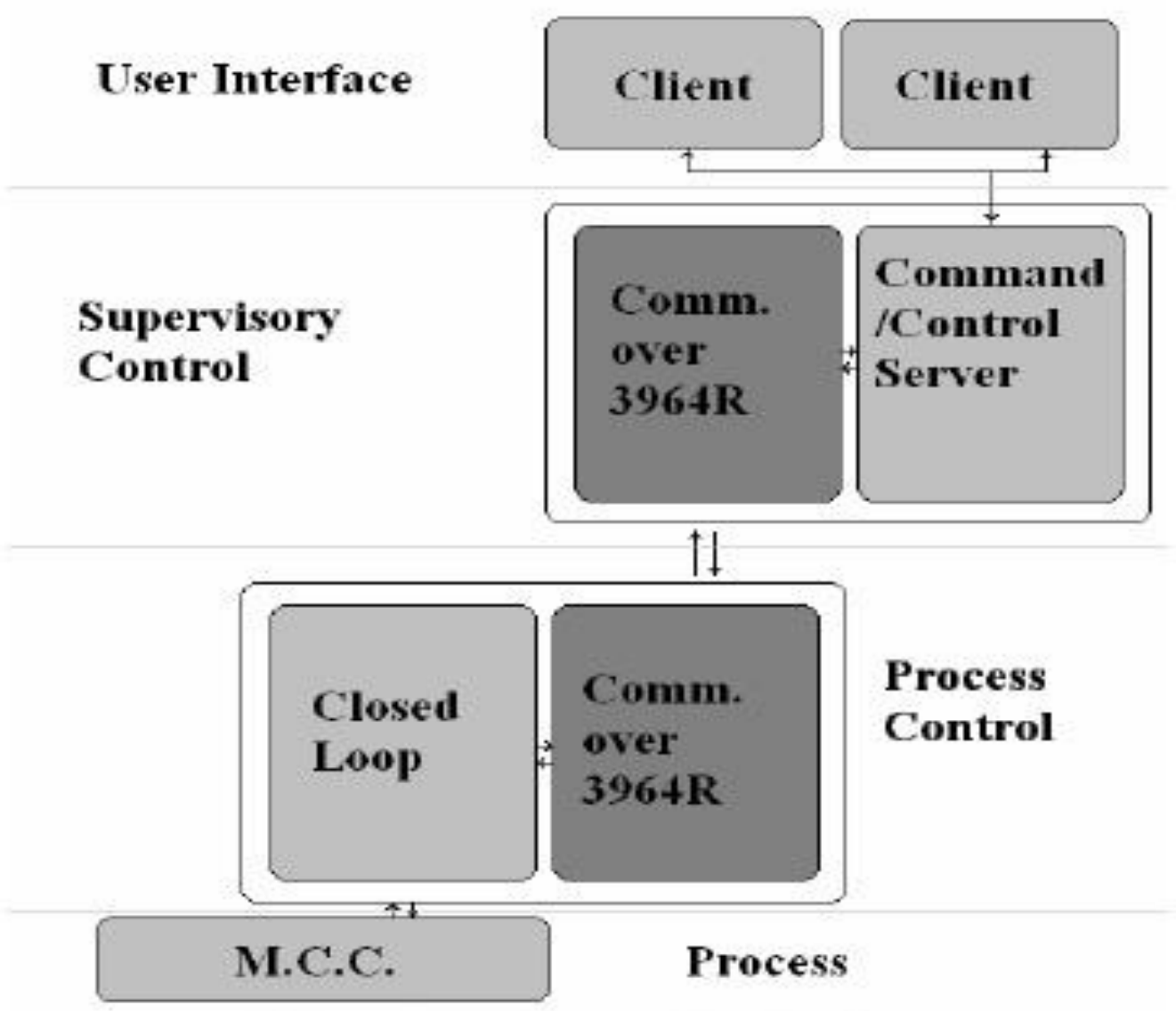

**Figure 34 Hardware-Software Compatibility**

So with the above illustrations we have shown how we have configured our vacuum measurement system. Now we are ready with our hardware! The last thing that remains now is the software implementation. We are going to discuss that in detail in the next chapter.

# 8

# Software design

At last, we come to the main part of the whole project. Being from a non software oriented background this was hardest part of the project for us. In this chapter first we are going to give an overview of our approach to the software design and follow it up with the necessary code and HMI interface details using NI LabVIEW.

## 8.1 Overview

Our first challenge to designing a HMI in LabVIEW was to solve the problem associated with the format in which data was received from the ADAM module as well as communicating with it. If you have gone through chapter 3 (*Communication Protocols*) you would know that the data is sent and received in Hexadecimal format. Now the main problem with that is we cannot work with hexadecimal numbers. So we need to figure out some way in which we can convert the hexadecimal format into decimal form. Fortunately NI LabVIEW allows for creation of user defined sub VIs, the use of which was necessary as you would come to know in a while. We searched through the whole LabVIEW database to find a suitable block which would solve this problem but could not find one. At last we ourselves created a sub VI 'HEX Convertor' and embedded that in our main project VI.

So the software design runs likes this, first the command is issued to the TCP/Modbus protocol to send the "fetch" data command to the ADAM module which sends the data. From then on, first data is taken up from the ethernet line and is directly given to our HMI real time. The hexadecimal number which is the sensors output (all combined) is first broken down into the required number of individual parts and is converted into decimal form with the help of our sub VI. It is then compared with the threshold voltage level which we had set. If that value is above the preset threshold then data is logged onto the database otherwise we map the threshold value in the database indicating that the value is limited to the threshold.

## 8.2 Sub VI

In this section, we are going to discuss the sub VI which we created. Now let us try to analytically think how the conversion of hexadecimal can be done into decimal form. First we store the hexadecimal number in an array and separate out the number in two parts, we multiply the first part by 256 and the second part by 1.Then we add the two parts and subtract 32,770 (half of 65,536) and then divide it by 65,536; lastly we multiply it by 20 and put in the formula for volt to vacuum conversion to get the final vacuum value. The above procedure is done because as we divide the two parts and multiplying one by 256 (2^8=256) and the other by one, we get our decimal number back. Similarly, now to map number to voltage we make use of the following linear analysis:

| Y (Volt) | X (Number Input) |
|---|---|
| -10V | 0 |
| 0 | 32,770 |
| 10V | 65,536 |

From the straight line equation,

$$\frac{Y-0}{0-10} = \frac{x-32,770}{32,770-65,365}$$

this simplifies to

$$Y = 10(\frac{X}{32,770} - 1)$$

Thus, we a voltage value which is converted into a vacuum value by the following equation,

$$Z = 10^{((1.667*y)-11.33)}$$

### 8.2.1 Front Panel of the Sub VI

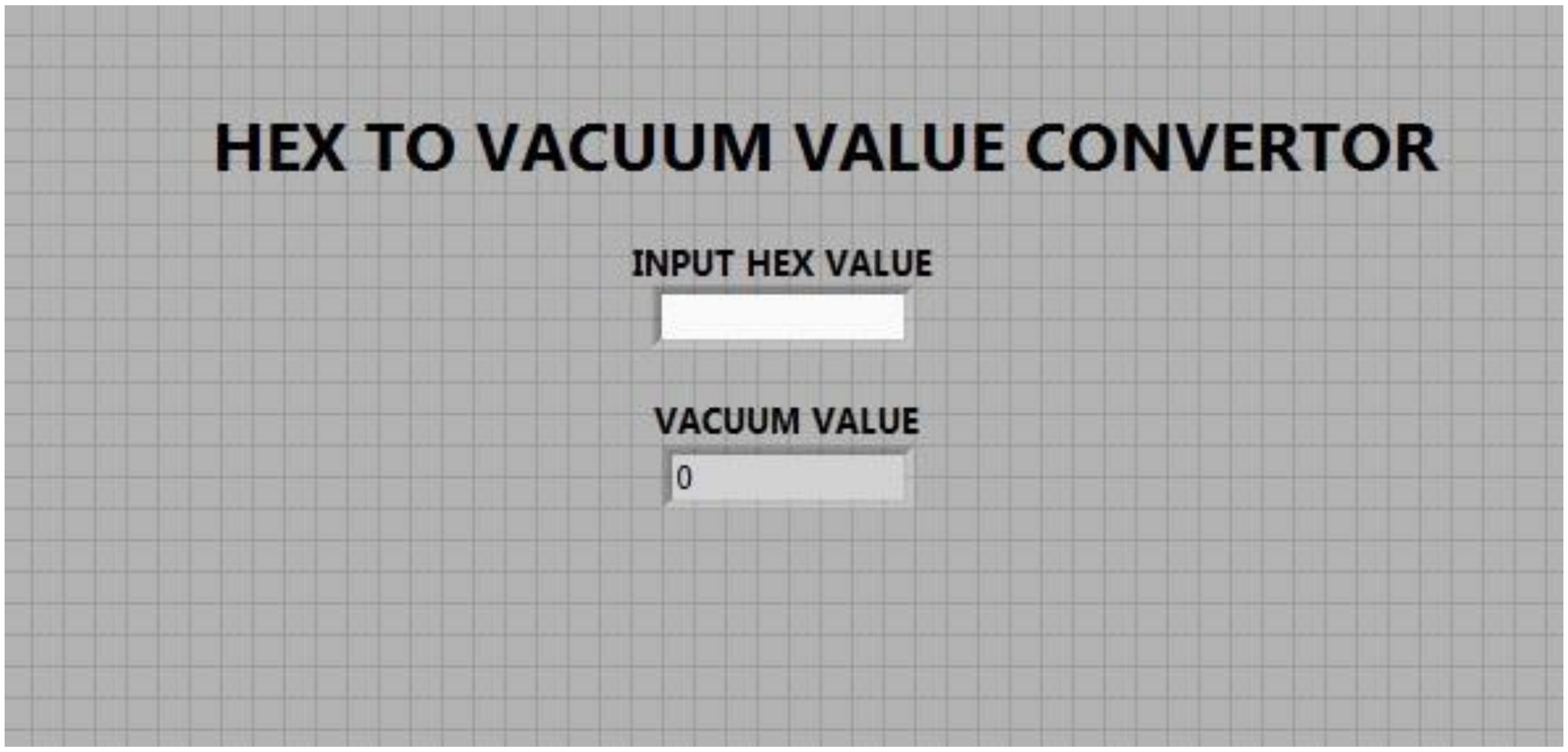

Figure 35 Front Panel of Sub VI

## 8.2.2 Block Diagram of Sub VI

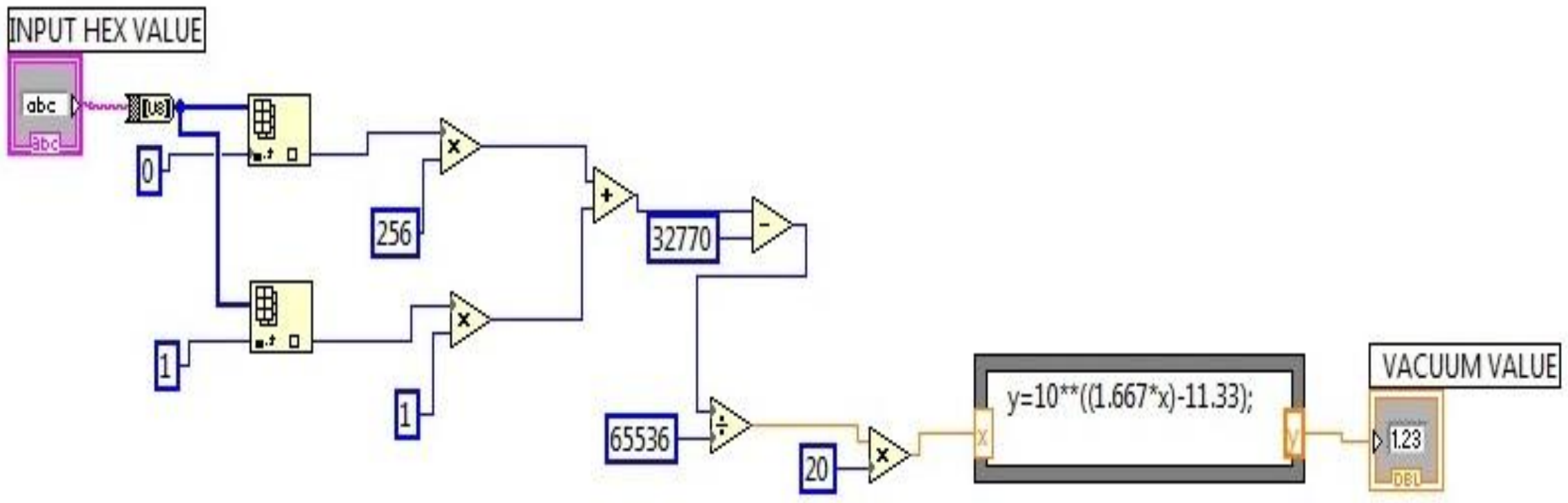

Figure 36 Block Diagram of Sub VI

## 8.3 Main VI (HMI)

In the main project VI, first the command is issued to the TCP/Modbus protocol to send the "fetch" data command to the ADAM module which sends the data. From then on, first data is taken from the ethernet line and is directly given to our HMI real time. The hexadecimal code which consists of the sensors output (all combined) is first broken down into the required number of individual parts and is converted into decimal form with the help of our sub VI. It is then compared with the threshold voltage level which we had set. If that value is above the preset threshold then data is logged onto the database otherwise we map the threshold value in the database indicating that the value is limited to the threshold.

Please note the following very carefully:

**Local IP of the host machine: 172.16.4.156**
**Timeout Interval: 1000**
**Modbus command to fetch data: 0000 0000 0006 0104 0000 0006**

Here, the IP address and the timeout interval will vary depending on where we are implementing the software. However the Modbus command would remain the same.

0000 0000 0006 -> Number of sensors to read.
0104 -> Read input coil status and the corresponding input register.
0006-> Number of data segments to read.

### 8.3.1 Front panel of HMI

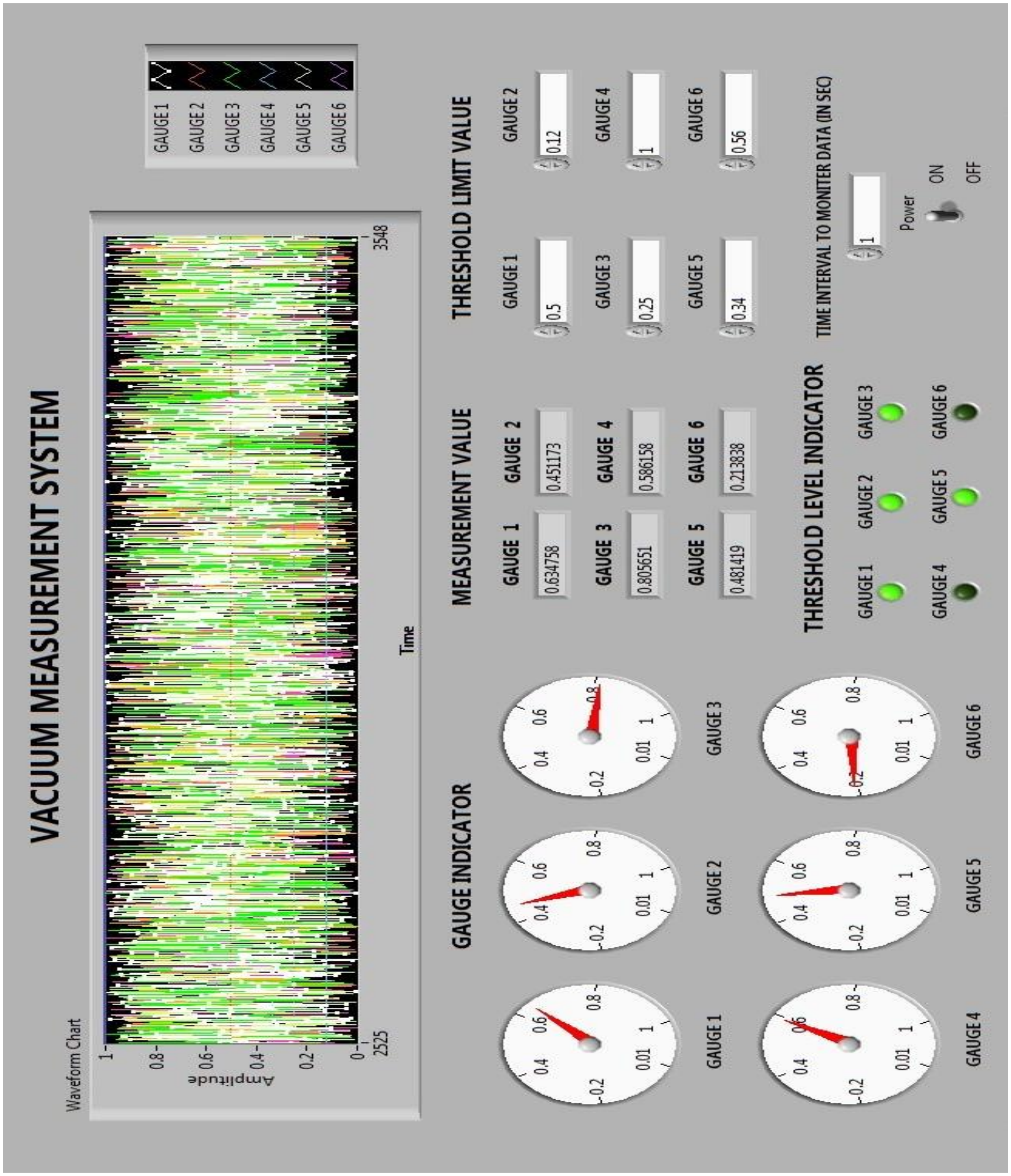

Figure 37 Front panel of HMI

# 8.3.2 Block Diagram of HMI

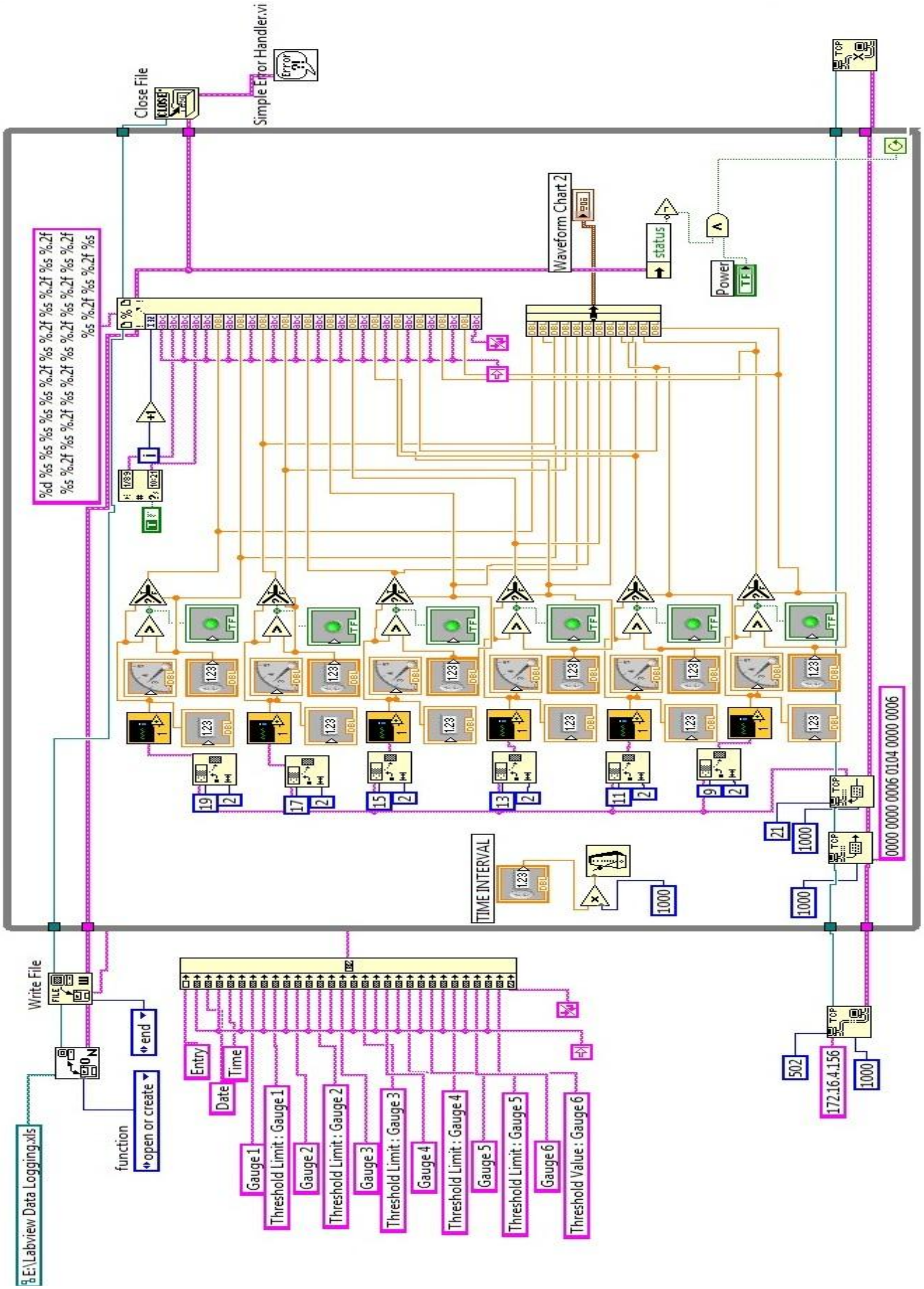

Figure 38 Block Diagram of HMI

# 9

# Overview & Discussion

In this section we are going to discuss some of the answers to the questions which might have crossed your minds while going through the previous chapters. We would also be discussing the advantages of our proposed system and provide insights on further improvements which could be made to the system towards the end of the chapter.

Well, the three most important questions would be:

1. Why was there a need to have a DAQ implementation when we could have implemented the following scheme using a PLC?
2. Why was Ethernet based TCP/Modbus protocol chosen when the following could be implemented using only RS232 port?
3. Why was NI LabVIEW chosen over so many of application oriented Software making tools?

## 9.1 Answer to Question 1:

Well, let's proceed to answer the first question which was raised in our discussion. We agree to the fact that a PLC could be used but at the same time we can argue that a PLC implementation in the following case would not be feasible due to the following reasons:

- ✓ PLC was not used because of its high price.
- ✓ It is suitable for large number of I/O and complex logic but in this case we are only required to work with only one parameter i.e. vacuum.
- ✓ Microprocessor and microcontroller is complex to interface with.
- ✓ Interfacing is more time consuming for development.
- ✓ Complex implementation paradigm.

So, we could definitely see that we are at a great disadvantage if we go for a PLC implementation in this case. Rather a DAQ implementation is more feasible. ADAM Module, a well known DAQ module was used which had the following advantages:

- ✓ Here the parameters to be controlled were only pressure so ADAM was used to avoid complexity with reduced instruction set in MODBUS protocol.
- ✓ Easy to interface with the Modbus protocol scheme.
- ✓ Use of ADAM module allows us to install it directly in the field and then connect with the computer through LAN connection. It can be set up anywhere and data can be fetched and transmitted from any place.
- ✓ Basically it can be physically and functionally distributed and governed by supervisory software (GUI).
- ✓ ADAM module can also be used to account for upgradation capabilities.
- ✓ Any change in the number of inputs and outputs of the system can be easily dealt with in the ADAM module.
- ✓ It is one of the cheapest solutions too.
- ✓ Supports a wide variety of communication protocols.

Unlike the difficulties we would have faced in implementing a PLC, use of the ADAM DAQ module saved us from unnecessary circuit complexity and deployment hazards. Because of this we choose to go with a DAQ implementation.

## 9.2 Answer to Question 2:

The answer to this is also quite application specific. Quite a valid question really! If we go with the convention of the previous solution where we implemented DAQ because it was easy to work with, it would be suicidal to go for Ethernet Protocol when RS232 would have easily done the job for us. We agree with all of you who would have reasoned it out like that. But let us see why we decided to go with the more complex Ethernet protocol.

Firstly, we had been asked to implement a vacuum measurement system. Now, a vacuum measurement system consists of a gas chamber to which the pressure gauges are attached. The gases inside the chamber are actually radioactive gases used in nuclear experiments. So to explain in simple terms monitoring could not be done very close to the vacuum measurement chamber. But the apparatus required for measurement i.e. the Maxiguage Controller have to be kept very close to the vacuum gauges. Since monitoring requires a human user and exposure to radioactive emissions is harmful for them, we needed to look for ways in which our objective of remote monitoring could be achieved at the same time also preserving the health safety of the human operator.

Secondly, RS232 could be extended to a maximum of 10m after which the received signal considerably degrades linearly as the function of the distance. But here we are looking for a range of at least 40-500m.So we needed to look for something else. In this range, the most suitable is the Ethernet communication protocol i.e. TCP/Modbus. Using a single cable the signal can be made to travel long distances without much degradation in the signal quality.

Thirdly, one could argue with the use of Wi-fi for the purpose. But again interfacing Wi-fi with an ADAM module is a bit complex and Wi-fi also suffers from the problem of reduced data rates and security problems. The data rates offered by a Wi-fi network is at the most 540Mbps which is quite small as compared to the 10Gbps offered by the Ethernet.

The following constraints made us resort to a rather complex (compared to RS232) ethernet communication protocol.

### 9.3 Answer to Question 3:

This is one such question the answer to which fully depends on the persons who are actually handling the project. Their software skill levels along with their requisite software language skill sets. So the answer to this question may or may not satisfy the reader. But we will try to justify as much as we can. Being from an Electronics & Communication engineering background, we do not have much of computer language skills except just some preliminary basics. So making full fledged software by us is not only a formidable task but also an impossible one. But still we managed to build one Human –Machine Interface with the help of LabVIEW. The choice behind using LabVIEW was purely a choice based on the amount of time we were given, what resources we got and whatever software skills we had. Readers can also argue using other tools like Microsoft Visual Studio, Visual Basic, and Matlab etc for developing a HMI. Their argument in this case is quite valid also but given our software skill background we found NI LabVIEW to be quite easy to grasp and implement.

NI LabVIEW is a graphical programming language and we can't argue with the fact that graphical images are more easy to grasp than writing plain text of code. In this area LabVIEW won it with us. Otherwise there was no or would be no difficulty in using other development platforms.

The answers we provided to the main questions which could have been raised in your mind hopefully will satisfy you. For the last question, the choice fully depends on the persons' skill set associated with the project

## 9.4 Advantages of Our Proposed Model

The following could be stated as the advantage of our proposed model for vacuum measurement of radioactive gases:

1. Easy to implement as no complexity is involved which are generally associated with PLC implementations, which work with a lot parameters simultaneously.
2. Easy to interface with available hardware.
3. Can be used for data logging on an ordinary computer which is a huge advantage.
4. The system can be remotely controlled from practically anywhere in the world where ethernet lines can reach.
5. The HMI is suited to also control the Maxiguage controller remotely.
6. Signal conditioning is improved with the use of DAQ module.
7. HMI is open source, hence can be tweaked to add more features.
8. The DAQ module provides extra slots unlike PLC, hence adding or removing a number of vacuum gauges is easy.

## 9.5 Future Scope of the Project

Due to the time constraints there were some things which could not be taken care of, hence it would be best to include them in this section:

1. The HMI lacks the facility to indicate whether a gauge is connected or not. This becomes pretty difficult to tell as when we set the threshold to some value, the system would indicate 0 value if vacuum is under the threshold (if designed in such a way) but at the same time it would also indicate 0 if no gauge is connected. So definitely this creates an ambiguity as to when the value is 0 we do not know due to which of the following reason, the 0 is shown. This needs to be taken care of.
2. The HMI also suffers from the drawback of indicating the preset threshold when the vacuum value is under threshold. But in essence we require the value to be retained to the previous value which was above threshold.
3. Another major drawback is that we cannot remotely put the system in ON or OFF mode.
4. During data logging if the measurement file is open, the data cannot be logged i.e. real time data updation is not quite yet fully implemented. This needs to be worked on.
5. The HMI needs to have good security encryption mechanisms so that unauthorized access of the data can be stopped. This part also needs to be looked into.

# 10

# Conclusion

The duration of the project was limited to about a month. If it had not been so, this project could have been further developed. But nevertheless I had learnt to work in a competitive and realistic environment. I am really thankful to everyone for this opportunity which had been given to me. It has been a privilege to be a part of the ongoing curriculum of VECC. The one month tenure has helped me enrich my engineering knowledge. I had learnt how to attack and solve a customized and unique problem and also how to approach to its solution in stipulated time. I had started right from scratch and primarily constructed a block diagram of the project. Then I started to give attention to very minute details and select each and every component related to my project. I had to gather idea about actual hardware and software and their selection criteria. This was the time when I got the real taste of engineering in R&D sector. Some of the theoretical concepts were changed and showed me how to tackle real life problems in engineering. This one month period was, indeed, a great learning experience and I gave our best effort to exploit this opportunity to the best, but there is still scope of development and improvement on this project which can be carried out further. Top of all I had a rare experience of working with a real system which is in use now and will be used in the centre in years to come.

# APPENDIX A:
# TROUBLESHOOTING

## 1. No change in pressure reading:

- ✓ Check the mains voltage(Recommended:230VAC, 50Hz)
- ✓ Check the LAN connection.
- ✓ Check that the gauges are properly connected.
- ✓ Reset the program.

## 2. Power supply output is not proportional:

- ✓ Check ADAM analog output
- ✓ Check supply of 230V to power supply
- ✓ Check local mode calibration

## 3. No communication between ADAM & Computer:

- ✓ Check if IP conflicts
- ✓ Check physical cable communication
- ✓ Check the different LAN points

## 4. Fluctuations in readings:

- ✓ Check the connections of the ADAM module.
- ✓ Check the LAN connections.

## 5. Other precautionary measures:

- ✓ The ADAM should not be supplied with more than 24 V DC.

# APPENDIX B:
# LIST OF TABLES

**TABLE** **PAGE NO.**

# APPENDIX C:
# LIST OF FIGURES

**FIGURES** **PAGE NO.**